\documentclass[11pt,a4paper]{article}

\usepackage[
            top=1.0in,
            bottom=1.0in,
            left=1.0in,
            right=1.0in
            ]{geometry}
            
\usepackage[utf8]{inputenc}

\usepackage[natbib=true,
		style=ieee,
		sorting=none,
		citestyle=numeric-comp,
		]{biblatex}
	
\usepackage{subcaption}
\usepackage{longtable}
\usepackage{graphicx}
\usepackage{booktabs}
\usepackage{amsmath}
\usepackage{authblk}
\usepackage{float}

\title{Machine Learning Metallic Glass Critical Cooling Rates Through Elemental and Molecular Simulation Based Featurization}

\author[a*]{Lane E. Schultz}
\author[a]{Benjamin Afflerbach}
\author[a]{Paul M. Voyles}
\author[a]{Dane Morgan}

\affil[a]{University of Wisconsin-Madison, 1500 engineering Drive, Madison, WI 53706, USA}
\affil[*]{Corresponding author. E-mail address: lsschultz@wisc.edu (L. E. Schultz)}

\date{\today}

\begin{document}

\maketitle

\begin{abstract}

We have developed a machine learning model for critical cooling rates for metallic glasses based on computational properties. We compare results for features derived from easy-to-compute functions of elemental properties to more complex physically motivated properties using ab initio, machine-learning potentials, and empirical potential molecular dynamics methods. The established approach enables property acquisition across a diverse range of alloys. Analysis of various features for 34 alloys from 20 chemical systems shows that the best model for critical cooling rates was learned from one elemental property-based feature and three simulated features. The elemental property-based feature is an ideal entropy value based on alloy stoichiometry. The simulated features were acquired from estimates of energies above the convex hull, changes in heat capacity, and the fraction of icosahedra-like Voronoi polyhedra. Models were assessed through a demanding cross validation test based on repeatedly leaving out full chemical systems as test sets and had an $R^2$ of 0.78 and a mean average error of 0.76 in units of $[log_{10}(K/s)]$. We demonstrate with Shapley additive explanation analysis that the most impactful features have physically reasonable influence on model predictions. The established methodology can be applied to other high-throughput studies of material properties of diverse compositions.

\end{abstract}
\section{Introduction}
\label{introduction}

Unlike their crystalline counterparts, metallic glasses are a type of metallic material characterized by a lack of long-range atomic order. Metallic glasses can have exceptional hardness, increased strength, superior corrosion resistance, smaller magnetic hysteresis, and higher resistivity when compared to their crystalline counterparts. These properties have led to a range of applications, including structural supports within the human body \cite{Li2016, Jafary-Zadeh2018}, voltage transformers \cite{Herzer2013}, and golf clubs \cite{Wang2004}. Each of these applications depend on the composition of the material in question and the ability for that composition to form a glass instead of a crystal during processing.

One significant obstacle to the more widespread adoption of metallic glasses for engineering applications is their size limitation, which is primarily due to the greater thermodynamic stability of their crystalline counterparts. When metallic glasses are manufactured through melt-quench methods, heat must be quickly dissipated from samples or else crystalline phases are formed, which limits the maximum size the quenched samples can reach and remain a glass. The slowest rate at which a molten material can be cooled to produce a glass is called the critical cooling rate (denoted by $R_{c}$) and is an intrinsic measure of glass forming ability (GFA). The size of a glass that can be produced is closely related to $R_{c}$. This size is typically quantified by the maximum potential length of the shortest sample section of a produced glass before nucleation of crystalline phases occurs (often represented as $Z_{max}$). If the geometry is cylindrical, then the diameter, $D_{max}$, can be referenced specifically. Any length above $Z_{max}$ will lead to the emergence of crystalline phases. $Z_{max}$ and $D_{max}$ are also widely used as measures of GFA, although they can depend on the quenching method and are not as intrinsic to the material as $R_{c}$.

Many previous efforts to quantify GFA relied on arithmetic operations of the glass transition ($T_{g}$), crystallization ($T_{x}$), and the liquidus ($T_{l}$) temperatures \cite{Long2009, Schultz2021}. A significant limitation of these methods is the prerequisite formation of a glass to measure $T_g$ and $T_x$. Nonetheless, models derived from these temperatures can offer valuable insights into the potential size of metallic glass samples, once they are established as glass formers. The rheology of a material, typically characterized by viscosity or relaxation time, is suggested by other studies to play a pivotal role in explaining GFA \cite{Johnson2016, Dai2019-2, Jaiswal2016}. Relaxation time and viscosity are frequently used interchangeably because they are directly proportional to each other \cite{Gao2020}. As a material cools from a molten state and transitions into a glass, atoms with increasingly restrained movement, reflected by higher viscosity at lower temperatures, tend to remain amorphous. The change in viscosity (or relaxation time) with respect to temperature when the material is near $T_{g}$ is called fragility, $m$, \cite{Angell1995}. The conventional understanding is that high $m$ indicates poor glass formers, while low $m$ suggests favorable glass formers. There are alternate properties that correlate with $m$, as highlighted in the work by Gangopadhyay et al. \cite{Gangopadhyay2017}. Two noteworthy properties include a cross-over from Arrhenius behavior temperature during cooling to non-Arrhenius behavior, $T_{a}$, and the temperature at which materials reach a chosen viscosity cutoff, $T_{c}$, as described in Ref.~\cite{Gangopadhyay2017}. Ratios of the aforementioned temperatures and $T_{g}$ were shown to have significant correlations with $m$. However, $m$ alone does not describe GFA. As discussed in Ref.~\cite{Kube2022}, $m$ sometimes exhibits the opposite relationship with GFA than anticipated. In the studied $MgCuY$ system, many compositions displayed high GFA despite having high $m$ values. Additionally, Ref.~\cite{Johnson2016} uses $T_{g}$, $T_{l}$, and $m$ to predict $D_{max}$ because none of these properties cannot robustly predict $D_{max}$ alone. Together, however, these properties produced a model for $log_{10}(D_{max}^{2})$ with a coefficient of determination of $R^{2}=0.980$. While these correlations are often extremely good, any correlation based on experimentally measured values has a severe limitation of needing to make and characterize the material to predict $R_{c}$. Several efforts have been made to correlate GFA to descriptors that can be trivially computed from properties of elements or properties that can be computed from molecular simulations.

Among these properties are those based on pre-tabulated values of simple alloying properties. These properties are nearly instantaneous to evaluate and are very easy to use. Some authors have combined experimentally produced physical descriptors with easy to compute elemental properties (i.e., properties built from chemical information of each element). In Ref.~\cite{Long2023}, the experimental descriptors of $T_{g}$, $T_{x}$, and $T_{l}$ were combined with elemental features to produce a model for $D_{max}$ with a score of $R^{2}=0.763$. While this model shows promise, the use of experimental characteristic temperatures introduces the limitations noted above. More general models that use only composition as features to predict GFA have been produced. A notable example is from Ward et al. \cite{Ward2018} who built a $D_{max}$ model with a correlation coefficient (not $R^{2}$) of $R=0.89$ when assessing their model through 10-fold cross validation (CV). Their mean average error ($MAE$) was 0.21 $mm$. However, 10-fold CV produces an overly confident assessment of model performance due to similar compositions appearing in both the training and test sets. When the authors assess their model by iteratively excluding binary systems from training, the $MAE$ rises to 0.81 $mm$ for tests (a 286\% increase). A MAE of 0.81 $mm$ accounts for 68\% of the mean absolute deviation of test sets, so there is considerable room for improvement. Afflerbach et al., combined experimental data containing $Z_{max}$, $D_{max}$, $R_{c}$, $T_{g}$, $T_{x}$, $T_{l}$, and melt-spun ribbon information to produce approximate $R_{c}$ values (denoted as $R_{a}$ here) for 2,125 materials \cite{Afflerbach2022}. When manufactured, ribbons can form fully amorphous, partially amorphous, or fully crystalline samples. The phases present in a ribbon are indicative of the quality of the GFA. The authors assigned $R_c$ values of $10^{5.5}$ and $10^7$ $[K/s]$ for partially amorphous and fully crystalline ribbons, respectively. However, fully amorphous sheets were not included in their analysis. Through systematically excluding complete chemical systems as test sets (i.e., materials were grouped by chemical systems and iteratively left out as the test set), they were able to produce random forest regression models of $log_{10}(R_{a})$ with a $MAE$ of 0.82 $[log_{10}(K/s)]$. A lower $MAE$ of 0.36 $[log_{10}(K/s)]$ was obtained from a 5-fold CV assessment because it is a less demanding test. When applying 5-fold CV, many of the test compositions may be similar to those included in the training set (e.g., $Zr_{49}Cu_{51}$ in training and $Zr_{50}Cu_{50}$ in test). The features used to build the model were from elemental properties generated with the Materials Simulation Toolkit for Machine Learning (MAST-ML) \cite{Jacobs2020}. The studies in Refs.~\cite{Ward2018} and \cite{Afflerbach2022} highlight the challenge of GFA features being applicable across chemical systems, as model predictions deteriorate markedly for compositions significantly different from those used in training. As demonstrated in Ref.~\cite{Liu2023}, models constructed using physically motivated features exhibit greater extrapolation capabilities compared to models built with features acquired through numerical brute force. The authors examined models trained separately on compositions, elemental features, and three physically motivated features to predict whether a composition is likely to produce a large metallic glass. When assessed through 10-fold CV, all models showed similar prediction ability. When assessed with chemistries more different from those used for training models, they find that models built from physically motivated features correctly classified 20 times more materials of interest than models built from elemental features. Models built from compositions alone identified none. This result leads us to heavily emphasize the physical relevance of features used to build models in this present work.

The computational approaches mentioned previously either rely on pre-existing or easy to generate data for constructing models. More sophisticated computational methods exist for extracting physical properties from materials and present an opportunity to enhance GFA prediction. From a structural standpoint, exploring the short-range order of glasses through Voronoi polyhedra (VPs) could provide valuable insights into their GFA. To construct VPs bisecting planes are positioned between an atom and its nearest neighbors. The type of VP is determined by the number of faces and their corresponding edges that enclose the atom. The work in Ref.~\cite{Wang2019} argues that the variance of VPs from a high temperature liquid (at 1,478 $[K]$ and above $T_{l}$) can be used as an indicator of GFA. Follow up work in Ref.~\cite{Weeks2022} uses clustering techniques on VPs. Studies have also focused on VPs resembling icosahedra (ICO-like, characterized by having at least 10 faces with 5 edges each \cite{Bokas2016}), demonstrating that ICO-like VPs exhibit slow movement during cooling, consequently enhancing GFA \cite{Bokas2016, Bokas2018}. However, none of these works were assessed on a large database of GFA for their ability to provide accurate regression or classification predictions.

Structural features in the form of VPs, among many others, were studied in Ref.~\cite{Afflerbach2021} in a regression task. Molecular dynamics (MD) was used to measure $R_{c}$ by studying crystallization rates, and a set of simulated material properties was used to predict $log_{10}(R_{c})$. Elemental features were also included from the elemental properties generated with MAST-ML. While differences exist between experimentally acquired $R_{c}$ and its simulated counterpart (in particular, the simulated values were limited by computational constraints to above $10^{11}$ $[K/s]$), the results appear promising given that the model validation yielded an $R^{2}$ of 0.769. Features from VPs, rheology, and elemental properties were found to significantly impact the model. In a similar spirit to Ref.~\cite{Afflerbach2021}, characteristic temperatures obtained through rheology, temperature, energies, and phase diagrams were used to construct regression and classification models within a limited composition space \cite{Schultz2022}. Regression models yielded unsatisfactory results for $D_{max}$ prediction, while classification metrics for high or low $D_{max}$ were more promising. Both Ref. \cite{Afflerbach2021} and \cite{Schultz2022} relied on classical interatomic potentials, which are difficult and time-consuming to construct. The need to have acceptably accurate potentials available for each system being studied greatly limited the chemical space accessible to the studies. Ab initio methods can be used for many chemical compositions, but several of the studied properties were impractical to obtain through ab initio methods because of the computational expense associated with the necessary system sizes and simulation time scales. In this work, we take advantage of recent developments in machine learning potentials (MLPs), which allow for the rapid construction of accurate potentials. These MLPs were used to study 34 systems with diverse chemical compositions. Employing MLPs constructed from ab initio data with machine learning can enable practical simulations at the necessary length and time with near ab initio accuracy \cite{Zuo2020, Novikov2021}.

Here, we integrate the best practices of previous efforts to quantify GFA and use them to guide developing descriptors from material properties that are computationally accessible. We extend the previous chemical domain of properties that can be obtained with MLPs. Then we evaluate the ability of these descriptors to predict $R_c$. The properties we use include those that are trivial to compute but also ones derived from simulations of rheological behavior, energy differences across phases, and many other physically motivated aspects. We obtain encouraging results, with a cross validation test based on repeatedly leaving out full chemical systems yielding an $R^2$ of 0.78 and $MAE$ of 0.76 in units of $[log_{10}(K/s)]$. We use SHapley Additive exPlanations (SHAP) values to extract correlations of $R_c$ with physically motivated features and show that these correlations align with physical expectations. SHAP values quantify the contribution of each feature to a machine learning model's prediction, based on game theory principles \cite{shap}.
\section{Methods}
\label{methods}

\subsection{Software}
\label{software}

The MLPs in this study were constructed using the Machine-Learning Interatomic Potentials 2 (MLIP-2) package, which fits a moment tensor potential (MTP) \cite{Novikov2021}. For all other machine learning tasks, MAST-ML was employed along with scikit-learn \cite{Jacobs2020, scikit-learn}. Ab initio calculations were carried out using the Vienna Ab initio Simulation Package (VASP) \cite{VASP} and classical MD simulations utilized the Large-scale Atomic/Molecular Massively Parallel Simulator (LAMMPS) \cite{LAMMPS}. OVITO was used for the calculation of VP types through its python application program interface (API) \cite{Stukowski2010}.

\subsection{Database}
\label{data}

Two sources of data for $R_{c}$ were used in this study. One dataset contained 299 entries (where some are duplicates) gathered with the assistance of large language models \cite{polak2023extracting, polak2023flexible}. Another set of data was taken from a metallic glass database containing $R_{c}$ and other properties from Ref.~\cite{mg_foundry}. These last data were curated by researchers sifting through various publications manually. If duplicates existed across both sources of data, then the entry from Ref.~\cite{mg_foundry} was kept and the other was discarded. For duplicate compositions within each dataset, the mean $R_{c}$ was taken to produce a one-to-one mapping between composition and $R_{c}$. The merged data set contained 177 entries. For the portion of the study that includes fitting MLPs and computationally expensive simulations, only 34 out of the 177 entries were studied. The subset of 34 compositions was acquired from Ref.~\cite{mg_foundry}, and none of them originated from the data obtained using large language models.

\subsection{Summary of Explored Properties}
\label{summary}

The properties studied in this work include those acquired using MLPs and those gathered from elemental properties. Due to the extensive number of features and potential for confusion in feature nature and naming conventions, we present a comprehensive summary of features and their corresponding descriptions in Table~\ref{features}. Secs.~\ref{properties_mtp}, \ref{properties_dft}, and \ref{properties_easy} will elaborate on the acquisition of features and which set of features and their connection to the different feature sets (i.e., $X_{i}$ where $i \in \{ long, mtp, crys, mastml \}$).

\begin{table}[H]
    \centering
    \caption{The features included in this study are tabulated below. Formulas are provided for features that are further discussed from other sources.}
    \label{features}
    \begin{tabular}{l l p{12cm}}
    \toprule
    Feature & Set & Description \\
    \midrule
    $\bar{x}$ & $X_{long}$ & Sum of the multiple of Pauling electronegativity and element fractions \cite{Long2023} \\
    $\gamma$ & $X_{long}$ & A function of atomic radii \cite{Long2023} \\
    $H_{mix}$ & $X_{long}$ & A function of molar mixing enthalpy \cite{Long2023} \\
    $\delta$ & $X_{long}$ & A function of atomic radii and element fractions \cite{Long2023} \\
    $\bar{r}$ & $X_{long}$ & Sum of the multiple of atomic radii and element fractions \cite{Long2023} \\
    $S_{mix}$ & $X_{long}$ & $-R\sum_{i=1}^{N}c_i ln(c_i)$, where $c$ is the atomic fraction, $R$ is the gas constant, and $i$ is the element in question  \cite{Long2023} \\
    $\Delta C$ & $X_{mtp}$ & Difference in $dE/dT$ between a frozen glass and liquid \\
    $T_{f}$ & $X_{mtp}$ & The fictive temperature \\
    $T_{s}$ & $X_{mtp}$ & High temperature dynamical transition \\
    $T_{a\_diff}$ & $X_{mtp}$ & The cross-over temperature denoting the start of non-Arrhenius behavior for self-diffusion \\
    $T_{a\_visc}$ & $X_{mtp}$ & The cross-over temperature denoting the start of non-Arrhenius behavior for viscosity \\
    $T_{g\_diff}$ & $X_{mtp}$ & The temperature where a material reaches a self-diffusion of $10^{-12}$ $[\text{\r{A}}/ps]$ \\
    $T_{g\_visc}$ & $X_{mtp}$ & The temperature where a material reaches a viscosity of $10^{12}$ $[Pa/s]$ \\
    $T_{c\_diff}$ & $X_{mtp}$ & The temperature where a material reaches a self-diffusion of $10^{-6}$ $[\text{\r{A}}/ps]$ \\
    $T_{c\_visc}$ & $X_{mtp}$ & The temperature where a material reaches a viscosity of $10^{4}$ $[Pa/s]$ \\
    $D_{a}$ & $X_{mtp}$ & The MYEGA fitting parameter $a$ in Eq.~\ref{myega} for self-diffusion \\
    $D_{b}$ & $X_{mtp}$ & The MYEGA fitting parameter $b$ in Eq.~\ref{myega} for self-diffusion \\
    $D_{c}$ & $X_{mtp}$ & The MYEGA fitting parameter $c$ in Eq.~\ref{myega} for self-diffusion \\
    $\eta_{a}$ & $X_{mtp}$ & The MYEGA fitting parameter $a$ in Eq.~\ref{myega} for viscosity \\
    $\eta_{b}$ & $X_{mtp}$ & The MYEGA fitting parameter $b$ in Eq.~\ref{myega} for viscosity \\
    $\eta_{c}$ & $X_{mtp}$ & The MYEGA fitting parameter $c$ in Eq.~\ref{myega} for viscosity \\
    $D_{m}$ & $X_{mtp}$ & The MD fragility index for self-diffusion \\
    $\eta_{m}$ & $X_{mtp}$ & The MD fragility index for viscosity \\
    ICO-like & $X_{mtp}$ & The fraction of Voronoi polyhedrons where number of faces with 5 edges are greater than 9 at 1500 $[K]$ \cite{Bokas2016} \\
    $var(VP)$ & $X_{mtp}$ & The variance of Voronoi polyhedrons at 1500 $[K]$ \cite{Wang2019} \\
    $E_{form}$ & $X_{crys}$ & The formation energy for the amorphous phase \\
    $E_{above}$ & $X_{crys}$ & The energy above the convex hull \\
    Many & $X_{mastml}$ & Algebraic operations of elemental properties (not explored in feature selection) \\
    Many & $X_{long}$ & Features from Ref.~\cite{Long2023} \\
    Many & $X_{mtp}$ & Features derived from simulation with MTPs only \\
    Many & $X_{crys}$ & Features derived from simulation with MTPs and ab initio \\
    Many & $X_{tot}$ & $X_{long} \cup X_{mtp} \cup X_{crys}$ \\
    Many & $X_{best}$ & $X_{best} \subset X_{tot}$ \\
    \bottomrule
    \end{tabular}
\end{table}

\subsection{Creation and Validation of MLP Fitting Methodology}
\label{mtp_validation}

Prior to calculating computationally expensive properties for learning $R_{c}$, MLPs were fit and the approach carefully validated. Our properties of interest stem from the relationships between viscosity, self-diffusion, and potential energies with respect to temperature. Any MLP must have almost the same behavior across these properties when compared to first principles MD for its training to be considered accurate. For validation of our approach, we needed a ground truth system for which all properties could be robustly determined. This ruled out validating on ab initio data since some key properties, like viscosity, were not practical to compute fully with ab initio. Therefore, we fit MLPs to data from embedded atom method (EAM) potentials and compared the MLP predictions to those from EAM potentials \cite{Sheng, Sheng2012, eam1, eam2, eam3, eam4, eam5}. Specifically, the potential for the systems of $AlCuZr$, $PdSi$, and $NiP$ are from Refs.~\cite{eam2}, \cite{PdSiPotential2024, Sheng}, and \cite{Sheng2012}, respectively. While the EAM potentials may not be highly accurate vs. experiments, they can still be used as a physically realistic materials system for validation. We assessed our fitting of MLPs using atomic positions, energies, forces, and stresses (i.e., the set of which form one configuration) generated through classical interatomic potentials for three materials: $Ni_{80}P_{20}$, $Pd_{75}Si_{25}$, and $Al_{10}Cu_{40}Zr_{50}$. 

For producing data to train potentials, LAMMPS was used to simulate the aforementioned compositions. The data generation process was chosen so that it was practical for ab initio molecular dynamics (MD) and therefore includes fairly short MD simulation times. Materials were created randomly in a cubic configuration, then melted at 3,000 $[K]$ for 1 $[ps]$. Then, materials were held for another 1 $[ps]$ at 3,000 $[K]$ and iteratively cooled by 300 $[K]$ with 1 $[ps]$ holds until a final temperature of 300 $[K]$ was reached. At the beginning of each hold, 61 configurations with volumes expanded and contracted by $\pm$ 15\% the cubic side length were taken. We conducted these expansions and contractions to ensure that the training data for potentials encompassed a broad range of energies and forces. While our choices regarding the number of configurations and the extent of contractions and expansions may not be optimal, we find them sufficient for our work because they yield excellent agreement for properties of interest. The 61 configurations making out the equation of state might seem particularly excessive but we found that just using a modest number like five configurations significantly degraded the results. All simulation was done with a time step of 1 $[fs]$. Energies, forces, and stresses were taken from all the volumetric contraction/expansion data along with every 100th frame from isothermal holds to produce the data used to fit level 08 MTPs (the level 08 and other levels are described in Ref.~\cite{Novikov2021}). The final number of MD frames used for training was 724. All configurations used for building MLPs were from amorphous phases (liquid and solid). None of the potentials have crystalline information used in the fitting. Constructed MTPs were validated with separate MD. We performed a new set of melt quench runs following the approach previously discussed but without the volume compression/expansion grid using EAM potentials. For each set of atomic positions, we used MTPs to predict the energies and forces. We then compared them directly to values from classical EAM potentials. All simulation described so far in this section included system sizes of 100 atoms in the NPT ensemble.

Subsequently, MD simulations were conducted for both the classical potentials and MLPs to collect data on potential energy, self-diffusion, and viscosity as functions of temperature (methodology covered in Sec.~\ref{properties_mtp}). While MLPs are employed to acquire other properties in our study, we find that acquiring these properties was particularly challenging, and their validation serves as an effective check. Trends for these properties were compared between the MLP and EAM potentials for simulated systems of 1,000 atoms. The comparison was done with values of potential energy, self-diffusion, and viscosity averaged over 10 independent runs separately for each potential to assume robust converged values. We show in Sec.~\ref{errors_from_mlps} that potential energy, self-diffusion, and viscosity were  modeled accurately with the MLP vs. EAM potentials and therefore apply a similar methodology to MLPs fit with ab initio data in Sec.~\ref{mtp_34}.

\subsection{Applying MLP Fitting Methodology to Materials of Interest}
\label{mtp_34}

We selected 34 compositions from 20 different chemical systems for fitting MLPs. The number of explored compositions was decreased from 177 because of the computational cost of fitting potentials and simulation of properties. The $R_{c}$ for the chosen compositions span eight orders of magnitude and exhibit variations in the number of constituent element types. The smallest systems studied comprised two elements, while the largest system investigated involved six elements. The total number of chemical species considered in this study amounted to 19 and formed 20 chemical systems (with some chemical systems being subsets of others). Training data for MLPs were created with the use of ab initio MD in VASP. We employed a 1x1x1 $\Gamma$ mesh for the k-points in our calculations. The energy cutoffs were determined by selecting the largest value of ENMAX defined in the POTCAR file for each composition (the default for VASP). Ab initio calculations were conducted for the 34 compositions following as closely as possible to the methods described for the EAM and MLP comparisons (Sec.~\ref{mtp_validation}). The only notable difference between the ab initio and EAM training approaches was a switch from the NPT to the NVT ensemble. The Nose-Hoover thermostat was used for NVT. The only thermostat compatible with NPT simulation in VASP is the Langevin thermostat, which requires defining friction coefficients for each atomic species. However, determining these coefficients for all atomic species across the 34 studied compositions was impractical for generalized approaches. For holds under the NVT ensemble, the grid of volumetric contractions/expansions were used to find the zero pressure volume that was used for each respective isothermal hold. Some MLPs trained on the ab initio data obtained in this manner were unstable, possibly because of the change in ensemble from the previous validation on EAM potentials to the approach used in fitting with ab initio. NPT allows for many more perturbations to volumes in the system and may provide more information for potentials. The issue of MLP instability was solved with the use of active learning. NPT runs described in Sec.~\ref{mtp_validation} with isothermal holds of 100 $[ps]$ instead of 1 $[ps]$ were simulated using the MLIP-2 interface with LAMMPS. Whenever a configuration of atoms was considered uncertain based on the default criterion outlined in Ref.~\cite{Novikov2021}, an additional ab initio calculation was carried out on that configuration. The information obtained from these supplementary ab initio calculations was subsequently incorporated into the training data. MD was repeated until potentials were stable. Because of this process, differing number of training configurations were used for each composition. Additional data were acquired to train MLPs compared to the method outlined in Sec.~\ref{mtp_validation}, so we anticipate improvements in behavior across properties. Errors in forces and energies were acquired in the same manner as in Sec.~\ref{mtp_validation}. The fitting errors are described in Sec.~\ref{errors_from_mlps}. All MD in this study was conducted with a time step of 1 $[fs]$.

\subsection{Features Generated from MLP MD}
\label{properties_mtp}

This section describes how we extracted properties from simulations of 34 compositions to be used to machine learn $R_{c}$. In Sec.~\ref{introduction}, we discussed the significance of viscosity and characteristic temperatures in prior GFA studies. As self-diffusion data can be obtained from the same MD simulations used for measuring viscosity and signifies the mobility of atoms within a system, we incorporate it into our set of properties of interest. For MD in this section, viscosity, self-diffusion, energies, etc. were acquired from the average of 5 independent simulations for each of the 34 studied materials. Each composition of interest underwent an initial equilibration at 2,000 $[K]$ for 100 $[ps]$ with a system size of 1,000 atoms. Subsequently, materials were iteratively quenched by decreasing the temperature by 50 $[K]$ instantaneous decrements and holding for 110 $[ps]$ until a final temperature of 100 $[K]$. The first 10 $[ps]$ were used for equilibration and were discarded from any data analysis. Each isothermal hold was continued from the previous hold. The average cooling rate was approximately 0.45 $[K/ps]$. As an example of one iteration, an isothermal hold at 1,500 $[K]$ would decrease instantly to 1,450 $[K]$ after 110 $[ps]$. All these processes were conducted under the NPT ensemble and the potential energy was measured for each temperature. A Nose-Hoover thermostat and barostat were used for the NPT ensemble. The average potential energy versus temperature were used to measure the fictive temperature, $T_{f}$, and a dynamic slowdown temperature, $T_{s}$ \cite{Sheng2012}. $T_{s}$ indicates the onset of non-Arrhenius relaxation for a liquid material \cite{Sheng2012}. Both $T_{f}$ and $T_{s}$ were determined by fitting the data with three optimal linear segments using the PieceWise Linear Functions (PWLF) package \cite{pwlf}. Essentially, the two most drastic changes in the slope of potential energy with respect to temperature denote $T_{f}$ and $T_{s}$, with $T_{f}$ occurring at a lower temperature than $T_{s}$. The difference between slopes of total energies with respect to temperature above and below $T_{f}$ effectively gives change in the heat capacity, $\Delta C$. $\Delta C$ was used as a feature for learning $R_{c}$. This feature was included based on arguments made in Ref.~\cite{biroli2009random}, which relate change in heat capacity to $m$ with random first order transition theory. Generally, an increase in $\Delta C$ results in a higher value of $m$ and vice versa (See Eq. 43 or Fig. 3b in Ref.~\cite{biroli2009random}).

We included two features regarding VPs by measuring atom positions at 1,500 $[K]$ for each composition. Note that the atom positions at 1,500 $[K]$ were near the temperature of 1,478 $[K]$ used in Ref~\cite{Wang2019} and is above all $T_{f}$ explored in our study. These positions were determined at the final step of the NPT quenching process at 1,500 $[K]$ specified at the beginning of this section. We measure the fraction of ICO-like Voronoi polyhedra (VP) and the variance of VP as defined in Ref.s~\cite{Bokas2016} and \cite{Wang2019}, respectively. ICO-like VP are VP whose number of faces with 5 edges are greater than 9. Variance of VP (denoted as $var(VP)$) is defined by

\begin{equation}
    var(VP) = \frac{1}{G}\sum^{G}_{i}(f_{i}-\mu)^2
\end{equation}

where $i$ is the type of VP, $f$ is the fraction of $i$ in the system, $G$ is the number of VP types, and $\mu$ is the average of all $f_{i}$. The python3 API from OVITO was used to calculate VP with an edge threshold of 0.1 $[\text{\AA}]$ \cite{Stukowski2010}. These are not necessarily the optimal types of VPs, temperatures to explore VPs, etc. for predicting GFA. However, such detailed exploration would involve an impractical level of effort given all the features being modeled and was therefore outside the scope of this work. 

For each of the explored temperatures above $1.1T_{f}$, simulations were extended for 10 $[ns]$ under the NVT ensemble and were used to measure self-diffusion and viscosity. We explored temperatures near $T_{f}$ but above because atoms are sufficiently mobile to acquire converged measurements of self-diffusion and viscosity. Viscosity can be determined using the Green-Kubo formalism \cite{Rapaport, Puosi2018}. The autocorrelation of pressure tensor components ($P_{ij} \in \{P_{xy}, P_{xz}, P_{yz}\}$) under the NVT ensemble was integrated, following Eq.~\ref{gbvisc}. Here, $\eta$ is the viscosity, $V$ represents the system volume, $T$ is the temperature, $k_{B}$ denotes Boltzmann's constant, and $t$ represents a specific time value. The integrals of $P_{xy}$, $P_{xz}$, and $P_{yz}$ were averaged together.

\begin{equation}\label{gbvisc}
    \eta = \frac{V}{k_{B}T} \int_{0}^{\infty} \left < P_{ij}(t_{0})P_{ij}(t_{0}+t)\right >dt 
\end{equation}

To measure self-diffusion, the long-time limit of mean squared displacement (MSD) was taken (Eq.~\ref{eq:diffusion}) \cite{Rapaport}. In Eq.~\ref{eq:diffusion}, $D$ is self-diffusion, $N$ is the number of atoms, $t$ is the time, $r$ is the position, and $i$ indicates the atom in question. We measured self-diffusion because it describes how rapidly atoms move in a system and is acquired from the same simulations needed to compute viscosity.

\begin{equation}\label{eq:diffusion}
    D=\lim_{t \to \infty} \frac{1}{6Nt} \left < \sum_{i=1}^{N} \left [ r_{i}(t)-r_{i}(t=0) \right ]^{2} \right >
\end{equation}

Both high temperature self-diffusion and viscosity, as functions of temperature, were fitted using the MYEGA equation \cite{Mauro2009, Reis2012}. The MYEGA function is a three parameter model that has superior extrapolation properties compared to the more commonly used Vogel-Fulcher-Tammann (VFT) equation. In Eq.~\ref{myega}, $\nu_{a}$, $\nu_{b}$, and $\nu_{c}$ are fitting parameters to map a temperature, $T$, to a given property, $\nu$. $\nu=D$ for discussions on self-diffusion and $\nu=\eta$ for discussions on viscosity.

\begin{equation}\label{myega}
	\log_{10}(\nu) = \nu_{a}+\frac{\nu_{b}}{T}e^{\nu_{c}/T}
\end{equation}

From the rheological data, we derived various features. The fitting parameters from the MYEGA function for self-diffusion and viscosity for each composition were obtained. Subsequently, we employed MYEGA to extrapolate self-diffusion values to $10^{-12}$ $[\text{\r{A}}/ps]$ and viscosity to $10^{12}$ $[Pa \cdot s]$. The temperatures corresponding to these extrapolated values were designated as the glass transition temperatures ($T_{g_visc}$ and $T_{g_diff}$ for viscosity and self-diffusion, respectively), based on experimental results and approaches. In particular, the glass transition temperature is determined experimentally when a material attains a viscosity of $10^{12}$ $[Pa \cdot s]$ \cite{Angell1995}. Regarding diffusion, we utilized the self-diffusion value observed at a glass transition temperature in Ref.~\cite{Chen2016}. The slopes immediately preceding $T_{g\_visc}$ and $T_{g\_diff}$ were adopted as a proxy for experimental $m$. Another feature derived from rheology is $T_{c}$, which is determined at the temperature where materials reach a cutoff in viscosity \cite{Gangopadhyay2017}. We extracted the temperatures at which self-diffusion and viscosity reach values of $10^{-4.0}$ $[\text{\r{A}}/ps]$ and $10^{2.0}$ $[Pa \cdot s]$, respectively for each composition. The chosen values ensured that the functions from Eq.~\ref{myega} were not significantly extrapolated beyond the temperature range used to fit the equation's parameters for most compositions. Additionally, deviations from Arrhenius behavior for self-diffusion and viscosity were measured. For viscosity, this matches the definition of the temperature denoted as $T_{a}$. For diffusion, we are not aware of any specific naming convention for the temperature at which it is measured. However, we included this temperature as a feature because it seemed plausible that it could provide useful information. The features discussed so far will be denoted as $X_{mtp}$ as they all come from the MTPs.

\subsection{Features Generated by Mixing MLP Potential and Ab Initio Energies}
\label{properties_dft}

Enthalpy of formation, $E_{form}$, and energy above the convex hull, $E_{above}$, constitute the feature set $X_{crys}$ and are an approximation of the stability of the amorphous phase (given in Table~\ref{features}). Differences between energies of phases have previously been shown to be important for GFA models as seen by the top features in Ref.~\cite{Afflerbach2021}. We generated $X_{crys}$ to capture information about crystalline metals through a combination of MLPs and ab initio. We added pure ab initio because the MLPs were not fit to any crystalline structures and might be unreliable for those atomic arrangements. All simulations were conducted with 100 atoms. Amorphous structures were obtained through simulated quenches from 3,000 $[K]$ down to 100 $[K]$ at 0.5 $[K/fs]$ using MLPs. The resulting structures were relaxed with ab initio approaches to obtain the enthalpy of amorphous structures (denoted as $E_{amorph}$). This process was repeated 3 times for each material to give an average $E_{amorph}$. $E_{form}$ was calculated with a face-centered cubic (FCC) crystal reference for each element in the compound (each of which is denoted by $E_{i}$ for a total of $N$ chemical species). Specifically, $E_{form}$ was determined with Eq.~\ref{formation}.

\begin{equation}\label{formation}
    E_{form} = E_{amorph}-\sum^{N}_{i} E_{i}
\end{equation}

For each chemical system, an estimated convex hull was generated using the Materials Project database of crystal structures \cite{Jain2013}. Comparing the enthalpy of formation for each specific composition to the hull energy gives the resulting $E_{above}$ for each amorphous material, which can be taken as an approximation to the driving force for crystallization.

\subsection{Elemental Property Based Features}
\label{properties_easy}

Here we discuss the more easily computed features based on elemental properties. For an alloy, one can take functions of the constituent elements in ways that represent the alloy \cite{Ward2016}. Specifically, simple to generate features based on minimum, maximum, averaged, and composition weighed averaged properties of elements were acquired through MAST-ML. 409 elemental properties included, such as thermal conductivity, atomic sizes, and electronegativities. We refer to these features as $X_{mastml}$. Additional descriptors involving specific physically motivated functions of atomic sizes, enthalpies, Pauling electronegativity, and mole fractions have shown promise in the work of Long et al.~\cite{Long2023}. These descriptors were also used in the present study, following the formulae and naming conventions provided in Ref.~\cite{Long2023}. We refer to these features as $X_{long}$. Table~\ref{features} provides the formulation for features from $X_{long}$ that were identified as important in the SHAP analysis (see Sec.~\ref{feature_selection}). Features from $X_{mastml}$ and $X_{long}$ can be easily generated for all 177 materials in Sec.~\ref{data}.

\subsection{Regression Models and Corresponding Assessment}
\label{feat_compare}

An eXtreme Gradient Boosting (XGBoost) model type was employed throughout the study \cite{xgboost} due to its ability to efficiently and accurately learn complex relationships. The target of regression models was $log_{10}(R_{c})$ and were evaluated through a Chemical System Leave-Out (CSLO) CV. For CSLO CV, we group each material by its chemical system. We then perform leave one group out CV. As an illustration, assume a dataset has material entries with elements $e_{1}$, $e_{2}$, and $e_{3}$. The following are all possible combinations of left out sets of materials: $\{e_{1}\}$, $\{e_{2}\}$, $\{e_{3}\}$, $\{e_{1}, e_{2}\}$, $\{e_{1}, e_{3}\}$, $\{e_{2}, e_{3}\}$, and $\{e_{1}, e_{2}, e_{3}\}$. As an illustration of 1 iteration, an evaluation focused on the $ZrCuAl$ system could use data from the systems of $LaAu$ and $ZrCu$ for training and predict $ZrCuAl$ as left out validation data. Note that $ZrCu$ is in the $ZrCuAl$ system but has a different number of components and is therefore considered to be in another group. Test metrics were gathered across all iterations of leaving chemical systems out.

Various metrics were employed to evaluate models, including the coefficient of determination ($R^{2}$), mean average error ($MAE$), root mean squared error ($RMSE$), and $RMSE$ normalized by the target standard deviation ($RMSE/\sigma_{y}=\sqrt{1-R^2}$). In comparing model assessments, $RMSE/\sigma_{y}$ was favored due to its intuitive nature. If $RMSE/\sigma_{y}$ is close to one, it suggests that the errors of the model are comparable to a naive model that predicts only the mean of the property of interest. $RMSE/\sigma_{y}$ is defined by

\begin{equation}\label{rmse_sigma}
    RMSE/\sigma_{y} = \sqrt{\frac{\sum^{N}_{i=1} (y_{i}-\hat{y}_{i})^{2}}{\sum^{N}_{i=1} (y_{i}-\bar{y})^{2}}}
\end{equation}

where $i$ denotes a specific material, $y$ denotes a property of interest, $\hat{y}$ denotes the prediction of $y$, and $\bar{y}$ denotes the mean value calculated from all $y$.

\subsection{Comparison of Feature Effectiveness}
\label{feature_selection}

XGBoost models constructed using $X_{mastml}$, $X_{long}$, and a union of both (i.e. $X_{mastml} \cup X_{long}$) were assessed with CSLO CV for 177 materials, with results shown in Sec.~\ref{elemental_feature_comparison}. Models fit with $X_{long}$ alone tended to outperform models fit to $X_{mastml}$ or $X_{mastml} \cup X_{long}$. Consequently, only $X_{long}$ was kept for subsequent analyses and compared to more intricate, simulated features. To assess the value added by our set of simulated features to GFA models, we train XGBoost models on 34 materials (i.e., materials that have MLPs) with the combined set of $X_{long} \cup X_{mtp} \cup X_{crys}$. We will refer to $X_{long} \cup X_{mtp} \cup X_{crys}$ as $X_{tot}$ for simplicity. We also train models on $X_{long}$ alone as a comparison.

SHAP values were used to compare the relative importance of physically motivated features for $X_{tot}$ \cite{shap}. These values assess feature contributions to predictions for a given model, with principles rooted in game theory. We create a feature learning curve, ordered in importance by their SHAP values, to select the subset of $X_{tot}$ with the lowest $RMSE/\sigma_{y}$ which we will call $X_{best}$. CSLO CV was used in feature selection because we desire models that can predict GFA across chemical systems. To demonstrate that the selection of $X_{best}$ was not due to chance, given the similar number of features and observations, we repeated the feature selection process with shuffled $log_{10}(R_c)$ values. This approach demonstrated that achieving an $RMSE/\sigma_{y}$ as low as the model from the original feature selection is improbable when randomly fitting a model to $log_{10}(R_c)$ with features unrelated to GFA (see Sec.~\ref{selection_results}).
\section{Results}
\label{results}

\subsection{Errors from MLPs}
\label{errors_from_mlps}

\begin{figure}
    \centering
    \begin{subfigure}[b]{0.49\textwidth}  
        \centering 
        \includegraphics[width=\textwidth]{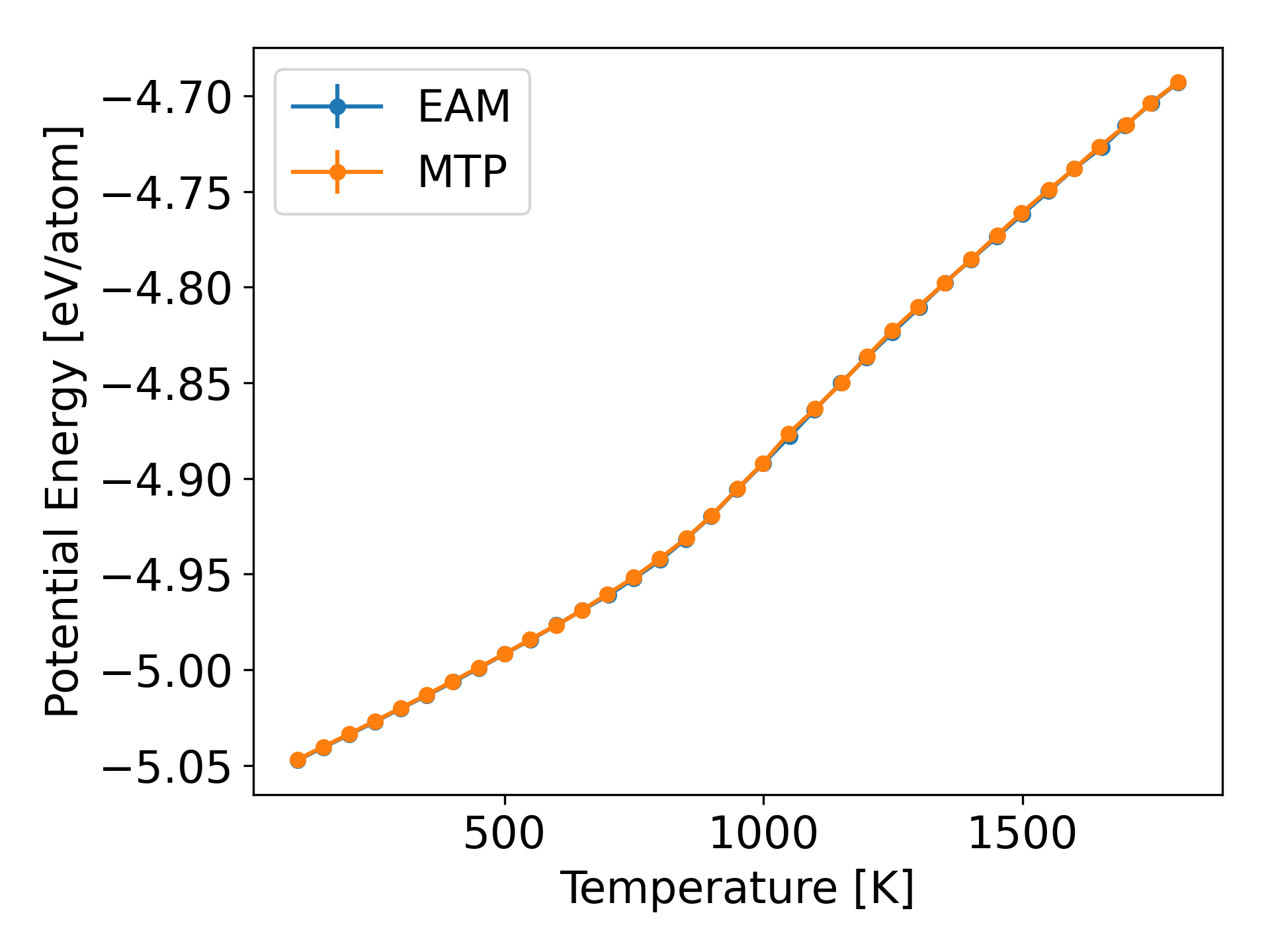}
        \caption{Potential Energy}
    \end{subfigure}
    
    \vspace{1cm}
    
    \begin{subfigure}[b]{0.49\textwidth}   
        \centering 
        \includegraphics[width=\textwidth]{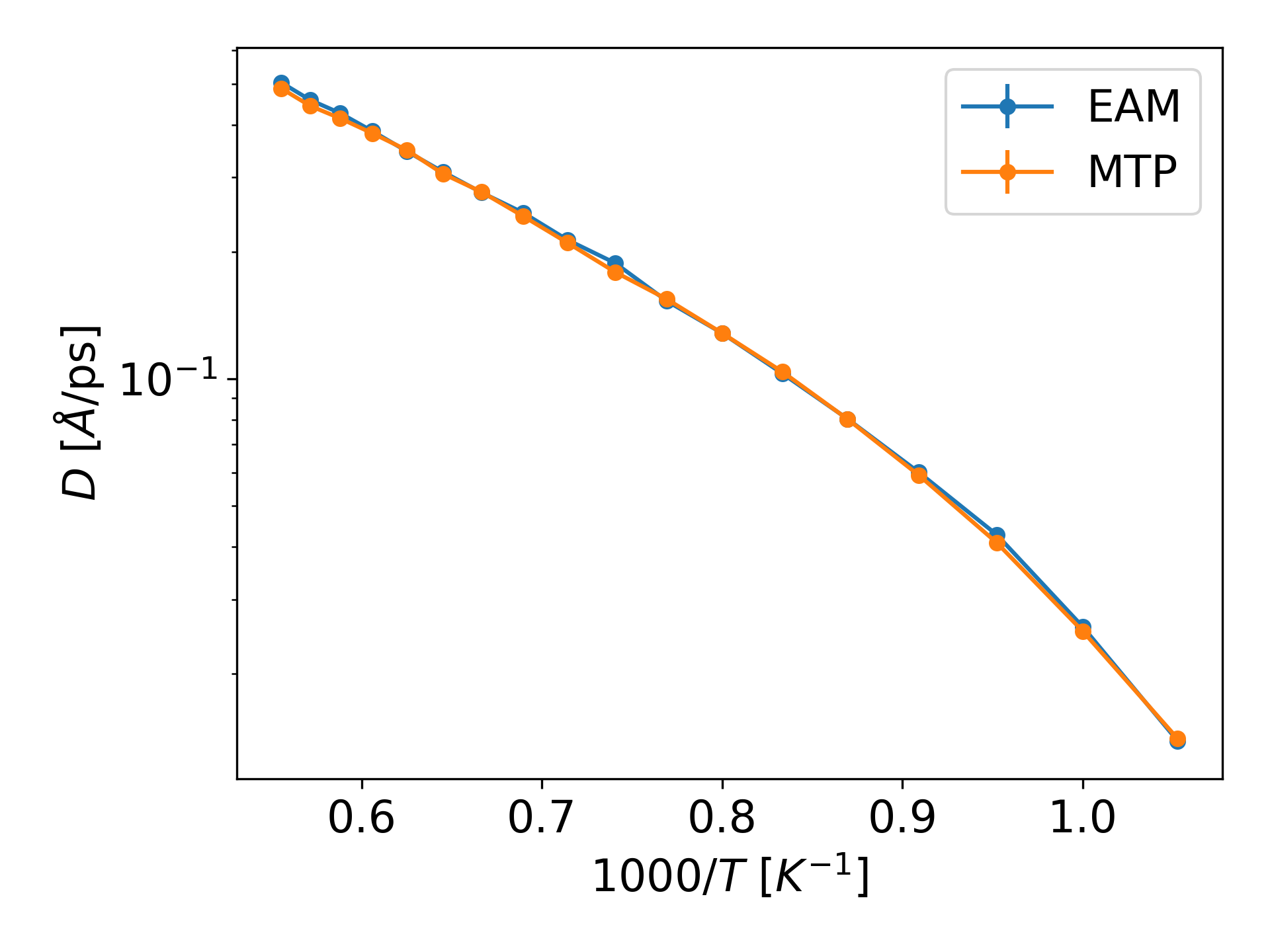}
        \caption{Self-Diffusion}
    \end{subfigure}
    \hfill
    \begin{subfigure}[b]{0.49\textwidth}   
        \centering 
        \includegraphics[width=\textwidth]{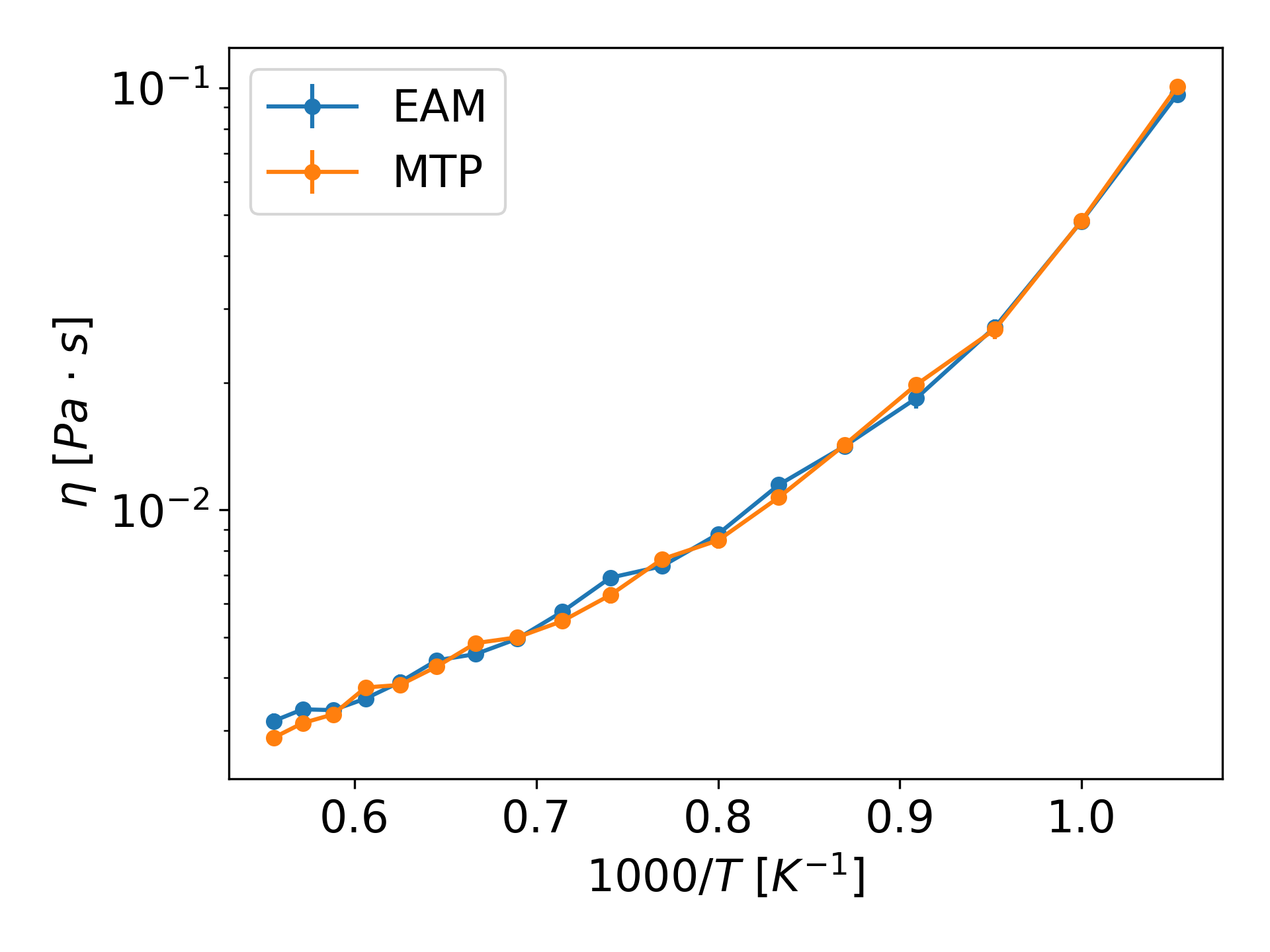}
        \caption{Viscosity}
    \end{subfigure}
    
\caption{Potential energy, self-diffusion, and viscosity for $Al_{10}Cu_{40}Zr_{50}$ as functions of temperature are shown. MTP and classical EAM potentials show excellent agreement across all shown properties. Note that simulated properties were averaged across 10 independent runs.}

\label{mtp_val}
\end{figure}

For the three test systems where we are comparing the MLPs to EAM potentials, the fitting methodology outlined in Sec.~\ref{mtp_validation} produced the $RMSE$ for forces and energies (shown in the Supplemental Materials). Our $RMSE$ values are generally comparable to other MLPs on metal alloys (~0.1 $[eV/\text{\r{A}}]$ for forces and ~0.01 $[eV/atom]$ for energies), except maybe for the force errors being a bit higher. However, the MLPs are sufficiently accurate to produce properties of interest. Fig. \ref{mtp_val} shows an example temperature dependence of potential energy, self-diffusion, and viscosity for $Al_{10}Cu_{40}Zr_{50}$. Parallel data for the other materials may be found in the Supplemental Materials. All three properties are similar between EAM potentials and MLPs. In other words, we can simulate properties of interest with sufficient accuracy with the potential errors in Table~\ref{eam_to_mtp_table}.

For the 34 MLPs used to extract features for machine learning, the errors in energies and forces were acquired with the method outlined in Sec.~\ref{mtp_validation} (see Supplemental Materials for tabulated data). Generally, chemical systems with more types of elements require more training configurations. The maximum obtained $RMSE$ across all MLPs for energies and forces was 0.031 $[eV/atom]$ and 0.290 $[eV/\text{\AA}]$, respectively. These values are larger than ideal and larger than our worst test case from Sec.~\ref{errors_from_mlps}, but only by up to at most 2-3 times. We therefore believe that these MLPs will produce similarly accurate model results as those used in the EAM comparison. While more accurate MLPs can certainly be developed using more complex potential functions and additional training data, such refinements were beyond the scope of this work, given the large number of systems studied.

\subsection{Visualization of Some Properties Attained Through MD}

\begin{figure}[H]
    \centering
    
    \begin{subfigure}{0.49\textwidth}
        \includegraphics[width=\linewidth]{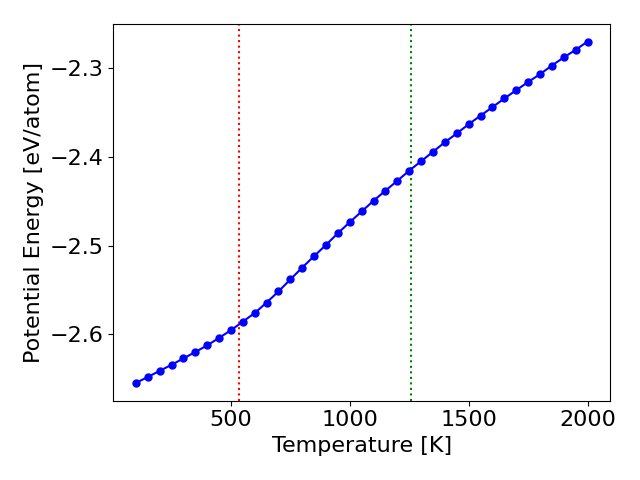}
        \caption{Potential Energy}
        \label{tf}
    \end{subfigure}
    
    \vspace{1cm}
    
    \begin{subfigure}{0.49\textwidth}
        \includegraphics[width=\linewidth]{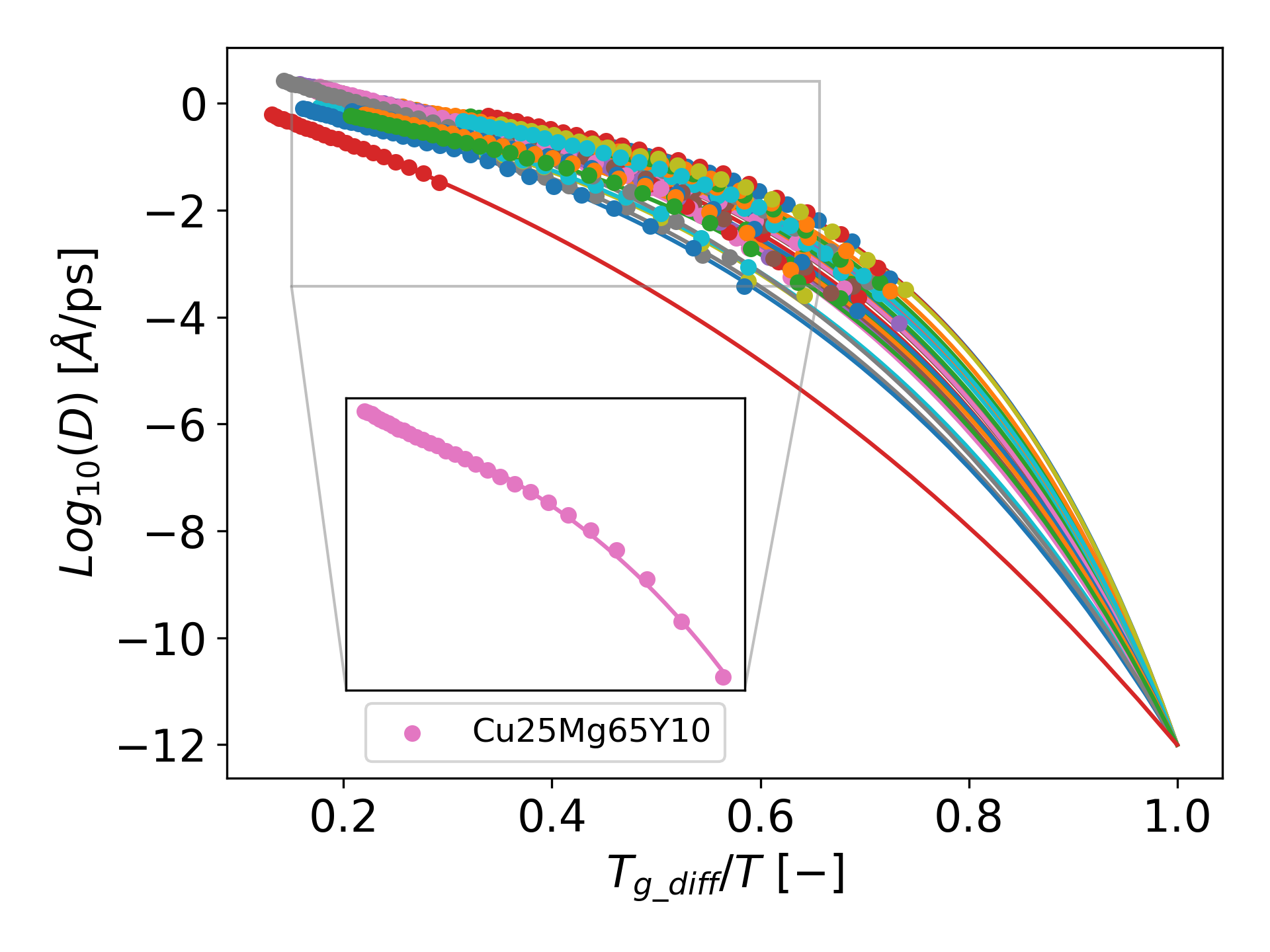}
        \caption{Self-Diffusion}
        \label{diff_subfig}
    \end{subfigure}
    \hfill
    \begin{subfigure}{0.49\textwidth}
        \includegraphics[width=\linewidth]{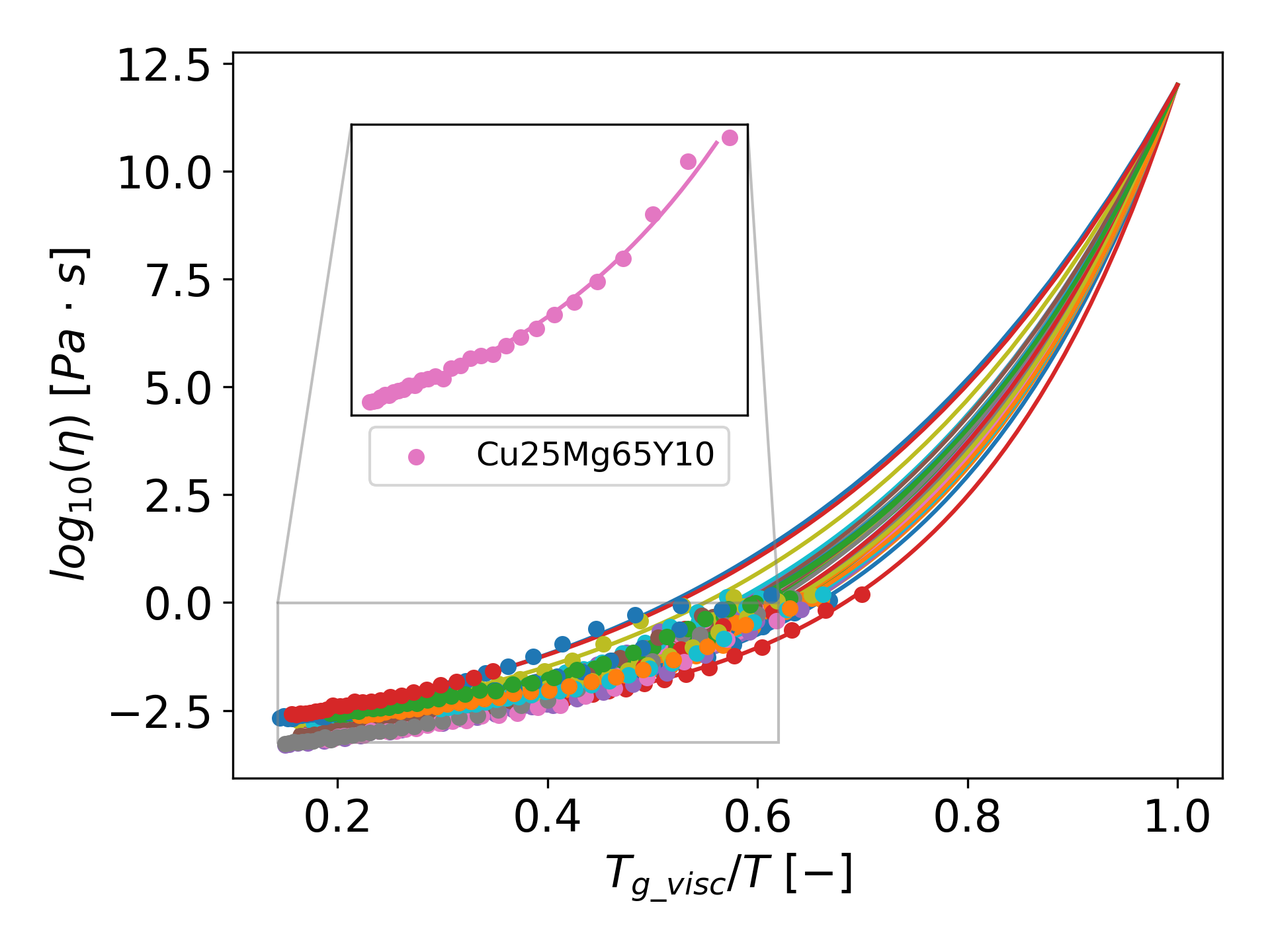}
        \caption{Viscosity}
        \label{visc_subfig}
    \end{subfigure}
    
    \caption{Potential energy (Fig.~\ref{tf}), self-diffusion (Fig.~\ref{diff_subfig}), and viscosity (Fig.~\ref{visc_subfig}) versus temperature for 34 compositions are shown. The potential energy relationship with respect to temperature is shown for $Cu_{25}Mg_{65}Y_{10}$ only. There are high and low temperature transitions denoted by $T_{f}$ and $T_{s}$. As materials cool in Figs.~\ref{diff_subfig} and \ref{visc_subfig}, they experience restrained movement at varying degrees with respect to $T_{g\_diff}$ or $T_{g\_visc}$. The points represent measured viscosity values and solid lines are the MYEGA fits. The MYEGA function fit to points was extrapolated to -12 and 12 on the vertical axis for self-diffusion and viscosity, respectively. Note that all fits converge at $T_{g\_visc}/T=1$ and $T_{g\_diff}/T=1$ because of our definitions of $T_{g\_visc}$ and $T_{g\_diff}$ for viscosity and diffusion, respectively.}
    \label{residual_test_results}
\end{figure}

The potential energy versus temperature curve for $Cu_{25}Mg_{65}Y_{10}$ is shown in Fig.~\ref{tf}. Figures for all other compositions are available as described in the Data Availability section. Fig.~\ref{tf} also shows the values of $T_{f}$ and $T_{s}$. Self-diffusion data for 34 compositions are shown in Fig.~\ref{diff_subfig}. Viscosity values are shown in Fig.~\ref{visc_subfig}. The data points denote isothermal holds where averaging was performed, and continuous lines show the MYEGA fits to the data.

\subsection{Exploring Models Fit to Elemental Features}
\label{elemental_feature_comparison}

For 177 compositions, we evaluated the $RMSE/\sigma_{y}$ for $log_{10}(R_{c})$ across three feature sets, $X_{mastml}$, $X_{long}$, and $X_{mastml} \cup X_{long}$. XGBoost models were evaluated with CSLO CV. The feature set that yielded the lowest $RMSE/\sigma_{y}$ was found to be $X_{long}$. Parity plots are supplied in the Supplemental Materials. Models fit to $X_{long}$ had an $RMSE/\sigma_{y}$ of 0.85 whereas models fit to $X_{mastml}$ and $X_{mastml} \cup X_{long}$ had $RMSE/\sigma_{y}$ of 1.00 and 0.92, respectively. None of these models are particularly accurate, as can be seen by the fact that $RMSE/\sigma_{y}$ for all feature sets are close to 1.00, which is the score that would be obtained by a naive model that predicts just the mean of $y$. Therefore, it is useful to explore other properties that can improve our $log_{10}(R_c)$ model.

\subsection{Exploring Models Fit with Simulated Features}

XGBoost models were assessed using the CSLO CV for 34 materials that had MLPs. Models were built by using $X_{long}$ alone and the combined feature set of $X_{tot}$ (defined in Sec.~\ref{feature_selection} and Table~\ref{features}). The resulting $RMSE/\sigma_{y}$ was 0.76 and 0.60 for $X_{long}$ and $X_{tot}$ respectively (parity plots provided in the Supplemental Materials). These results suggest that the inclusion of simulated features improve a model's ability to predict GFA. We refined our model by implementing feature selection via SHAP values to show which features in $X_{tot}$ are most important for predicting $log_{10}(R_{c})$.

\subsection{Ranking Features Based with Feature Selection}
\label{selection_results}

\begin{figure}[H]
	\centering
	\includegraphics[width=0.95\textwidth]{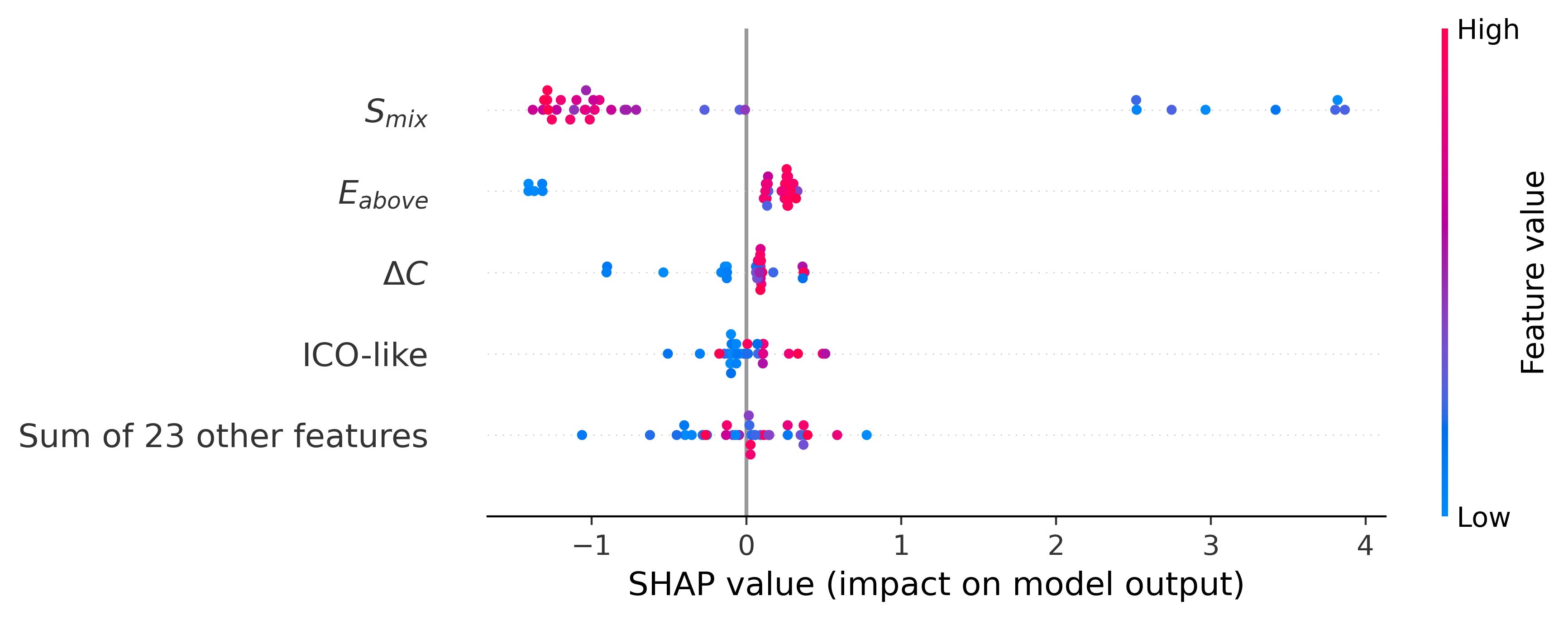}
	\caption{The SHAP values for XGBoost is shown for $X_{tot}$. Higher values of $log_{10}(R_{c})$ are denoted by higher values of model output on the horizontal axis and vice versa.}
	\label{shap}
\end{figure}

Shapley Additive Explanations (SHAP), a method based on game theory principles, provides valuable insights into feature contributions. Fig.~\ref{shap} illustrates the SHAP values for $X_{tot}$ with respect to an XGBoost model. The vertical axis of Fig.~\ref{shap} represents the features, while the horizontal axis displays the SHAP values, indicating whether the prediction of a model using that feature increases or decreases the predicted value of $log_{10}(R_{c})$. The mean absolute SHAP values for each feature in Fig.~\ref{shap} determine the importance and ranking of the features. Higher values signify greater influence on a model's prediction. Each dot from each feature value corresponds to a specific material (e.g., $Al_{25}Co_{5}Cu_{10}La_{55}Ni_{5}$). By examining the color bar, we can assess whether the feature value for a specific material is high or low and then observe if those values drive the model's predictions to be higher or lower. Note that out of the 4 features shown, only $S_{mix}$ was derived from the easy-to-compute elemental features in $X_{long}$ and is ranked first. Other features derived from more complex simulations significantly contribute to a model's prediction on $log_{10}(R_{c})$.

\begin{figure}[H]
	\centering
	\includegraphics[width=0.5\textwidth]{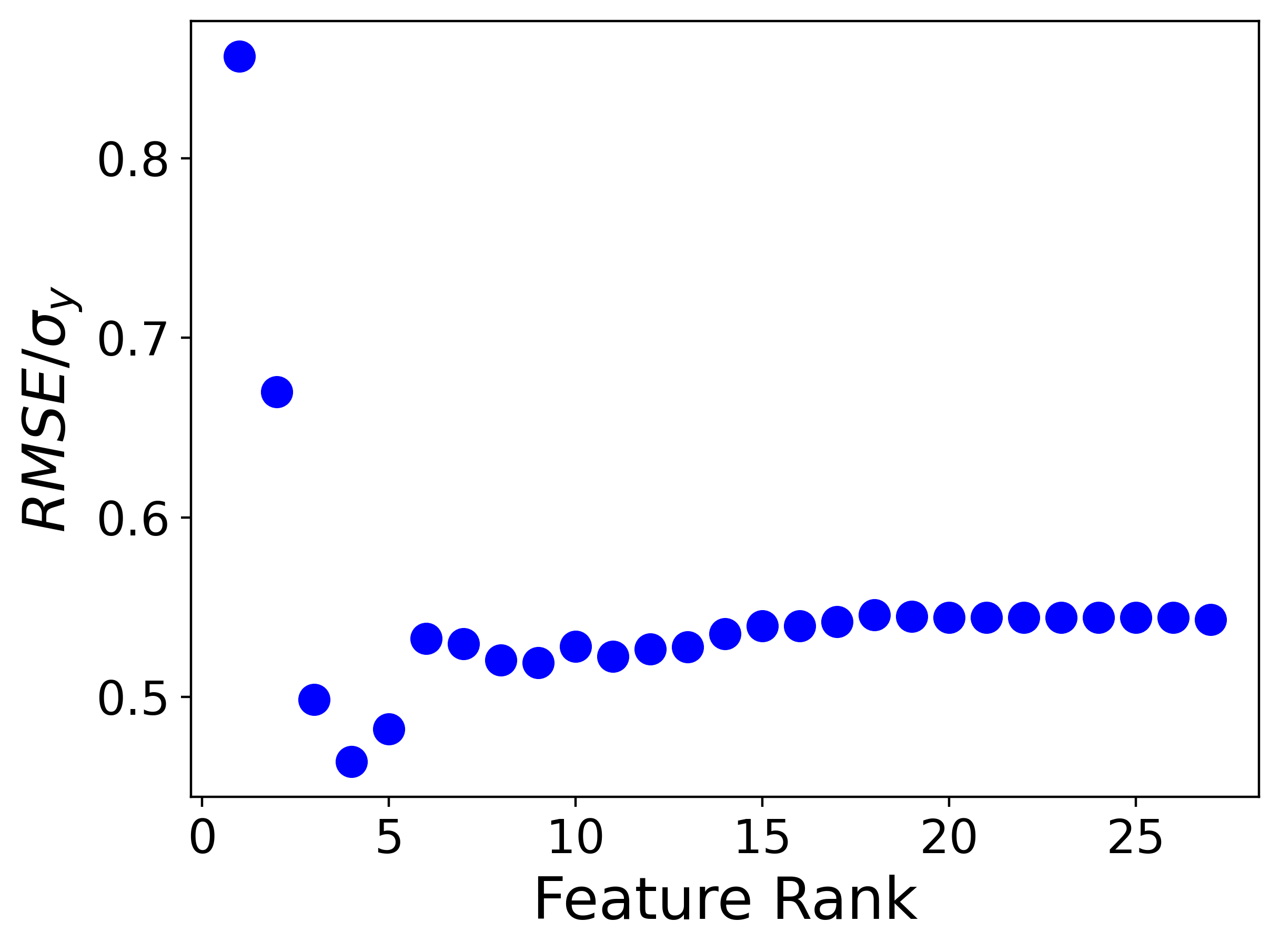}
	\caption{$RMSE/\sigma_{y}$ as a function included features sorted with SHAP values for XGBoost models are shown.}
	\label{selection}
\end{figure}

We show $RMSE/\sigma_{y}$ as a function of the number of features used when the features are ordered based on SHAP values in Fig.~\ref{selection}. Any feature after the top 4 did not lead to regression improvements. The addition of the fifth top feature arbitrarily increased model complexity which is concerning for a small data set of 34 points and causes overfitting. Overfitting can be seen by the increase of $RMSE/\sigma_{y}$ starting at 5 features and beyond. The top 4 features are denoted as $X_{best}$, which are the ones shown in Fig.~\ref{shap}.

\begin{figure}[H]
	\centering
	\includegraphics[width=0.5\textwidth]{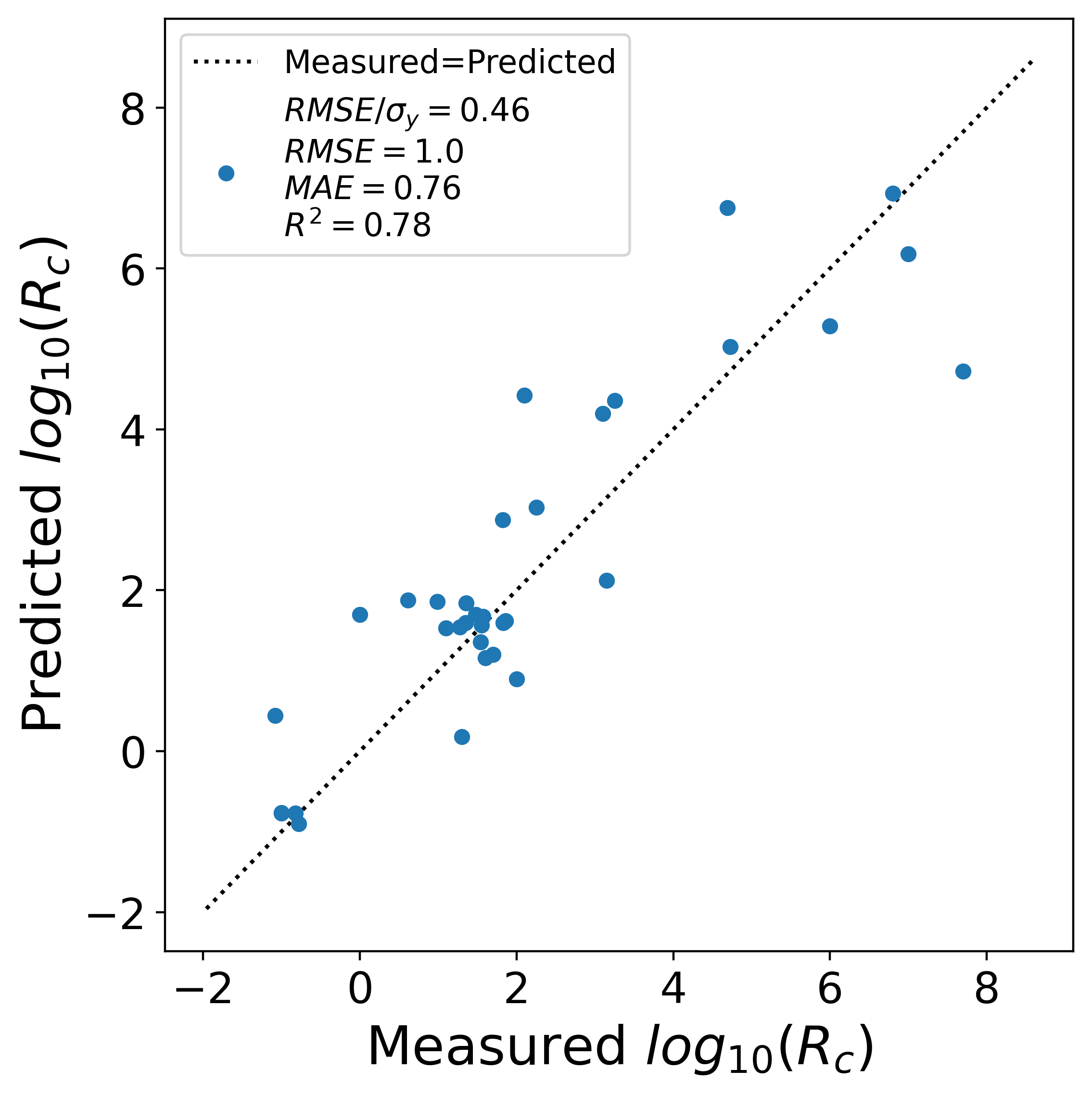}
	\caption{The parity plot using $X_{best}$ for the 34 compositions that had MLPs are shown. The model appears to predict $R_{c}$ across chemical systems well.}
	\label{top_selected}
\end{figure}

Assessing models with $X_{best}$ yields an $RMSE/\sigma_{y}$ ($R^{2}$) of 0.46 (0.78). The associated parity plot is shown in Fig.~\ref{top_selected}. It is worth noting that the best cited $RMSE/\sigma_{y}$ is an overestimation of what might be expected on test data due to feature selection being conducted without an explicit test set. Nevertheless, the number of features in $X_{best}$ is much smaller than the 34 studied materials and the assessment is tested with CSLO CV, which provides some protection against significant overfitting. The top four features are discussed in detail in Sec.~\ref{discussion}.

The lowest $RMSE/\sigma_{y}$ obtained from the shuffling feature selection procedure (see the end of Sec.~\ref{feature_selection}) is about 1.36 (see Supplemental Materials) and is significantly higher than the 0.46 achieved in Fig.~\ref{top_selected}. These results indicate that the selection of $X_{best}$ from our extensive feature set was not random and that $X_{best}$ is crucial for constructing a $log_{10}(R_{c})$ model.
\section{Discussion}
\label{discussion}

$X_{long}$ emerged as the feature set yielding the lowest model errors of $RMSE/\sigma_{y}$ among the 177 values of $R_{c}$ compared to $X_{mastml}$. It is perhaps not surprising that a set of physically motivated features would yield superior predictions compared to those derived from brute force arithmetic operations on elemental properties, although this does not always occur (e.g., see~\cite{Jacobs2024}). Nevertheless, models built from the 177 compositions using just $X_{long}$ have a high $RMSE/\sigma_{y}$ of 0.85.

SHAP value rankings show that both simulated properties and $X_{long}$ could produce viable models for predicting $log_{10}(R_c)$. The best performing models were built from only the top four features. The selected features were the mixing entropy ($S_{mix}$ defined in Sec.~\ref{properties_easy}), the energy of the amorphous phase above the convex hull ($E_{above}$ defined in Sec.~\ref{properties_dft}), the fraction of icosahedral-like local VP environments in the high-temperature liquid (ICO-like defined in Sec.~\ref{properties_mtp}), and the difference in heat capacity between the glass and liquid phases ($\Delta C$ defined in Sec.~\ref{properties_mtp}).

The important of $S_{mix}$ is not surprising and it likely reflects the confusion principle associated with GFA \cite{Wang2004}. In other words, including more chemical species in a material may hinder crystalline phase formation because atoms struggle to find suitable sites to form crystals, which shows up as a higher value of $S_{mix}$. Given this argument, we would expect higher values of $S_{mix}$ to correspond to lower values $log_{10}(R_c)$ (higher GFA), which is exactly the trend in Fig.~\ref{shap}.  

The importance of $E_{above}$ is also expected. $E_{above}$ is a measure of the stability of the amorphous phase of our final glass relative to stable crystal structures and is therefore a qualitative guide to the drive for crystallization, a crucial factor controlling $R_{c}$. Higher values of $E_{above}$ indicate less stable glasses with more driving force to crystallization, while lower values imply greater stability glasses with less driving force for crystallization. Given this argument we would expect higher $E_{above}$ values to correspond to higher values of $log_{10}(R_c)$ (lower GFA), which is exactly the trend in Fig.~\ref{shap}. 

The importance of $\Delta C$ is perhaps easiest to understand when we consider its correlation with fragility (see Ref.~\cite{biroli2009random}), which is known to correlate with measures of GFA such as the $R_{c}$. An increase in $\Delta C$ is expected to correlate with an increase in fragility, which, in turn, is expected to correlate with a decrease in GFA and higher values of $log_{10}(R_c)$. This trend is precisely what is observed in Fig.~\ref{shap}.

The importance of ICO-like is consistent with results of previous research that have demonstrated that these types of polyhedral structures slow the liquid and relate to GFA \cite{Cheng2011, Chen2011, Bokas2016, janqi2020}. However, previous studies suggest a positive correlation between the icosahedral VP and GFA, which implies a larger ICO-like value would give a lower $log_{10}(R_c)$, which is the opposite of the trend in Fig.~\ref{shap}. It is worth noting that the impact of this variable in the SHAP analysis is modest, the negative trend of $log_{10}(R_c)$ with ICO-like value is weak and the overall impact of this variable on the model accuracy in Fig.~\ref{selection} is very modest. Taken together it is reasonable to say that the impact of this variable on $log_{10}(R_c)$ is not well determined by the preset model and data, so disagreement with trends is not unexpected. However, it may also be that icosahedral fractions do not have simple correlations with GFA and $log_{10}(R_c)$ across so many widely varying chemistries, especially when evaluated at a single fixed temperature (e.g., rather than a fixed homologous temperature). Even the definition of VP that are considered ICO-like is somewhat ambiguous and can vary because of distortions \cite{ding2014}. In summary, when one looks deeply, the possibly somewhat unphysical behavior of the ICO-like variable correlations is not evidence of any significant issue with the model.

We highlight the significance of developing a model of experimental $log_{10}(R_c)$ from features developed computationally, especially since many of the features from $X_{best}$ are more simple to acquire computationally compared to their experimental counterparts. Take $E_{above}$ and $\Delta C$ as examples. While $\Delta C$ can be obtained through calorimetry, it remains challenging unless a large quantity of a high GFA alloy is accessible (i.e., this quantity cannot be easily experimentally measured for low GFA alloys). Determining $E_{above}$ experimentally is even more difficult due to the complexities in establishing a baseline for differential scanning calorimetry. Thus, leveraging computational models not only streamlines the process of measuring critical material properties but also offers an alternative to experimental characterization and their inherent challenges.

Together the 4 features yield an apparently fairly capable model of $log_{10}(R_c)$, with an $R^2$ of 0.78 (although we reiterate that this is likely an overestimation due to some data leakage). Regardless of model metrics, it is crucial to note that the model should not be relied upon for any significant materials screening. Further tuning of model parameters with a larger dataset than the current 34 observations is necessary to assure the model has a reasonable domain of applicability.  However, these initial results are quite encouraging. Furthermore, generating larger data sets is largely limited by computation, which will become much easier with increases in computational power and more efficient computational techniques. This study suggests a promising path for integrating elemental and simulated features for predicting GFA in metals and illustrates how researchers can explore and identify physically motivated features with the flexibility now provided by MLPs. 
\section{Conclusion}
\label{conclusion}

In this work we explored an extensive range of elemental and molecular simulation calculated properties to develop a simple model for predicting critical cooling rate, $R_{c}$ for metallic glasses. In particular, we employed machine-learned interatomic potentials to facilitate the simulation of materials properties inaccessible via traditional ab initio or classical interatomic potential methods and difficult or expensive to obtain through experiments, all while maintaining accuracy comparable to ab initio and computational efficiency comparable to potentials. We conducted simulations on 34 compositions to collect properties potentially linked to the glass forming ability in metallic systems. We investigated self-diffusion, viscosity, characteristic temperatures, short range order, and energy comparisons between material crystalline and amorphous phases. Additionally, we explored easily computed features based on elemental properties that do not require simulation. Out of this feature set, we identified that four features were particularly significant for constructing regression models to predict $R_{c}$. Notably, one of the features was obtained through trivial computation, while the other three required simulation with machine-learned interatomic potentials. Evaluated models could predict $log_{10}(R_{c})$ with an $R^{2}$ of 0.78 for chemical systems excluded from model training. While our model had an impressive $R^{2}$, caution is warranted due to potential overfitting from the feature selection process and the small data sets involved. More broadly, we have demonstrated a versatile approach to systematically explore previously inaccessible material properties across diverse chemical systems using machine learned potentials. Almost all properties used as features for our model followed physical intuition outlined by previous research. Other computational endeavors can benefit from adopting a similar approach to quantify desirable characteristics across multiple materials, rather than focusing solely on properties within individual chemical systems.
\section{Data Availability}
\label{data_availability}

The raw and processed data required to reproduce these findings are available to download from figshare at doi: 10.6084/m9.figshare.26142382.
\section{Acknowledgments}
\label{acknowledgments}

Lane E. Schultz is grateful for the Bridge to the Doctorate: Wisconsin Louis Stokes Alliance for Minority Participation National Science Foundation (NSF) award number HRD-1612530, the University of Wisconsin–Madison Graduate Engineering Research Scholars (GERS) fellowship program for the financial support for graduate student investigation, and the PPG Coating Innovation Center for financial support. Dr. Afflerbach gratefully acknowledges research support from the NSF through the University of Wisconsin Materials Research Science and Engineering Center (DMR-2309000). All authors gratefully acknowledge support from the NSF Collaborative Research: Framework: Machine Learning Materials Innovation Infrastructure award number 1931306. Machine learning was performed with the computational resources provided by XSEDE 2.0: Integrating, Enabling and Enhancing National Cyberinfrastructure with Expanding Community Involvement Grant ACI-1548562.

\pagebreak
\printbibliography[heading=bibintoc]

@article{Long2023,
    author = {Long, Tao and Long, Zhilin and Peng, Zheng},
    doi = {10.1007/s10853-023-08528-x},
    issn = {15734803},
    journal = {Journal of Materials Science},
    month = {05},
    number = {21},
    pages = {8833--8844},
    title = {{Rational design and glass-forming ability prediction of bulk metallic glasses via interpretable machine learning}},
    url = {https://link.springer.com/10.1007/s10853-023-08528-x},
    volume = {58},
    year = {2023}
}

@article{Johnson2016,
    author = {Johnson, W. L. and Na, J. H. and Demetriou, M. D.},
    doi = {10.1038/ncomms10313},
    issn = {20411723},
    journal = {Nature Communications},
    month = {12},
    number = {1},
    pages = {10313},
    publisher = {Nature Publishing Group},
    title = {{Quantifying the origin of metallic glass formation}},
    url = {http://www.nature.com/articles/ncomms10313},
    volume = {7},
    year = {2016}
}

@article{Dai2019-2,
    author = {Dai, R. and Ashcraft, R. and Gangopadhyay, A. K. and Kelton, K. F.},
    doi = {10.1016/j.jnoncrysol.2019.119673},
    issn = {00223093},
    journal = {Journal of Non-Crystalline Solids},
    number = {October},
    pages = {119673},
    publisher = {Elsevier},
    title = {{Predicting metallic glass formation from properties of the high temperature liquid}},
    url = {https://doi.org/10.1016/j.jnoncrysol.2019.119673},
    volume = {525},
    year = {2019}
}

@article{Jaiswal2016,
    author = {Jaiswal, Abhishek and Egami, Takeshi and Kelton, K F and Schweizer, Kenneth S and Zhang, Yang},
    doi = {10.1103/PhysRevLett.117.205701},
    title = {{Correlation between Fragility and the Arrhenius Crossover Phenomenon in Metallic, Molecular, and Network Liquids}},
    url = {https://journals.aps.org/prl/pdf/10.1103/PhysRevLett.117.205701},
    year = {2016}
}

@article{Kube2022,
    author = {Kube, Sebastian A. and Sohn, Sungwoo and Ojeda-Mota, Rodrigo and Evers, Theo and Polsky, William and Liu, Naijia and Ryan, Kevin and Rinehart, Sean and Sun, Yong and Schroers, Jan},
    doi = {10.1038/s41467-022-31314-3},
    issn = {2041-1723},
    journal = {Nature Communications},
    month = {12},
    number = {1},
    pages = {3708},
    title = {{Compositional dependence of the fragility in metallic glass forming liquids}},
    url = {https://www.nature.com/articles/s41467-022-31314-3},
    volume = {13},
    year = {2022}
}

@article{Sheng2012,
  author = {Sheng, H. W. and Ma, E. and Kramer, M. J.},
  year = {2012},
  title = {Relating Dynamic Properties to Atomic Structure in Metallic Glasses},
  journal = {JOM},
  volume = {64},
  number = {7},
  pages = {856--881},
  month = {7},
  issn = {1543-1851},
  url = {https://doi.org/10.1007/s11837-012-0360-y},
  doi = {10.1007/s11837-012-0360-y}
}

@article{Herzer2013,
    author = {Herzer, Giselher},
    doi = {10.1016/J.ACTAMAT.2012.10.040},
    issn = {1359-6454},
    journal = {Acta Materialia},
    month = {02},
    number = {3},
    pages = {718--734},
    publisher = {Pergamon},
    title = {{Modern soft magnets: Amorphous and nanocrystalline materials}},
    url = {https://www.sciencedirect.com/science/article/pii/S1359645412007872?via%3Dihub},
    volume = {61},
    year = {2013}
}

@article{Li2016,
    author = {Li, H. F. and Zheng, Y. F.},
    booktitle = {Acta Biomaterialia},
    doi = {10.1016/j.actbio.2016.03.047},
    issn = {18787568},
    title = {{Recent advances in bulk metallic glasses for biomedical applications}},
    year = {2016}
}

@article{Jafary-Zadeh2018,
    author = {Jafary-Zadeh, Mehdi and {Praveen Kumar}, Gideon and Branicio, Paulo and Seifi, Mohsen and Lewandowski, John and Cui, Fangsen},
    doi = {10.3390/jfb9010019},
    issn = {2079-4983},
    journal = {Journal of Functional Biomaterials},
    number = {1},
    pages = {19},
    title = {{A Critical Review on Metallic Glasses as Structural Materials for Cardiovascular Stent Applications}},
    url = {http://www.mdpi.com/2079-4983/9/1/19},
    volume = {9},
    year = {2018}
}

@article{Wang2004,
    title = {Bulk metallic glasses},
    journal = {Materials Science and Engineering: R: Reports},
    volume = {44},
    number = {2},
    pages = {45-89},
    year = {2004},
    issn = {0927-796X},
    doi = {https://doi.org/10.1016/j.mser.2004.03.001},
    url = {https://www.sciencedirect.com/science/article/pii/S0927796X04000300},
    author = {W.H. Wang and C. Dong and C.H. Shek},
}

@article{Long2009,
    title = {A new criterion for predicting the glass-forming ability of bulk metallic glasses},
    journal = {Journal of Alloys and Compounds},
    volume = {475},
    number = {1},
    pages = {207-219},
    year = {2009},
    issn = {0925-8388},
    doi = {https://doi.org/10.1016/j.jallcom.2008.07.087},
    url = {https://www.sciencedirect.com/science/article/pii/S0925838808012206},
    author = {Zhilin Long and Hongqin Wei and Yanhuan Ding and Ping Zhang and Guoqiang Xie and Akihisa Inoue},
}

@article{Gao2020,
    author = {Gao, Qian and Jian, Zengyun},
    doi = {10.1016/j.matchemphys.2020.123252},
    issn = {02540584},
    journal = {Materials Chemistry and Physics},
    number = {May},
    pages = {123252},
    publisher = {Elsevier B.V.},
    title = {{Fragility and Vogel-Fulcher-Tammann parameters near glass transition temperature}},
    url = {https://doi.org/10.1016/j.matchemphys.2020.123252},
    volume = {252},
    year = {2020}
}

@article{Afflerbach2022,
    author = {Afflerbach, Benjamin T. and Francis, Carter and Schultz, Lane E. and Spethson, Janine and Meschke, Vanessa and Strand, Elliot and Ward, Logan and Perepezko, John H. and Thoma, Dan and Voyles, Paul M. and Szlufarska, Izabela and Morgan, Dane},
    doi = {10.1021/acs.chemmater.1c03542},
    issn = {0897-4756},
    journal = {Chemistry of Materials},
    month = {03},
    pages = {acs.chemmater.1c03542},
    title = {{Machine Learning Prediction of the Critical Cooling Rate for Metallic Glasses from Expanded Datasets and Elemental Features}},
    url = {https://pubs.acs.org/doi/10.1021/acs.chemmater.1c03542},
    year = {2022}
}

@article{Afflerbach2021,
    author = {Afflerbach, Benjamin T. and Schultz, Lane and Perepezko, John H. and Voyles, Paul M. and Szlufarska, Izabela and Morgan, Dane},
    doi = {10.1016/j.commatsci.2021.110728},
    issn = {09270256},
    journal = {Computational Materials Science},
    month = {11},
    publisher = {Elsevier B.V.},
    title = {{Molecular simulation-derived features for machine learning predictions of metal glass forming ability}},
    volume = {199},
    year = {2021}
}

@article{Schultz2021,
    title = {Exploration of characteristic temperature contributions to metallic glass forming ability},
    journal = {Computational Materials Science},
    volume = {196},
    pages = {110494},
    year = {2021},
    issn = {0927-0256},
    doi = {https://doi.org/10.1016/j.commatsci.2021.110494},
    url = {https://www.sciencedirect.com/science/article/pii/S0927025621002196},
    author = {Lane E. Schultz and Benjamin Afflerbach and Carter Francis and Paul M. Voyles and Izabela Szlufarska and Dane Morgan},
}

@article{Schultz2022,
    title = {Molecular dynamic characteristic temperatures for predicting metallic glass forming ability},
    journal = {Computational Materials Science},
    volume = {201},
    pages = {110877},
    year = {2022},
    issn = {0927-0256},
    doi = {https://doi.org/10.1016/j.commatsci.2021.110877},
    url = {https://www.sciencedirect.com/science/article/pii/S0927025621005899},
    author = {Lane E. Schultz and Benjamin Afflerbach and Izabela Szlufarska and Dane Morgan},
}

@article{Jain2013,
    author = {Jain, Anubhav and Ong, Shyue Ping and Hautier, Geoffroy and Chen, Wei and Richards, William Davidson and Dacek, Stephen and Cholia, Shreyas and Gunter, Dan and Skinner, David and Ceder, Gerbrand and Persson, Kristin a.},
    doi = {10.1063/1.4812323},
    issn = {2166532X},
    journal = {APL Materials},
    number = {1},
    pages = {11002},
    title = {{The Materials Project: A materials genome approach to accelerating materials innovation}},
    url = {http://link.aip.org/link/AMPADS/v1/i1/p011002/s1%5C&Agg=doi},
    volume = {1},
    year = {2013}
}

@article{Ward2018,
    title = {A machine learning approach for engineering bulk metallic glass alloys},
    journal = {Acta Materialia},
    volume = {159},
    pages = {102-111},
    year = {2018},
    issn = {1359-6454},
    doi = {https://doi.org/10.1016/j.actamat.2018.08.002},
    url = {https://www.sciencedirect.com/science/article/pii/S1359645418306268},
    author = {Logan Ward and Stephanie C. O'Keeffe and Joseph Stevick and Glenton R. Jelbert and Muratahan Aykol and Chris Wolverton},
}

@article{Zuo2020,
    author = {Yunxing Zuo and Chi Chen and Xiangguo Li and Zhi Deng and Yiming Chen and Jörg Behler and Gábor Csányi and Alexander V. Shapeev and Aidan P. Thompson and Mitchell A. Wood and Shyue Ping Ong},
    doi = {10.1021/acs.jpca.9b08723},
    issn = {15205215},
    issue = {4},
    journal = {Journal of Physical Chemistry A},
    pages = {731-745},
    pmid = {31916773},
    title = {Performance and Cost Assessment of Machine Learning Interatomic Potentials},
    volume = {124},
    year = {2020},
}

@article{Novikov2021,
    author = {Ivan S Novikov and Konstantin Gubaev and Evgeny V Podryabinkin and Alexander V Shapeev},
    doi = {10.1088/2632-2153/abc9fe},
    issn = {2632-2153},
    issue = {2},
    journal = {Machine Learning: Science and Technology},
    pages = {025002},
    title = {The MLIP package: moment tensor potentials with MPI and active learning},
    volume = {2},
    year = {2021},
}

@article{polak2023extracting,
    author = {Maciej P. Polak and Dane Morgan},
    doi = {10.1038/s41467-024-45914-8},
    issn = {20411723},
    issue = {1},
    journal = {Nature Communications},
    month = {12},
    pmid = {38383556},
    publisher = {Nature Research},
    title = {Extracting accurate materials data from research papers with conversational language models and prompt engineering},
    volume = {15},
    year = {2024},
}

@article{polak2023flexible,
    title={Flexible, Model-Agnostic Method for Materials Data Extraction from Text Using General Purpose Language Models}, 
    author={Maciej P. Polak and Shrey Modi and Anna Latosinska and Jinming Zhang and Ching-Wen Wang and Shanonan Wang and Ayan Deep Hazra and Dane Morgan},
    year={2023},
    eprint={2302.04914},
    archivePrefix={arXiv},
    primaryClass={cond-mat.mtrl-sci}
}

@misc{mg_foundry,
    title = {Metallic Glasses and their Properties},
    author = {Paul M. Voyles and Lane E. Schultz and Dane Morgan and Carter Francis and Benjamin Afflerbach and Abdulrhman Hakeem},
    howpublished = {Data set},
    publisher = {Foundry-ML},
    doi = {10.18126/7yg1-osf2},
    url = {https://foundry-ml.org/#/datasets/10.18126%2F7yg1-osf2},
}

@article{Jacobs2020,
    author = {Ryan Jacobs and Tam Mayeshiba and Ben Afflerbach and Luke Miles and Max Williams and Matthew Turner and Raphael Finkel and Dane Morgan},
    doi = {10.1016/j.commatsci.2020.109544},
    issn = {09270256},
    issue = {October 2019},
    journal = {Computational Materials Science},
    title = {The Materials Simulation Toolkit for Machine learning (MAST-ML): An automated open source toolkit to accelerate data-driven materials research},
    volume = {176},
    year = {2020},
}

@inproceedings{xgboost,
    series={KDD ’16},
    title={XGBoost: A Scalable Tree Boosting System},
    url={http://dx.doi.org/10.1145/2939672.2939785},
    DOI={10.1145/2939672.2939785},
    booktitle={Proceedings of the 22nd ACM SIGKDD International Conference on Knowledge Discovery and Data Mining},
    publisher={ACM},
    author={Chen, Tianqi and Guestrin, Carlos},
    year={2016},
    month=aug, collection={KDD ’16}
}

@Article{LAMMPS,
    author = "A. P. Thompson and H. M. Aktulga and R. Berger and 
    D. S. Bolintineanu and W. M. Brown and P. S. Crozier and
    P. J. in 't Veld and A. Kohlmeyer and S. G. Moore and T. D. Nguyen and
    R. Shan and M. J. Stevens and J. Tranchida and C. Trott and S. J. Plimpton",
    title = "{LAMMPS} - a flexible simulation tool for
    particle-based materials modeling at the 
    atomic, meso, and continuum scales",
    journal = "Comp. Phys. Comm.",
    volume =  "271",
    pages =   "108171",
    year =    "2022",
    doi = "10.1016/j.cpc.2021.108171"
}

@article{VASP,
    title = {Ab initio molecular dynamics for liquid metals},
    author = {Kresse, G. and Hafner, J.},
    journal = {Phys. Rev. B},
    volume = {47},
    issue = {1},
    pages = {558--561},
    numpages = {0},
    year = {1993},
    month = {01},
    publisher = {American Physical Society},
    doi = {10.1103/PhysRevB.47.558},
    url = {https://link.aps.org/doi/10.1103/PhysRevB.47.558}
}

@book{Rapaport,
    author = {Rapaport, D.C.},
    publisher = {Cambridge University Press},
    title = {{The Art of Molecular Dynamics Simulation}},
    isbn = {9780521825689},
    year = {2007}
}

@article{Puosi2018,
    author = {Puosi, F. and Jakse, N. and Pasturel, A.},
    doi = {10.1088/1361-648X/aab110},
    issn = {1361648X},
    journal = {Journal of Physics Condensed Matter},
    month = {3},
    number = {14},
    publisher = {Institute of Physics Publishing},
    title = {{Dynamical, structural and chemical heterogeneities in a binary metallic glass-forming liquid}},
    volume = {30},
    year = {2018}
}

@article{Mauro2009,
    author = {John C Mauro and Yuanzheng Yue and Adam J Ellison and Prabhat K Gupta and Douglas C Allan},
    title = {Viscosity of glass-forming liquids},
    year = {2009},
}

@article{Reis2012,
    author = {Raphael M.C.V. Reis and John C. Mauro and Karen L. Geisinger and Marcel Potuzak and Morten M. Smedskjaer and Xiaoju Guo and Douglas C. Allan},
    doi = {10.1016/j.jnoncrysol.2011.11.029},
    issn = {00223093},
    issue = {3},
    journal = {Journal of Non-Crystalline Solids},
    pages = {648-651},
    publisher = {Elsevier B.V.},
    title = {Relationship between viscous dynamics and the configurational thermal expansion coefficient of glass-forming liquids},
    volume = {358},
    url = {http://dx.doi.org/10.1016/j.jnoncrysol.2011.11.029},
    year = {2012},
}

@article{Chen2016,
    author = {Chen, Yinshan and Zhang, Wei and Yu, Lian},
    doi = {10.1021/acs.jpcb.6b05658},
    issn = {15205207},
    journal = {Journal of Physical Chemistry B},
    number = {32},
    pages = {8007--8015},
    title = {{Hydrogen Bonding Slows Down Surface Diffusion of Molecular Glasses}},
    volume = {120},
    year = {2016}
}

@article{Angell1995,
    author = {Angell, Austen C.},
    doi = {10.1126/science.267.5206.1924},
    journal = {Science},
    number = {5206},
    pages = {1924--1935},
    title = {{Formation of Glasses from Liquids and Biopolymers}},
    url = {papers2://publication/uuid/46ED95A9-CC48-4242-A7C5-9438976B8C42},
    volume = {267},
    year = {1995}
}

@article{biroli2009random,
    title={The Random First-Order Transition Theory of Glasses: a critical assessment}, 
    author={G. Biroli and J. P. Bouchaud},
    year={2009},
    eprint={0912.2542},
    archivePrefix={arXiv},
    primaryClass={cond-mat.dis-nn}
}

@article{Gangopadhyay2017,
    author = {A.K. Gangopadhyay and K.F. Kelton},
    doi = {10.1557/jmr.2017.253},
    issn = {0884-2914},
    issue = {14},
    journal = {Journal of Materials Research},
    month = {7},
    pages = {2638-2657},
    publisher = {Cambridge University Press},
    title = {Recent progress in understanding high temperature dynamical properties and fragility in metallic liquids, and their connection with atomic structure},
    volume = {32},
    url = {https://www.cambridge.org/core/product/identifier/S0884291417002539/type/journal_article},
    year = {2017},
}

@incollection{shap,
    title = {A Unified Approach to Interpreting Model Predictions},
    author = {Lundberg, Scott M and Lee, Su-In},
    booktitle = {Advances in Neural Information Processing Systems 30},
    editor = {I. Guyon and U. V. Luxburg and S. Bengio and H. Wallach and R. Fergus and S. Vishwanathan and R. Garnett},
    pages = {4765--4774},
    year = {2017},
    publisher = {Curran Associates, Inc.},
    url = {http://papers.nips.cc/paper/7062-a-unified-approach-to-interpreting-model-predictions.pdf}
}

@article{CuZr_eam,
    author = {Mendelev, M. I. and Sun, Y. and Zhang, F. and Wang, C. Z. and Ho, K. M.},
    title = "{Development of a semi-empirical potential suitable for molecular dynamics simulation of vitrification in Cu-Zr alloys}",
    journal = {The Journal of Chemical Physics},
    volume = {151},
    number = {21},
    pages = {214502},
    year = {2019},
    month = {12},
    issn = {0021-9606},
    doi = {10.1063/1.5131500},
    url = {https://doi.org/10.1063/1.5131500},
    eprint = {https://pubs.aip.org/aip/jcp/article-pdf/doi/10.1063/1.5131500/13362744/214502\_1\_online.pdf},
}

@misc{Sheng,
    url={https://sites.google.com/site/eampotentials/},
    author = {Howard Sheng},
    journal={EAM potentials},
    author={Sheng, Howard}
}

@misc{PdSiPotential2024,
    title = {PdSi Potential Table},
    author = {Howard Sheng},
    journal={EAM potentials},
    howpublished = {\url{https://sites.google.com/site/eampotentials/table/pdsi?authuser=0}},
}

@article{eam1,
    title = {Highly optimized embedded-atom-method potentials for fourteen fcc metals},
    author = {Sheng, H. W. and Kramer, M. J. and Cadien, A. and Fujita, T. and Chen, M. W.},
    journal = {Phys. Rev. B},
    volume = {83},
    issue = {13},
    pages = {134118},
    numpages = {20},
    year = {2011},
    month = {04},
    publisher = {American Physical Society},
    doi = {10.1103/PhysRevB.83.134118},
    url = {https://link.aps.org/doi/10.1103/PhysRevB.83.134118}
}

@article{eam2,
    title = {Atomic Level Structure in Multicomponent Bulk Metallic Glass},
    author = {Cheng, Y. Q. and Ma, E. and Sheng, H. W.},
    journal = {Phys. Rev. Lett.},
    volume = {102},
    issue = {24},
    pages = {245501},
    numpages = {4},
    year = {2009},
    month = {06},
    publisher = {American Physical Society},
    doi = {10.1103/PhysRevLett.102.245501},
    url = {https://link.aps.org/doi/10.1103/PhysRevLett.102.245501}
}

@article{eam3,
    title = {Coupling between chemical and dynamic heterogeneities in a multicomponent bulk metallic glass},
    author = {Fujita, T. and Guan, P. F. and Sheng, H. W. and Inoue, A. and Sakurai, T. and Chen, M. W.},
    journal = {Phys. Rev. B},
    volume = {81},
    issue = {14},
    pages = {140204},
    numpages = {4},
    year = {2010},
    month = {04},
    publisher = {American Physical Society},
    doi = {10.1103/PhysRevB.81.140204},
    url = {https://link.aps.org/doi/10.1103/PhysRevB.81.140204}
}

@article{eam4,
    title = {Relationship between structure, dynamics, and mechanical properties in metallic glass-forming alloys},
    author = {Cheng, Y. Q. and Sheng, H. W. and Ma, E.},
    journal = {Phys. Rev. B},
    volume = {78},
    issue = {1},
    pages = {014207},
    numpages = {7},
    year = {2008},
    month = {07},
    publisher = {American Physical Society},
    doi = {10.1103/PhysRevB.78.014207},
    url = {https://link.aps.org/doi/10.1103/PhysRevB.78.014207}
}

@article{eam5,
    author = {Li, Qing-Jie and Sheng, Howard and Ma, Evan},
    title = {Strengthening in multi-principal element alloys with local-chemical-order roughened dislocation pathways},
    journal = {Nature Communications},
    year = {2019},
    volume = {10},
    number = {1},
    pages = {3563},
    month = {08},
    day = {08},
    doi = {10.1038/s41467-019-11464-7},
    issn = {2041-1723},
    url = {https://doi.org/10.1038/s41467-019-11464-7}
}

@article{scikit-learn,
    title={Scikit-learn: Machine Learning in {P}ython},
    author={Pedregosa, F. and Varoquaux, G. and Gramfort, A. and Michel, V.
     and Thirion, B. and Grisel, O. and Blondel, M. and Prettenhofer, P.
     and Weiss, R. and Dubourg, V. and Vanderplas, J. and Passos, A. and
     Cournapeau, D. and Brucher, M. and Perrot, M. and Duchesnay, E.},
    journal={Journal of Machine Learning Research},
    volume={12},
    pages={2825--2830},
    year={2011}
}

@article{Bokas2018,
    author = {G. B. Bokas and Y. Shen and L. Zhao and H. W. Sheng and J. H. Perepezko and I. Szlufarska},
    doi = {10.1007/s10853-018-2393-2},
    issn = {15734803},
    issue = {16},
    journal = {Journal of Materials Science},
    month = {8},
    pages = {11488-11499},
    publisher = {Springer New York LLC},
    title = {Synthesis of Sm–Al metallic glasses designed by molecular dynamics simulations},
    volume = {53},
    year = {2018},
}

@article{Jacobs2024,
    author = {Ryan Jacobs and Jian Liu and Harry Abernathy and Dane Morgan},
    doi = {10.1002/aenm.202303684},
    issn = {16146840},
    journal = {Advanced Energy Materials},
    publisher = {John Wiley and Sons Inc},
    title = {Machine Learning Design of Perovskite Catalytic Properties},
    year = {2024},
}

@article{Weeks2022,
    author = {W. Porter Weeks and Katharine M. Flores},
    doi = {10.1016/j.intermet.2022.107560},
    issn = {09669795},
    journal = {Intermetallics},
    month = {6},
    publisher = {Elsevier Ltd},
    title = {Using characteristic structural motifs in metallic liquids to predict glass forming ability},
    volume = {145},
    year = {2022},
}

@article{Wang2019,
    author = {Juan Wang and Anupriya Agrawal and Katharine Flores},
    doi = {10.1016/j.actamat.2019.04.001},
    issn = {13596454},
    journal = {Acta Materialia},
    month = {6},
    pages = {163-169},
    publisher = {Acta Materialia Inc},
    title = {Are hints about glass forming ability hidden in the liquid structure?},
    volume = {171},
    year = {2019},
}

@article{Bokas2016,
    author = {G. B. Bokas and L. Zhao and J. H. Perepezko and I. Szlufarska},
    doi = {10.1016/j.scriptamat.2016.06.045},
    issn = {13596462},
    journal = {Scripta Materialia},
    month = {11},
    pages = {99-102},
    publisher = {Elsevier Ltd},
    title = {On the role of Sm in solidification of Al-Sm metallic glasses},
    volume = {124},
    year = {2016},
}

@article{Stukowski2010,
    author = {Alexander Stukowski},
    doi = {10.1088/0965-0393/18/1/015012},
    issn = {09650393},
    issue = {1},
    journal = {Modelling and Simulation in Materials Science and Engineering},
    title = {Visualization and analysis of atomistic simulation data with OVITO-the Open Visualization Tool},
    volume = {18},
    year = {2010},
}

@article{Liu2023,
    author = {Guannan Liu and Sungwoo Sohn and Sebastian A. Kube and Arindam Raj and Andrew Mertz and Aya Nawano and Anna Gilbert and Mark D. Shattuck and Corey S. O'Hern and Jan Schroers},
    doi = {10.1016/j.actamat.2022.118497},
    issn = {13596454},
    journal = {Acta Materialia},
    month = {1},
    publisher = {Acta Materialia Inc},
    title = {Machine learning versus human learning in predicting glass-forming ability of metallic glasses},
    volume = {243},
    year = {2023},
}

@article{Ward2016,
  author = {Logan Ward and Ankit Agrawal and Alok Choudhary and Christopher Wolverton},
  title = {A general-purpose machine learning framework for predicting properties of inorganic materials},
  journal = {npj Computational Materials},
  volume = {2},
  number = {1},
  pages = {16028},
  year = {2016},
  doi = {10.1038/npjcompumats.2016.28},
  url = {https://doi.org/10.1038/npjcompumats.2016.28},
  issn = {2057-3960}
}

@article{janqi2020,
    title = {Microalloying effect in ternary Al-Sm-X (X=Ag, Au, Cu) metallic glasses studied by ab initio molecular dynamics},
    journal = {Computational Materials Science},
    volume = {185},
    pages = {109958},
    year = {2020},
    issn = {0927-0256},
    doi = {https://doi.org/10.1016/j.commatsci.2020.109958},
    url = {https://www.sciencedirect.com/science/article/pii/S0927025620304493},
    author = {J. Xi and G. Bokas and L.E. Schultz and M. Gao and L. Zhao and Y. Shen and J.H. Perepezko and D. Morgan and I. Szlufarska},
}

@article{Cheng2011,
    title = {Atomic-level structure and structure–property relationship in metallic glasses},
    journal = {Progress in Materials Science},
    volume = {56},
    number = {4},
    pages = {379-473},
    year = {2011},
    issn = {0079-6425},
    doi = {https://doi.org/10.1016/j.pmatsci.2010.12.002},
    url = {https://www.sciencedirect.com/science/article/pii/S0079642510000691},
    author = {Y.Q. Cheng and E. Ma},
}

@Manual{pwlf,
    author = {Jekel, Charles F. and Venter, Gerhard},
    title = {{pwlf:} A Python Library for Fitting 1D Continuous Piecewise Linear Functions},
    year = {2019},
    url = {https://github.com/cjekel/piecewise_linear_fit_py}
}

@article{ding2014,
    title = {Full icosahedra dominate local order in Cu64Zr34 metallic glass and supercooled liquid},
    journal = {Acta Materialia},
    volume = {69},
    pages = {343-354},
    year = {2014},
    issn = {1359-6454},
    doi = {https://doi.org/10.1016/j.actamat.2014.02.005},
    url = {https://www.sciencedirect.com/science/article/pii/S1359645414000858},
    author = {Jun Ding and Yong-Qiang Cheng and Evan Ma},
}

@article{Chen2011,
    author = {Chen, Mingwei},
    year = {2011},
    month = {9},
    title = {A brief overview of bulk metallic glasses},
    journal = {NPG Asia Materials},
    volume = {3},
    number = {9},
    pages = {82--90},
    issn = {1884-4057},
    url = {https://doi.org/10.1038/asiamat.2011.30},
    doi = {10.1038/asiamat.2011.30}
}
\pagebreak

\pagenumbering{arabic}
\setcounter{page}{1}
\setcounter{section}{0}

\renewcommand{\figurename}{Supplementary Figure}
\renewcommand{\tablename}{Supplementary Table}

\begin{center}
    {\huge Supplemental Materials for Machine Learning Metallic Glass Critical Cooling Rates Through Elemental and Molecular Simulation Based Featurization}
\end{center}

\section{Validation of Viscosity Computation}
\label{viscosity_validation}

Measuring viscosity via the Green-Kubo formalism is known to have long convergence times and can yield noisy results for individual runs. Therefore, we compared our acquisition of viscosity data with previously published work to ensure the quality of our calculations. For a single composition of $Cu_{50}Zr_{50}$, viscosities were measured through the same means described in the main text. We compared viscosity to Ref.~\cite{Puosi2018} for the potential in Ref.~\cite{CuZr_eam}. The only point of major disagreement was the lowest temperature point near $T_{f}$, which is reasonable given the sluggish kinetics nearing material freezing (Supplementary Figure~\ref{visc_compare}). Our data was averaged between 10 independent runs.

\begin{figure}[H]
	\centering
	\includegraphics[width=0.75\textwidth]{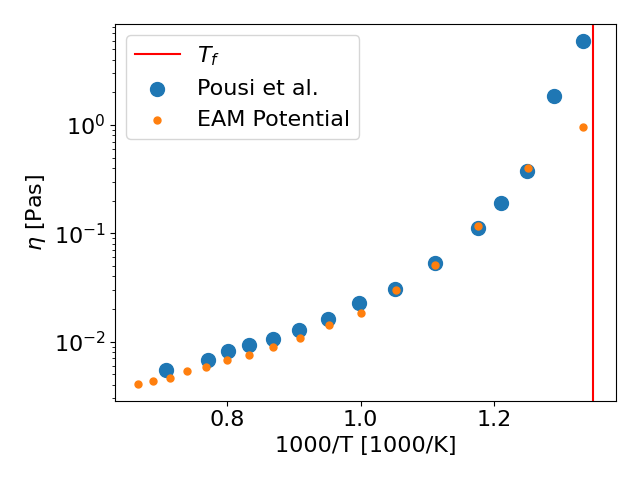}
	\caption{Comparison of viscosity.}
	\label{visc_compare}
\end{figure}

\section{Validation of MTP Fitting Methodology}

The energy and force errors from the MTP fitting methodology from EAM potentials are tabulated in Supplementary Table~\ref{eam_to_mtp_table}. Both classical potentials and MTPs were used to simulate the properties of potential energy, self-diffusion, and viscosity with respect to temperature and have been found to have excellent agreement (see figures below). The system size was 1,000 atoms when measuring these properties. Only $Ni_{80}P_{20}$ has slight, constant off-sets between the MLP and EAM potential for potential energy and viscosity with respect to temperature, and the offsets are small. More importantly, the relationship between temperature and potential energy, viscosity, and self-diffusion are similar. Properties were measured in the same manner described in Sec.~\ref{properties_mtp}.

\begin{table}[H]
\centering
\caption{The errors from MTPs compared with EAM potentials are tabulated below.}
\label{eam_to_mtp_table}
\begin{tabular}{llll}
\toprule
            Composition & Property &             Units &  RMSE \\
\midrule
        $Ni_{80}P_{20}$ &    Force & $eV/\text{\r{A}}$ & 0.187 \\
       $Pd_{75}Si_{25}$ &    Force & $eV/\text{\r{A}}$ & 0.124 \\
$Al_{10}Cu_{40}Zr_{50}$ &    Force & $eV/\text{\r{A}}$ & 0.080 \\
       $Pd_{75}Si_{25}$ &   Energy &         $eV/atom$ & 0.012 \\
        $Ni_{80}P_{20}$ &   Energy &         $eV/atom$ & 0.007 \\
$Al_{10}Cu_{40}Zr_{50}$ &   Energy &         $eV/atom$ & 0.004 \\
\bottomrule
\end{tabular}
\end{table}

\subsection{Potential Energy}

\begin{figure}[H]
	\centering
	\includegraphics[width=0.95\textwidth]{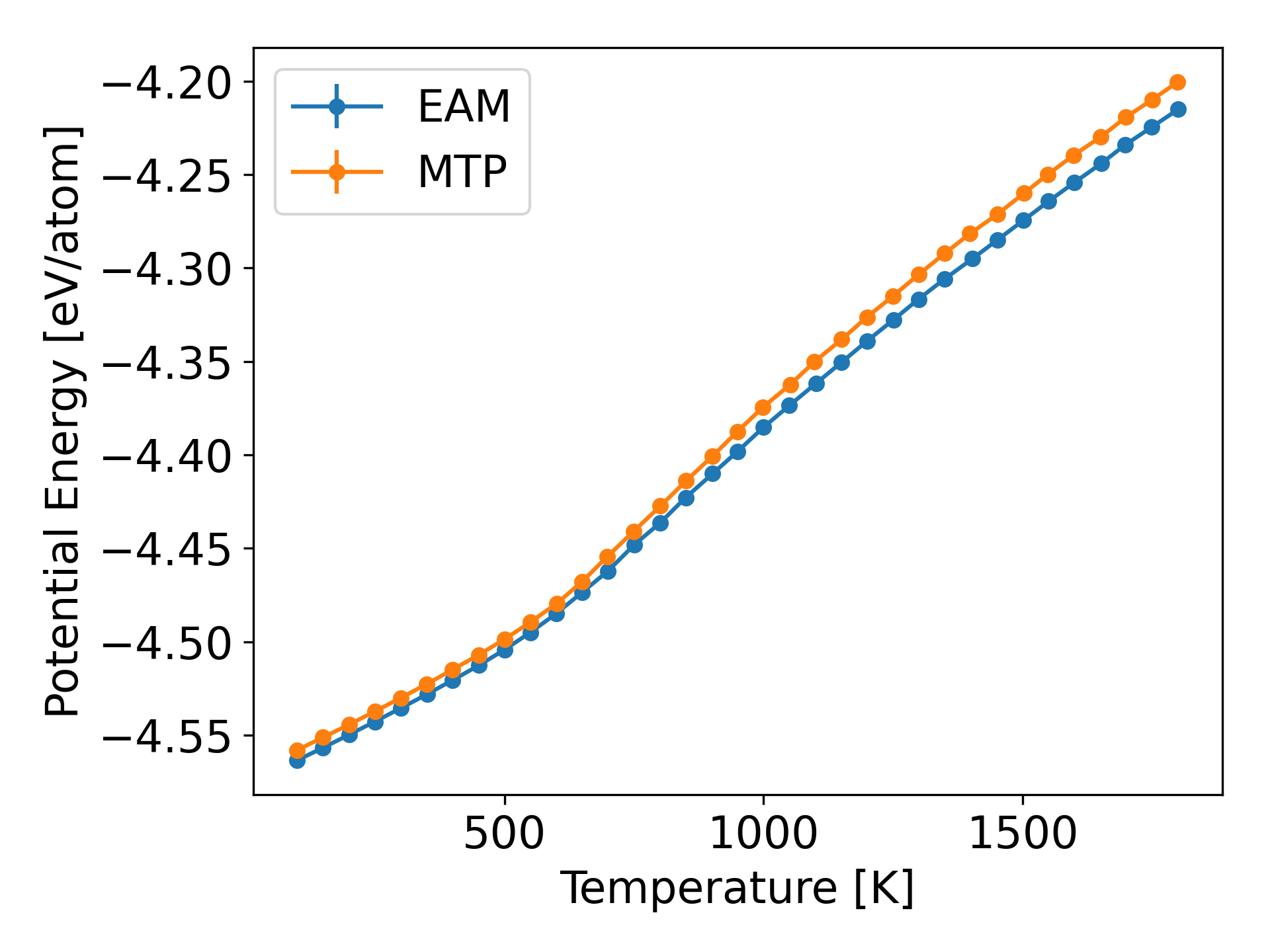}
	\caption{Comparison of potential energy between MLP and EAM potentials for $Ni_{80}P_{20}$.}
	\label{energy_EAM_MTP_Ni80P20}
\end{figure}

\begin{figure}[H]
	\centering
	\includegraphics[width=0.95\textwidth]{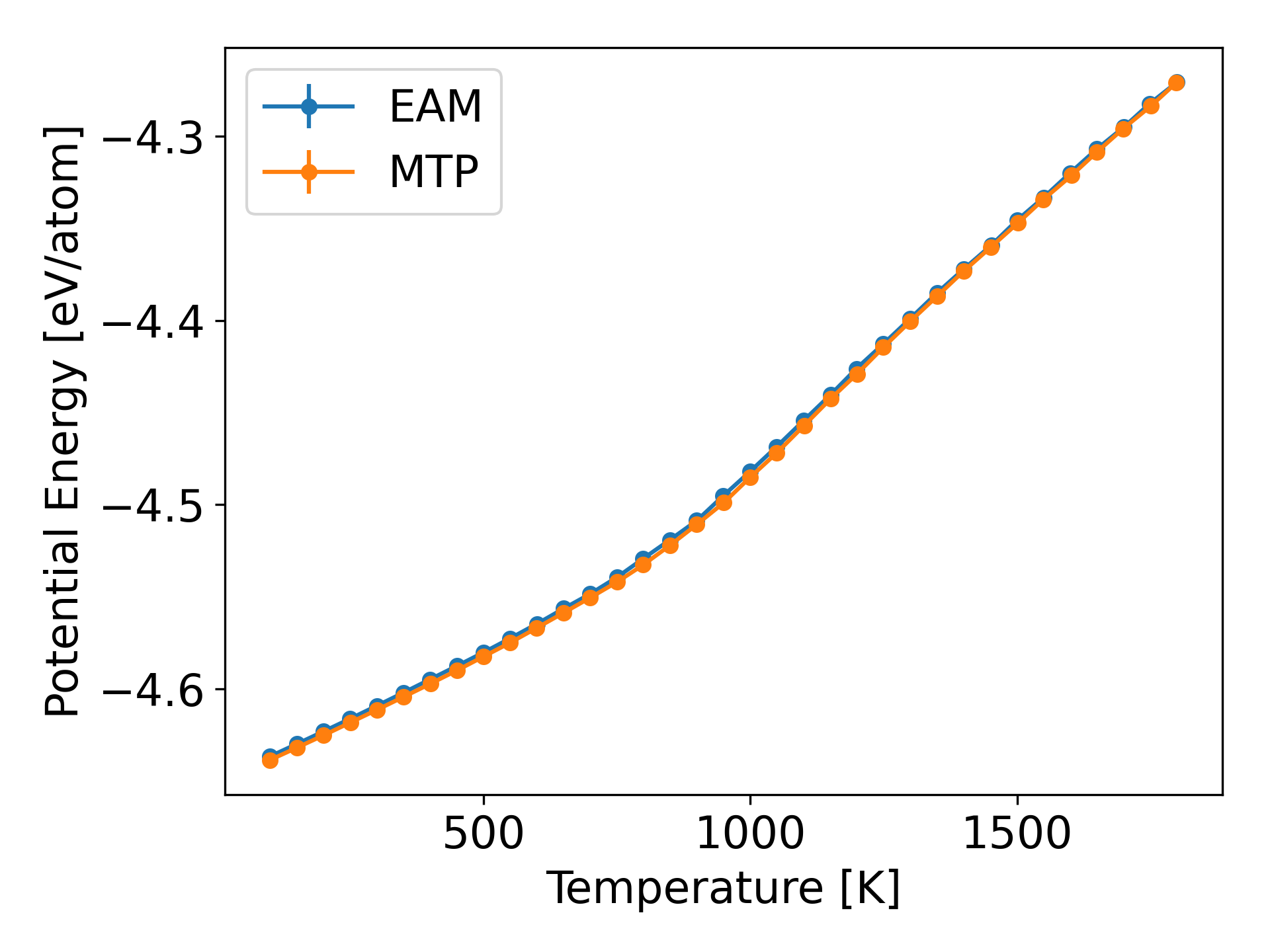}
	\caption{Comparison of potential energy between MLP and EAM potentials for $Pd_{75}Si_{25}$.}
	\label{energy_EAM_MTP_Pd75Si25}
\end{figure}

\begin{figure}[H]
	\centering
	\includegraphics[width=0.95\textwidth]{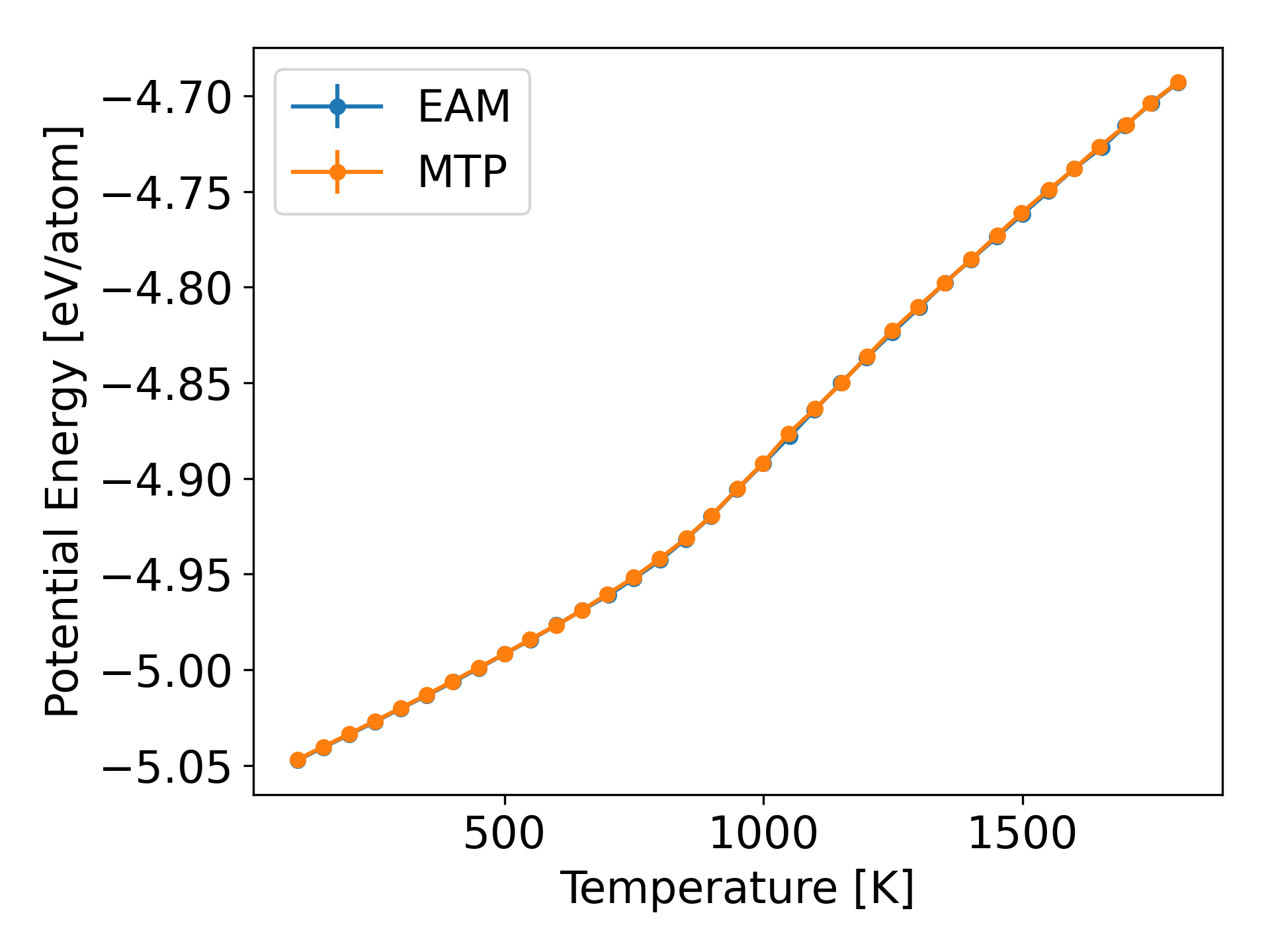}
	\caption{Comparison of potential energy between MLP and EAM potentials for $Al_{10}Cu_{40}Zr_{50}$.}
	\label{energy_EAM_MTP_Al10Cu40Zr50}
\end{figure}

\subsection{Self-Diffusion}

\begin{figure}[H]
	\centering
	\includegraphics[width=0.95\textwidth]{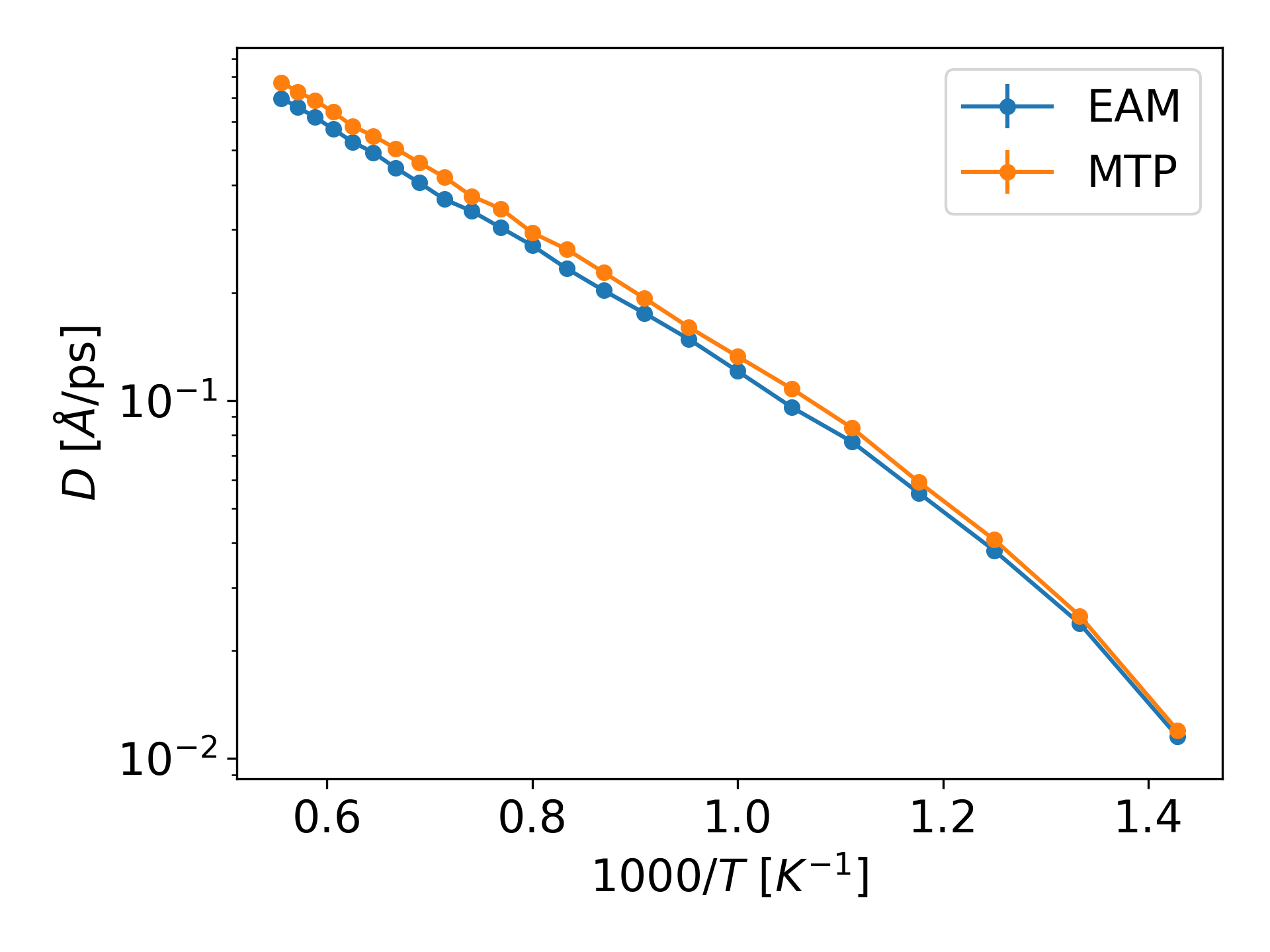}
	\caption{Comparison of self-diffusion between MLP and EAM potentials for $Ni_{80}P_{20}$.}
	\label{diff_EAM_MTP_Ni80P20}
\end{figure}

\begin{figure}[H]
	\centering
	\includegraphics[width=0.95\textwidth]{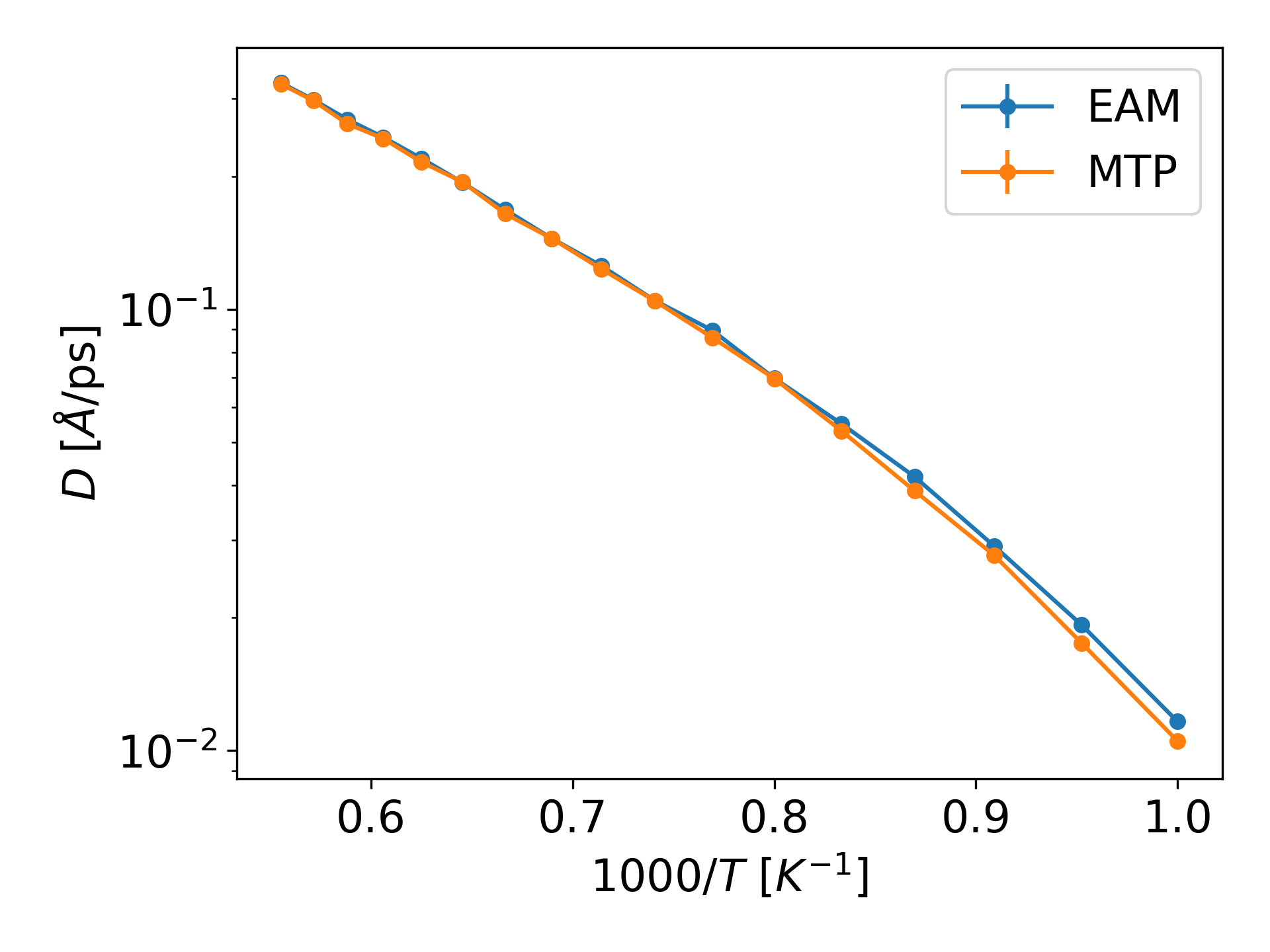}
	\caption{Comparison of self-diffusion between MLP and EAM potentials for $Pd_{75}Si_{25}$.}
	\label{diff_EAM_MTP_Pd75Si25}
\end{figure}

\begin{figure}[H]
	\centering
	\includegraphics[width=0.95\textwidth]{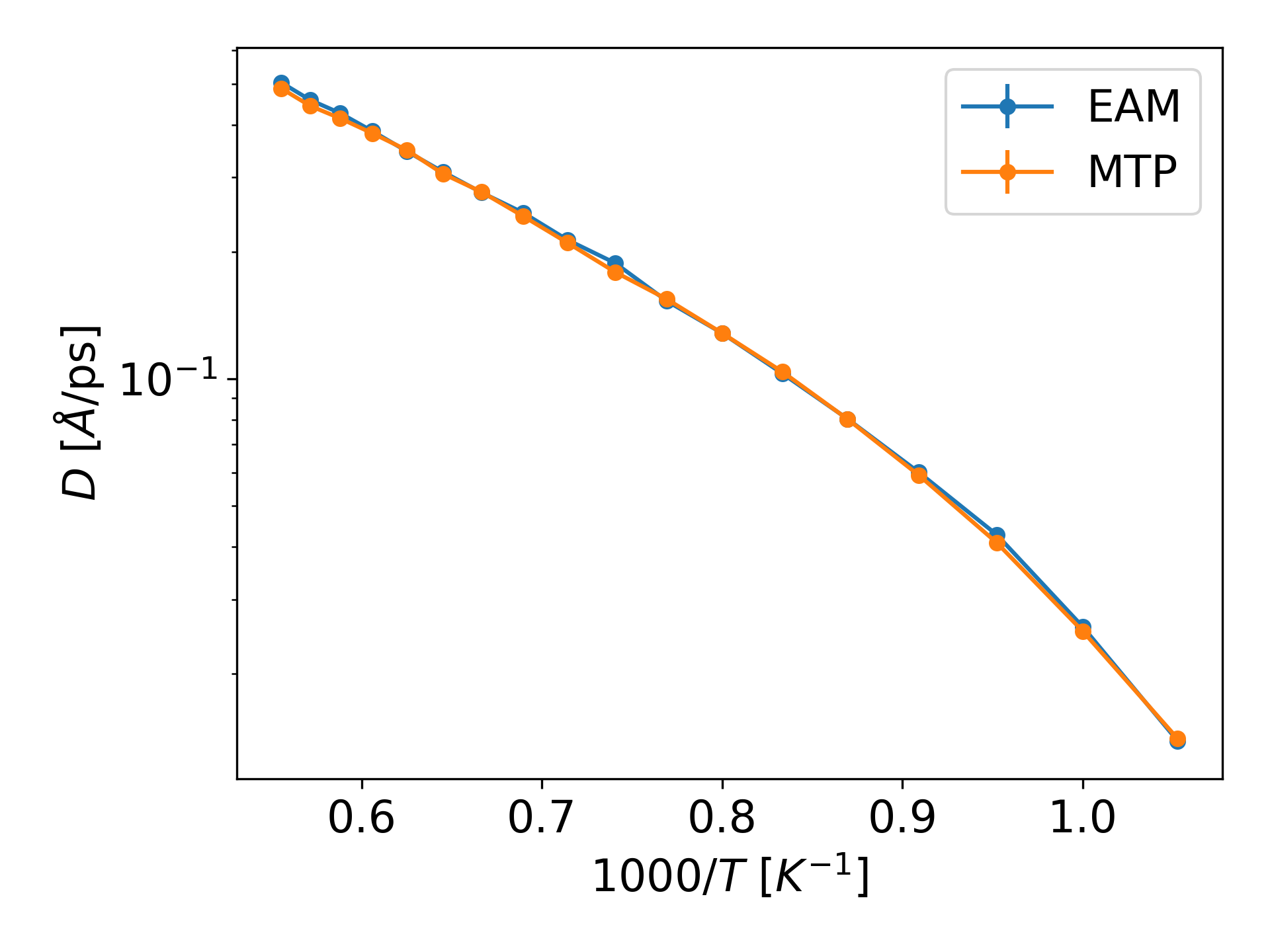}
	\caption{Comparison of self-diffusion between MLP and EAM potentials for $Al_{10}Cu_{40}Zr_{50}$.}
	\label{diff_EAM_MTP_Al10Cu40Zr50}
\end{figure}

\subsection{Viscosity}

\begin{figure}[H]
	\centering
	\includegraphics[width=0.95\textwidth]{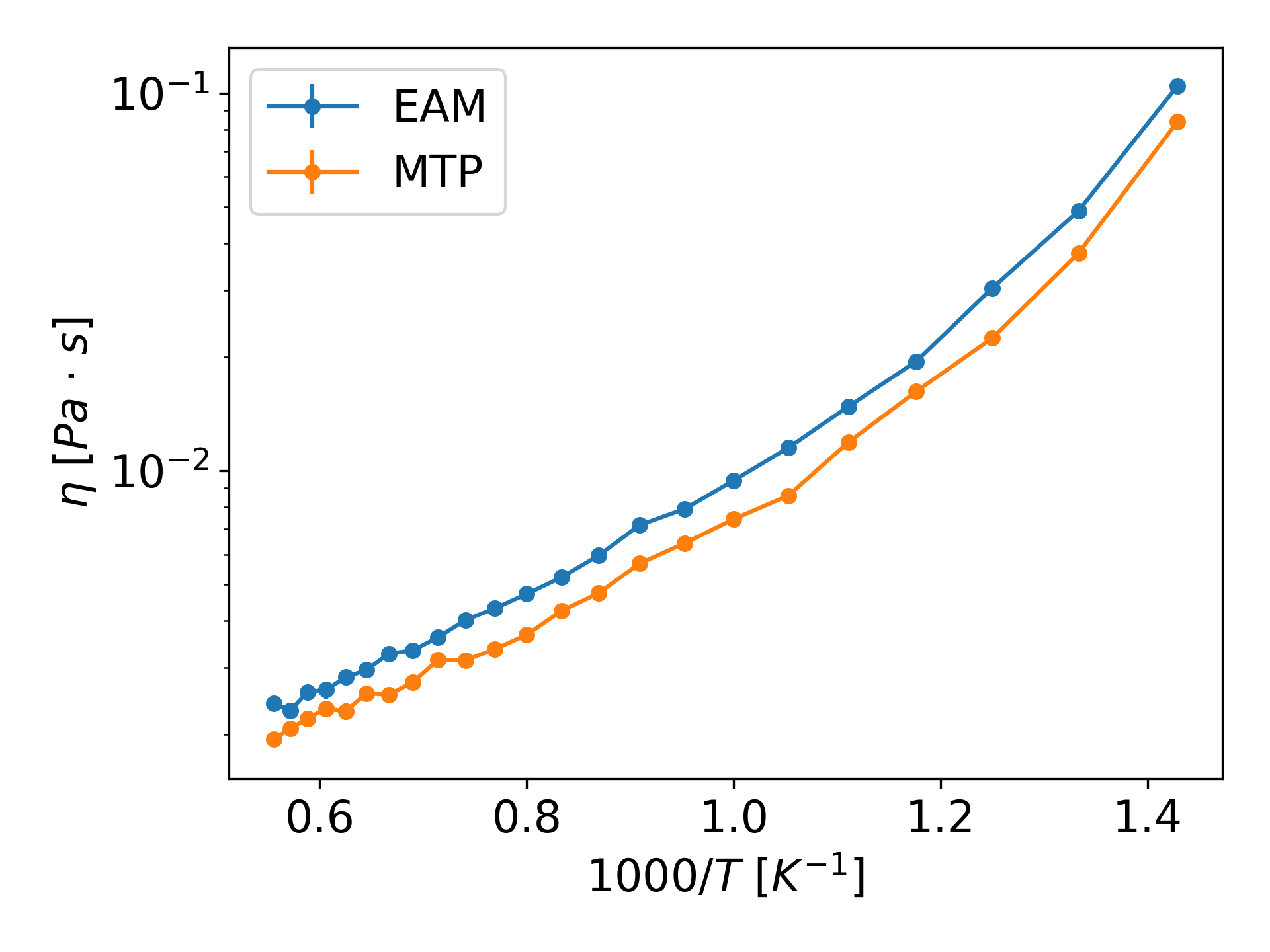}
	\caption{Comparison of viscosity between MLP and EAM potentials for $Ni_{80}P_{20}$.}
	\label{visc_EAM_MTP_Ni80P20}
\end{figure}

\begin{figure}[H]
	\centering
	\includegraphics[width=0.95\textwidth]{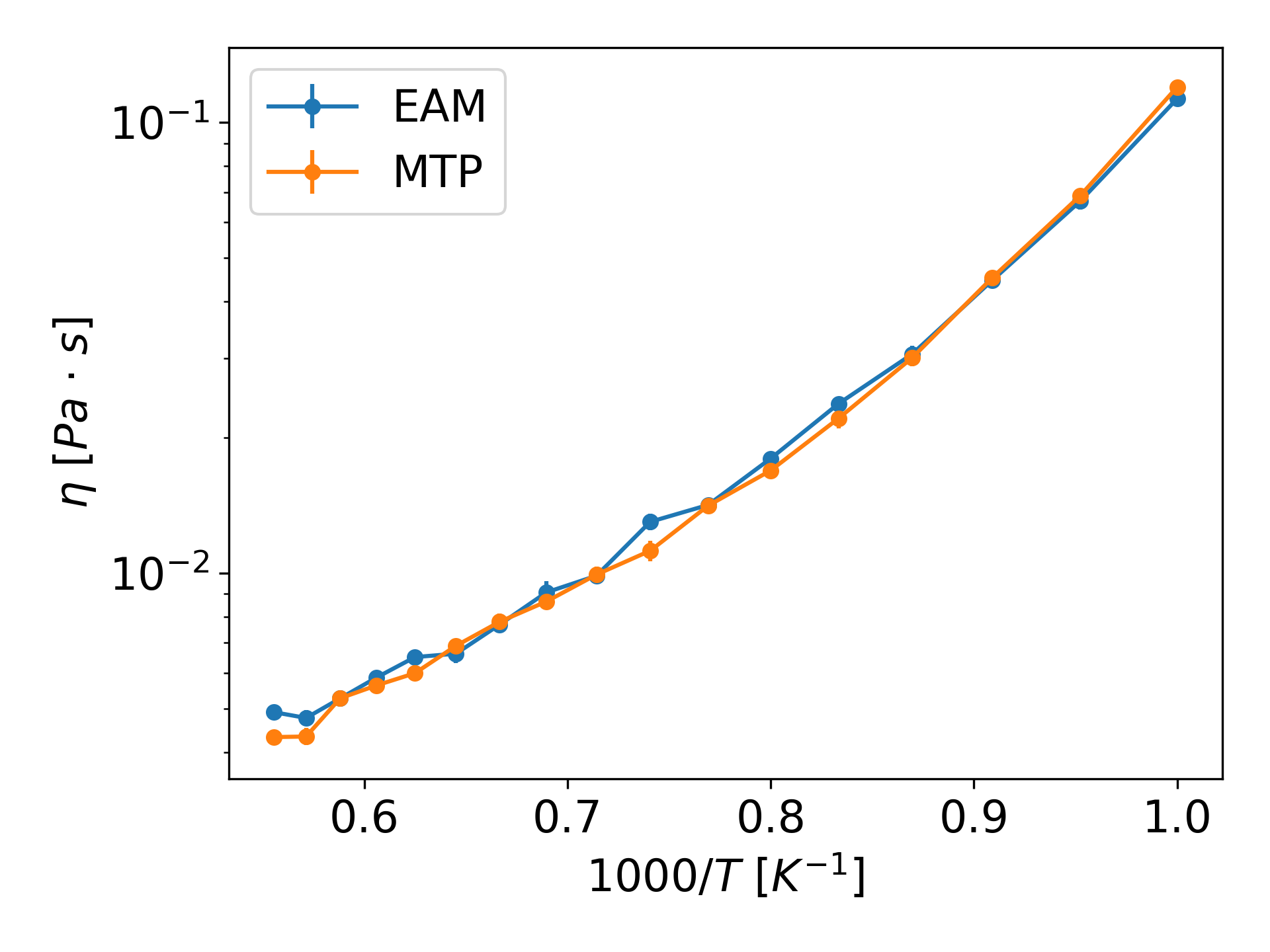}
	\caption{Comparison of viscosity between MLP and EAM potentials for $Pd_{75}Si_{25}$.}
	\label{visc_EAM_MTP_Pd75Si25}
\end{figure}

\begin{figure}[H]
	\centering
	\includegraphics[width=0.95\textwidth]{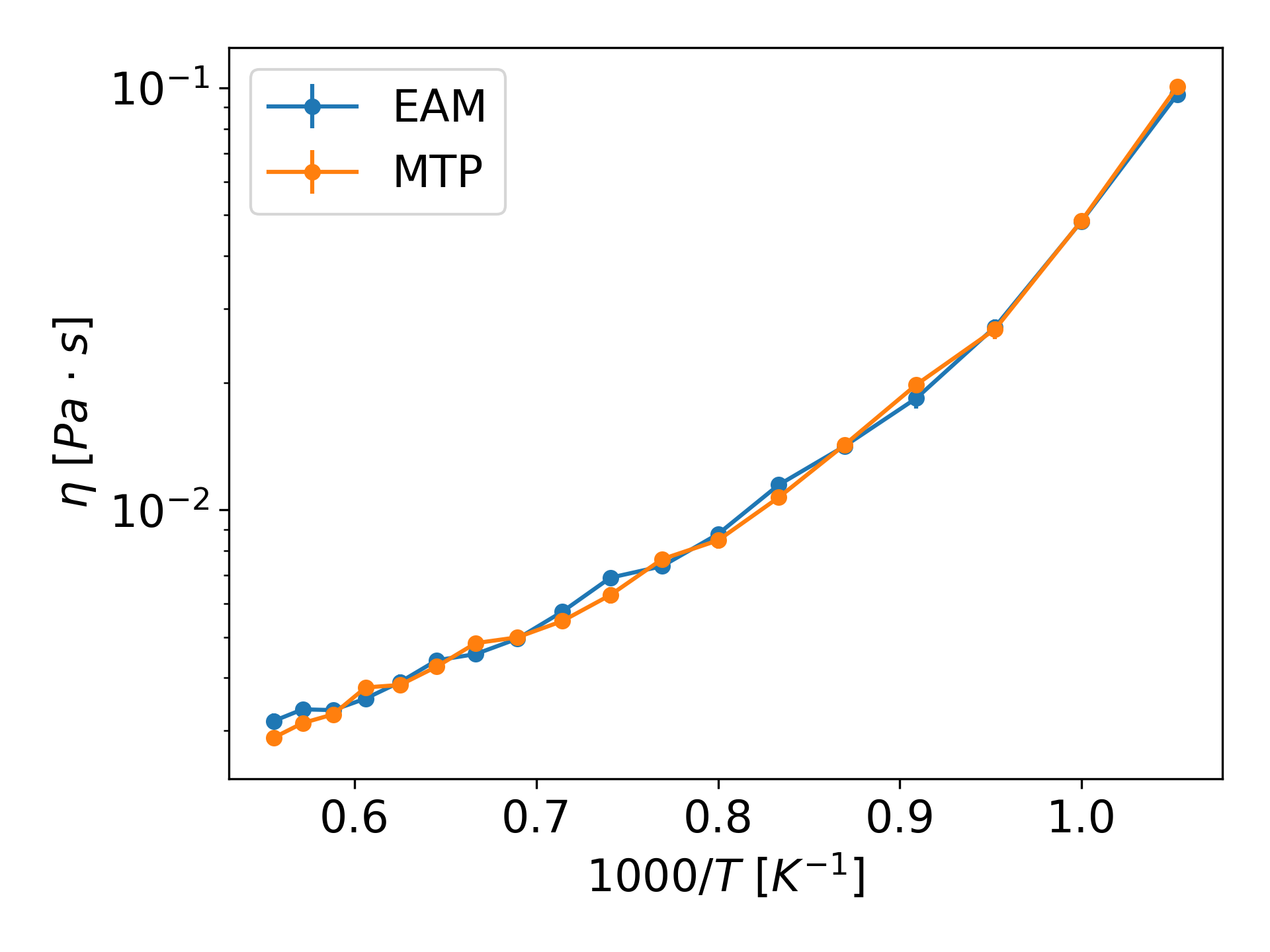}
	\caption{Comparison of viscosity between MLP and EAM potentials for $Al_{10}Cu_{40}Zr_{50}$.}
	\label{visc_EAM_MTP_Al10Cu40Zr50}
\end{figure}

\section{Energy and Force Errors from MLPs}

Each MLP has a corresponding energy and force $RMSE$ from the fitting methodology outlined in Sec.~\ref{mtp_validation} from the main text. The data are tabulated in Supplementary Table~\ref{mtp_errors}.

\begin{table}[H]
\centering
\caption{The energy $RMSE$, force $RMSE$, and number of training configurations from MTPs are tabulated below.}
\label{mtp_errors}
\begin{tabular}{lccr}
\toprule
                              Composition & Energy $[eV/atom]$ & Force $[eV/\text{\r{A}}]$ &  Configurations \\
\midrule
                    $Mg_{90}Nd_{5}Ni_{5}$ &              0.007 &                     0.084 &             992 \\
                   $Mg_{77}Nd_{5}Ni_{18}$ &              0.008 &                     0.092 &             990 \\
                          $Pd_{95}Si_{5}$ &              0.008 &                     0.109 &             893 \\
            $Cu_{30}Ni_{10}P_{20}Pd_{40}$ &              0.009 &                     0.143 &            1160 \\
                         $Pd_{82}Si_{18}$ &              0.009 &                     0.144 &             881 \\
                   $Ni_{40}P_{20}Pd_{40}$ &              0.009 &                     0.156 &            1039 \\
                   $Cu_{25}Mg_{65}Y_{10}$ &              0.010 &                     0.094 &             889 \\
                  $Cu_{25}Gd_{10}Mg_{65}$ &              0.010 &                     0.100 &             889 \\
                  $Mg_{65}Nd_{15}Ni_{20}$ &              0.010 &                     0.111 &            1010 \\
            $Cu_{25}Ni_{10}P_{20}Pd_{45}$ &              0.010 &                     0.145 &            1173 \\
            $Cu_{25}Ni_{15}P_{20}Pd_{40}$ &              0.010 &                     0.149 &            1172 \\
                  $Mg_{80}Nd_{10}Ni_{10}$ &              0.011 &                     0.098 &             978 \\
                  $Mg_{70}Nd_{15}Ni_{15}$ &              0.012 &                     0.113 &            1003 \\
                         $Pd_{75}Si_{25}$ &              0.012 &                     0.193 &             870 \\
$Nb_{7}Ni_{59}Si_{3}Sn_{2}Ti_{13}Zr_{16}$ &              0.012 &                     0.202 &            2541 \\
                  $Ca_{65}Mg_{15}Zn_{20}$ &              0.013 &                     0.113 &            1081 \\
                   $Cu_{6}Pd_{77}Si_{17}$ &              0.013 &                     0.156 &             917 \\
            $Cu_{47}Ni_{8}Ti_{34}Zr_{11}$ &              0.017 &                     0.198 &            1066 \\
             $Cu_{30}Ni_{5}P_{20}Pd_{45}$ &              0.019 &                     0.188 &            1046 \\
                         $Nb_{40}Ni_{60}$ &              0.019 &                     0.240 &             861 \\
                  $Al_{25}Cu_{20}La_{55}$ &              0.020 &                     0.158 &            1020 \\
             $Al_{9}Cu_{16}Ni_{9}Zr_{66}$ &              0.020 &                     0.230 &            1223 \\
            $Al_{25}Cu_{15}La_{55}Ni_{5}$ &              0.022 &                     0.167 &            1195 \\
            $Al_{8}Cu_{12}Ni_{14}Zr_{66}$ &              0.023 &                     0.239 &            1255 \\
                  $Al_{14}Cu_{20}La_{66}$ &              0.024 &                     0.162 &            1011 \\
    $Be_{25}Cu_{10}Ni_{10}Ti_{11}Zr_{44}$ &              0.024 &                     0.237 &            1510 \\
             $Al_{8}Cu_{7}Ni_{19}Zr_{66}$ &              0.025 &                     0.233 &            1203 \\
                         $Be_{35}Zr_{65}$ &              0.026 &                     0.250 &             870 \\
           $Al_{25}Cu_{10}La_{55}Ni_{10}$ &              0.028 &                     0.175 &            1230 \\
            $Al_{25}Cu_{5}La_{55}Ni_{15}$ &              0.028 &                     0.177 &            1247 \\
                  $Al_{25}La_{55}Ni_{20}$ &              0.029 &                     0.187 &             955 \\
                   $Al_{8}Ni_{26}Zr_{66}$ &              0.030 &                     0.235 &             946 \\
      $Al_{25}Co_{5}Cu_{10}La_{55}Ni_{5}$ &              0.031 &                     0.183 &            1773 \\
                         $Be_{37}Ti_{63}$ &              0.031 &                     0.290 &             903 \\
\bottomrule
\end{tabular}
\end{table}

\section{CSLO CV Chemical Systems}

The exact chemical system groups are tabulated in this section. For each iteration of CSLO CV, one group is selected and left out as a test set. For assessments on the 34 materials with MLP, entries from Table~\ref{mtp_chemical_systems_34} are used. CV for larger assessments of 177 materials use entries from Table~\ref{mtp_chemical_systems_177}.

\begin{table}[H]
\centering
\caption{The chemical systems considered for CSLO CV for the 34 materials with MLPs.}
\label{mtp_chemical_systems_34}
\begin{tabular}{lr}
\toprule
Chemical System &  Count \\
\midrule
         MgNdNi &      5 \\
        CuNiPPd &      4 \\
           PdSi &      3 \\
       AlCuNiZr &      3 \\
       AlCuLaNi &      3 \\
         AlCuLa &      2 \\
          NiPPd &      1 \\
   NbNiSiSnTiZr &      1 \\
           NbNi &      1 \\
         CuPdSi &      1 \\
       CuNiTiZr &      1 \\
          CuMgY &      1 \\
         CuGdMg &      1 \\
         CaMgZn &      1 \\
           BeZr &      1 \\
           BeTi &      1 \\
     BeCuNiTiZr &      1 \\
         AlNiZr &      1 \\
         AlLaNi &      1 \\
     AlCoCuLaNi &      1 \\
\bottomrule
\end{tabular}
\end{table}

\begin{longtable}[H]{lr}
\caption{The chemical systems considered for CSLO CV for 177 materials.}
\label{mtp_chemical_systems_177}\\
\toprule
Chemical System &  Count \\
\midrule
\endfirsthead
\caption[]{The chemical systems considered for CSLO CV for 177 materials.} \\
\toprule
Chemical System &  Count \\
\midrule
\endhead
\midrule
\multicolumn{2}{r}{{Continued on next page}} \\
\midrule
\endfoot

\bottomrule
\endlastfoot
        CuNiPPd &     15 \\
       AlCuNiZr &     15 \\
     BeCuNiTiZr &     11 \\
         MgNdNi &      7 \\
         CuTiZr &      6 \\
           CuZr &      5 \\
         CuPdSi &      4 \\
         CuGdMg &      4 \\
         CaMgZn &      4 \\
            BFe &      4 \\
         AlCuZr &      4 \\
     AlCuNiTiZr &      4 \\
       AlCuLaNi &      4 \\
           PdSi &      3 \\
           NiZr &      3 \\
           NbNi &      3 \\
    BCCoCrFeMoY &      3 \\
         AlFeNd &      3 \\
     AlCoCuLaNi &      3 \\
           NiPd &      2 \\
       CuNiTiZr &      2 \\
        CuNiPPt &      2 \\
     AlCuNiPdZr &      2 \\
     AlCuNbNiZr &      2 \\
         AlCuLa &      2 \\
         AlCoZr &      2 \\
     AgAuCuPdSi &      2 \\
              P &      1 \\
       NiSiTiZr &      1 \\
     NiSiSnTiZr &      1 \\
          NiPPt &      1 \\
          NiPPd &      1 \\
   NbNiSiSnTiZr &      1 \\
         GeSbTe &      1 \\
         FeNiZr &      1 \\
        CuPPdPt &      1 \\
          CuPPd &      1 \\
     CuNiSiTiZr &      1 \\
        CuMgYZn &      1 \\
          CuMgY &      1 \\
           CuLa &      1 \\
       CuHfTiZr &      1 \\
        CoCuPPt &      1 \\
           CFeP &      1 \\
           BeZr &      1 \\
           BeTi &      1 \\
     BeCuNbNiZr &      1 \\
            BZr &      1 \\
          BNiSi &      1 \\
          BFeSi &      1 \\
         BFeNiP &      1 \\
          BFeNi &      1 \\
      BFeNbNiSi &      1 \\
          BCoSi &      1 \\
      BCoFeNbSi &      1 \\
   BCoCrFeMoYZr &      1 \\
      BCCrFeMoY &      1 \\
  BCCrFeMnMoSiW &      1 \\
    BCCrFeGaMoP &      1 \\
         AuGeSi &      1 \\
         AlNiZr &      1 \\
           AlMg &      1 \\
         AlLaNi &      1 \\
       AlGdNiZn &      1 \\
       AlGdNiSn &      1 \\
       AlGdMnNi &      1 \\
       AlCuNbZr &      1 \\
       AlCuHfNi &      1 \\
       AlCuGdZr &      1 \\
       AlCuGdNi &      1 \\
         AlCoSm &      1 \\
      AlCoLaNiY &      1 \\
         AlCoGd &      1 \\
      AlCoFeNiY &      1 \\
       AlCoCuZr &      1 \\
     AlCoCuFeSm &      1 \\
         AlCeNi &      1 \\
         AgCuZr &      1 \\
    AgCuMgNiYZn &      1 \\
     AgCuGdMgPd &      1 \\
  AgCuGdMgNiYZn &      1 \\
       AgAlCuZr &      1 \\
     AgAlCuNiZr &      1 \\
\end{longtable}

\section{Shuffled Learning Curve}

The feature learning curve obtained from the shuffling feature selection procedure (see the end
of Sec.~\ref{feature_selection}) is shown in Fig.~\ref{shuffled_selection}. This is evidence that the selection of $X_{best}$ did not randomly produce a capable model for $log_{10}(R_{c})$ prediction.

\begin{figure}[H]
	\centering
	\includegraphics[width=0.5\textwidth]{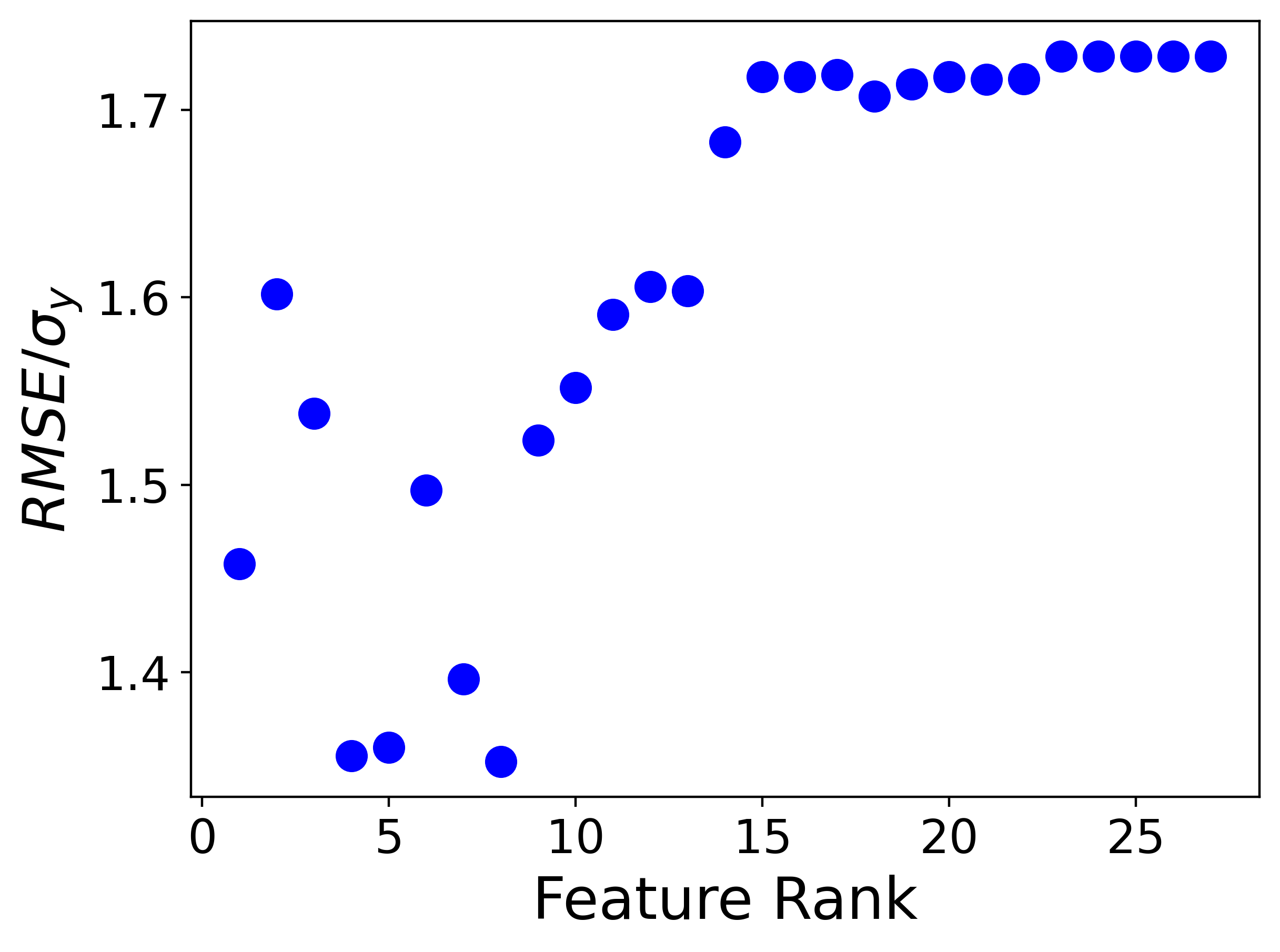}
	\caption{$RMSE/\sigma_{y}$ as a function included features sorted with SHAP values for XGBoost models are shown. This feature selection process was performed with shuffled $log_{10}(R_c)$.}
	\label{shuffled_selection}
\end{figure}

\section{Parity Plots}

We provide the parity plots for several sets of data and assessments in this section. We emphasized the use of CSLO CV throughout the main text because we wanted models that could predict well across chemical systems, and include all relevant data here. However, 5-fold CV (repeated 10 times) is a common way to assess models, so we provide those parity plots as well. Also, because of the small set of data with properties acquired from 34 MLPs, we provide plots for leave one out CV for the cases where only 34 data points are fit. The number of points in these data are small and will lead to poor performance in 5-fold CV.

\subsection{CSLO CV}

\begin{figure}[H]
	\centering
	\includegraphics[width=0.65\textwidth]{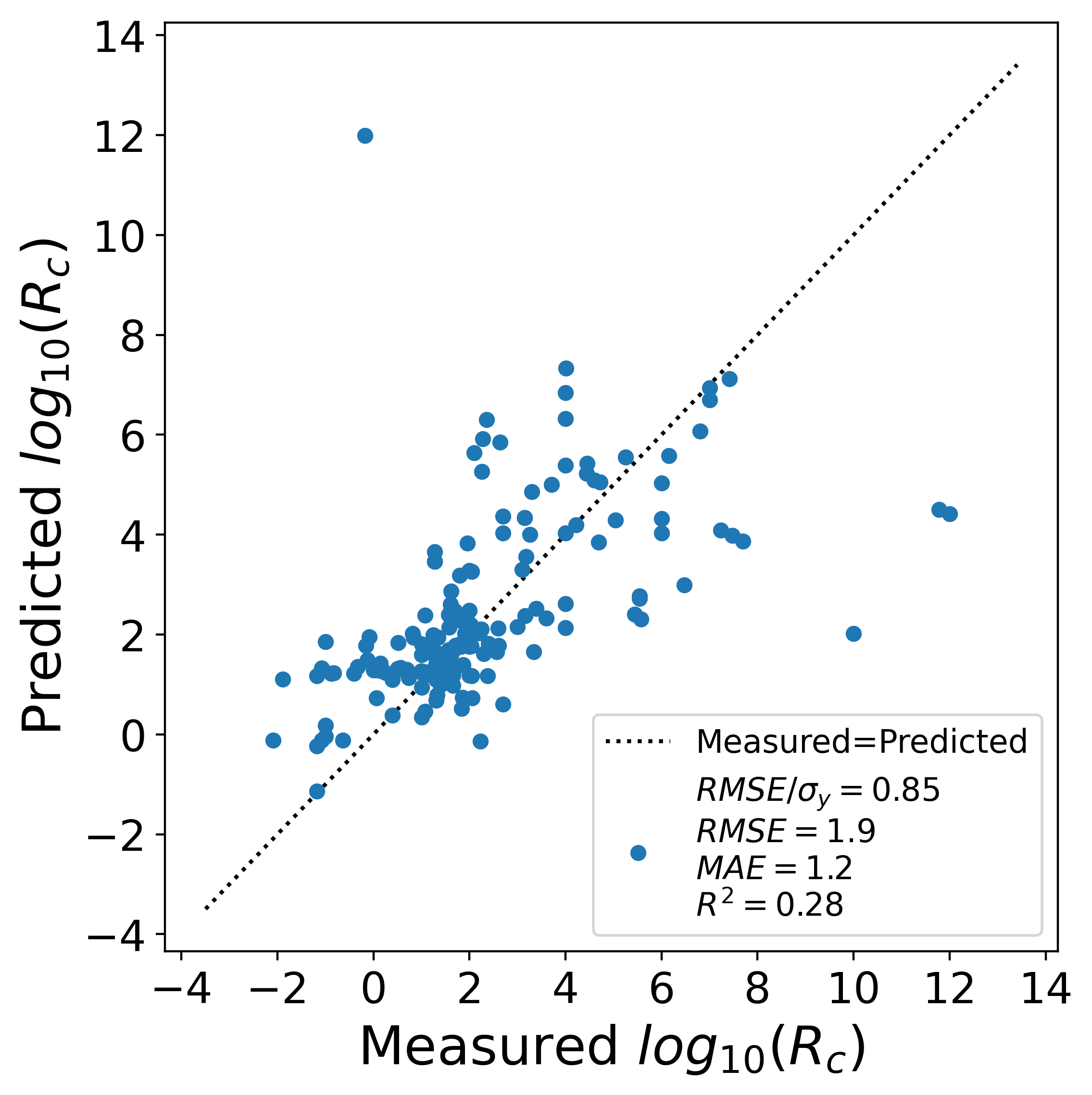}
	\caption{The parity plot for XGBoost models fit to $X_{long}$ for 177 materials. Note that test data were produced by CSLO CV.}
\end{figure}

\begin{figure}[H]
	\centering
	\includegraphics[width=0.65\textwidth]{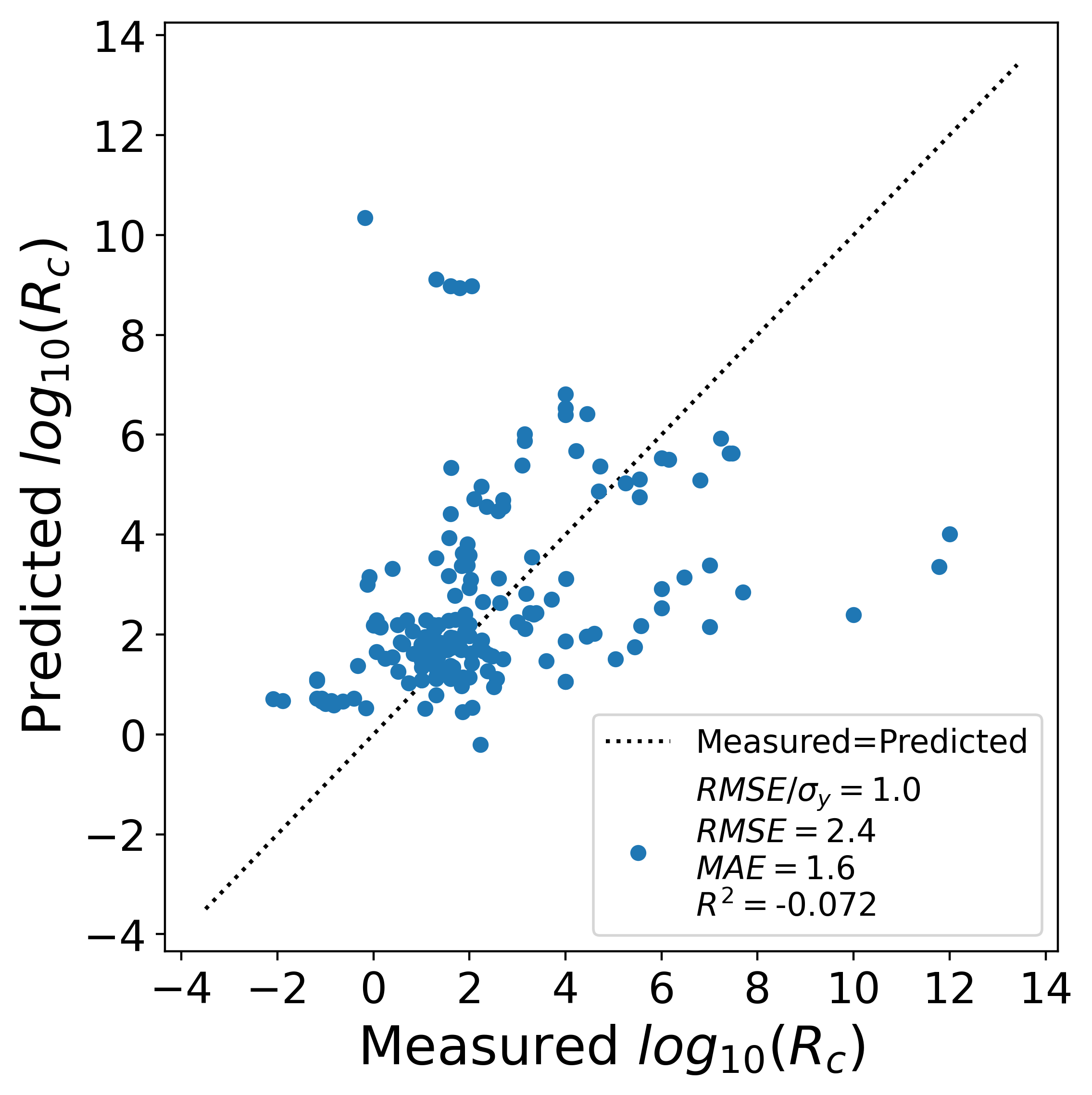}
	\caption{The parity plot for XGBoost models fit to $X_{mastml}$ for 177 materials. Note that test data were produced by CSLO CV.}
\end{figure}

\begin{figure}[H]
	\centering
	\includegraphics[width=0.65\textwidth]{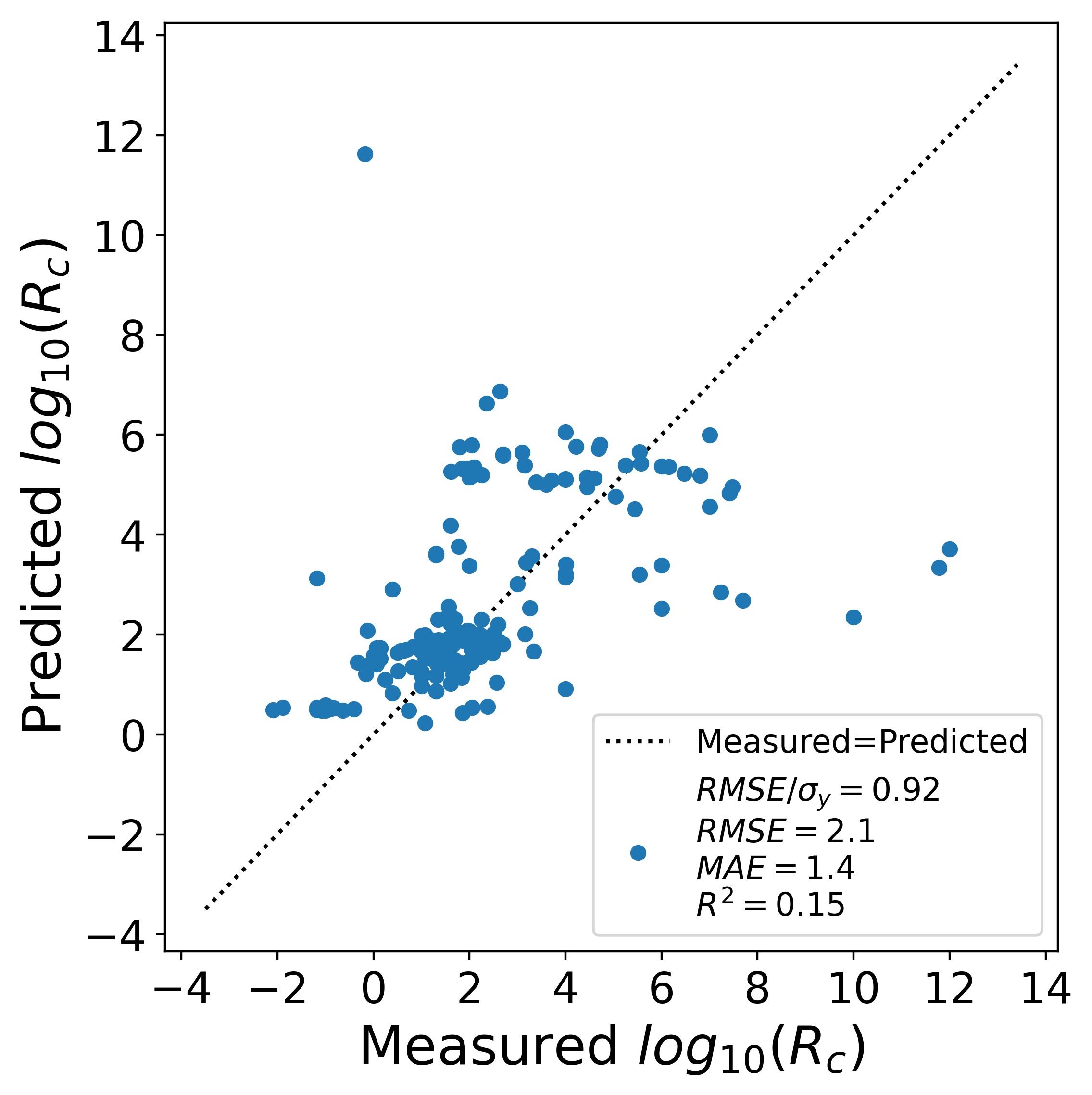}
	\caption{The parity plot for XGBoost models fit to $X_{mastml} \cup X_{long}$ for 177 materials. Note that test data were produced by CSLO CV.}
\end{figure}

\begin{figure}[H]
	\centering
	\includegraphics[width=0.65\textwidth]{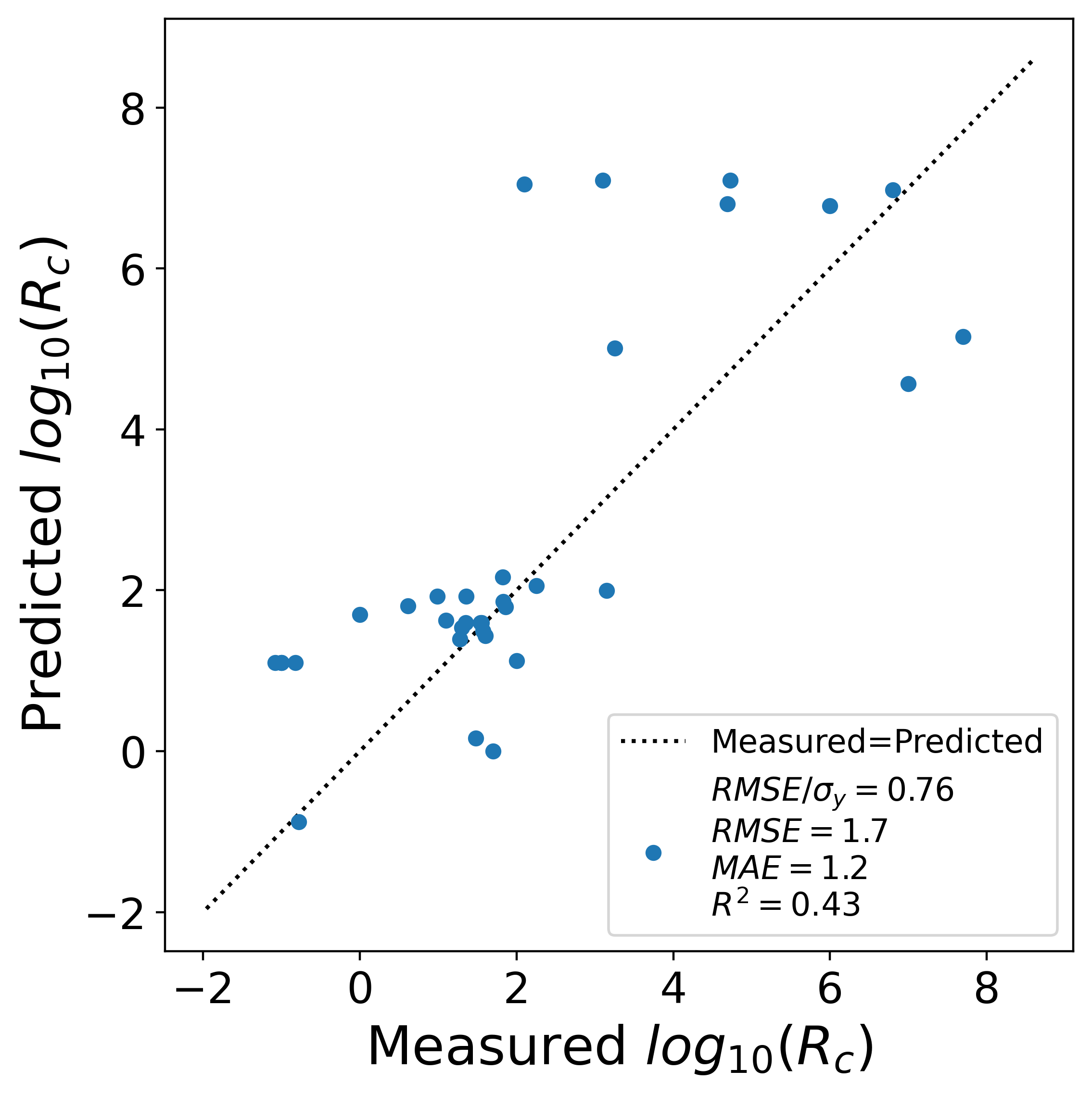}
	\caption{The parity plot for XGBoost models fit to $X_{long}$ for 34 materials. Note that test data were produced by CSLO CV.}
\end{figure}

\begin{figure}[H]
	\centering
	\includegraphics[width=0.65\textwidth]{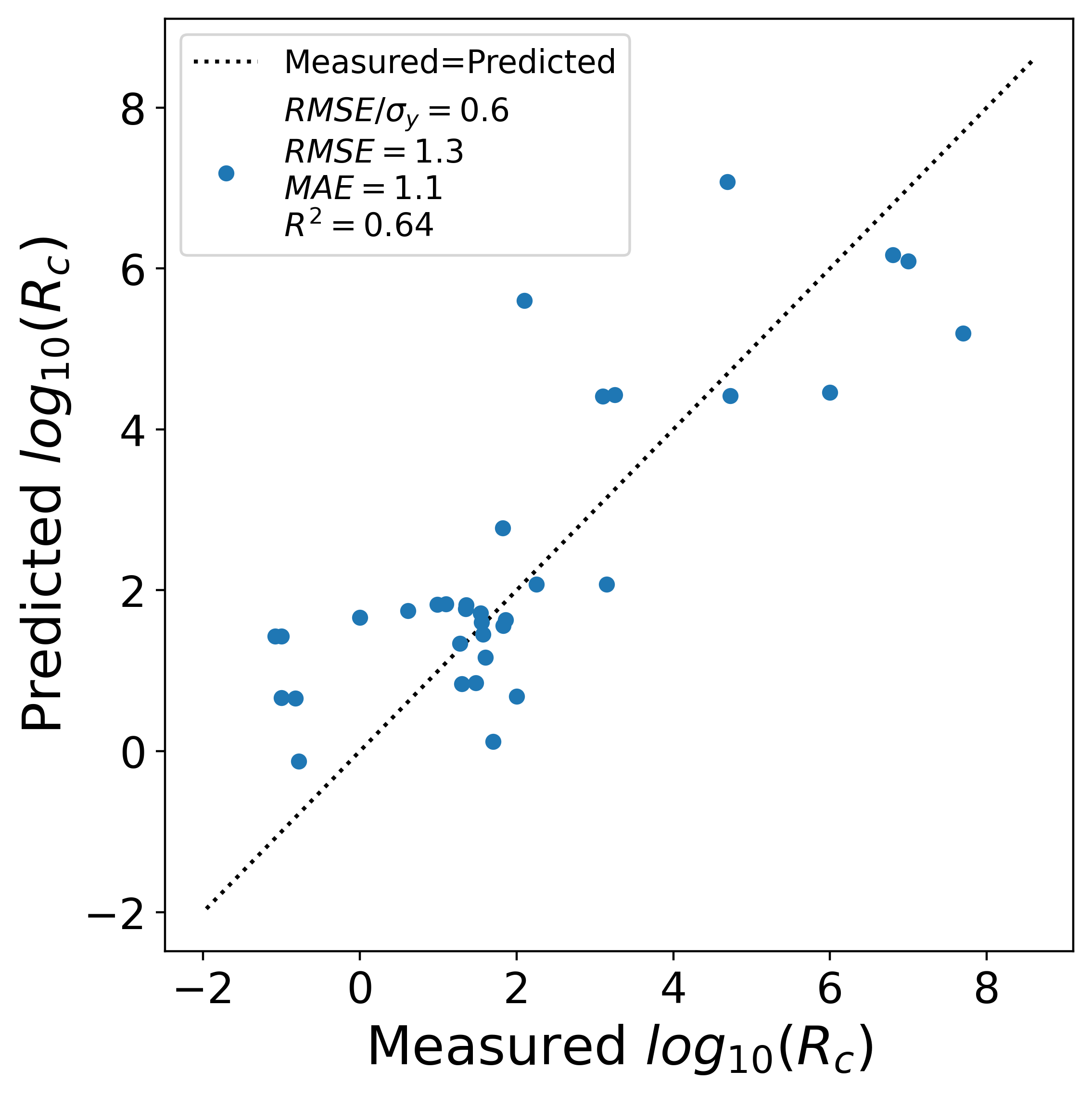}
	\caption{The parity plot for XGBoost models fit to $X_{tot}$ for 34 materials. Note that test data were produced by CSLO CV.}
\end{figure}

\subsection{5-Fold CV Repeated 10 Times}

\begin{figure}[H]
	\centering
	\includegraphics[width=0.65\textwidth]{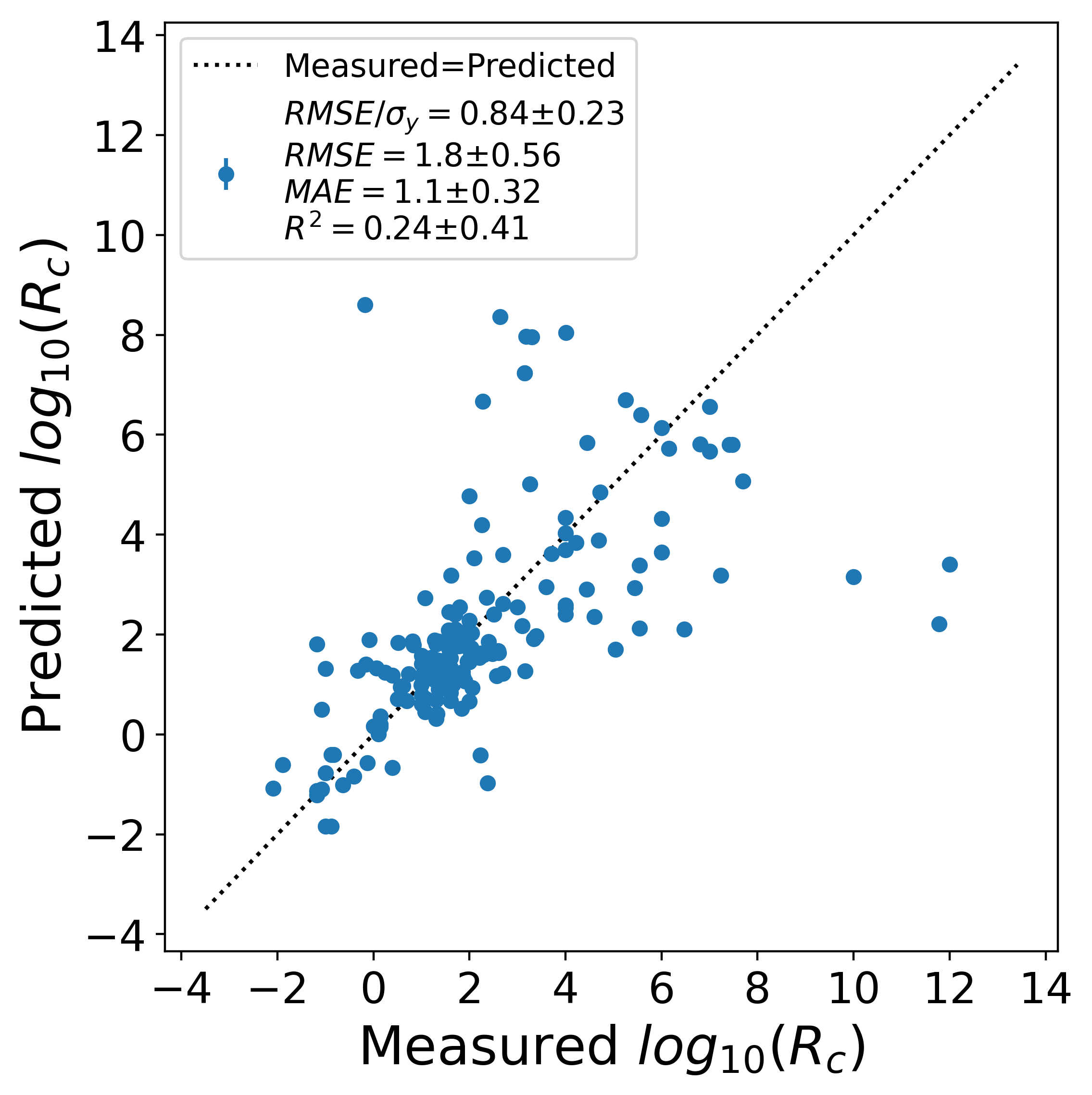}
	\caption{The parity plot for XGBoost models fit to $X_{long}$ for 177 materials. Note that test data were produced by 5-fold CV repeated 10 times.}
\end{figure}

\begin{figure}[H]
	\centering
	\includegraphics[width=0.65\textwidth]{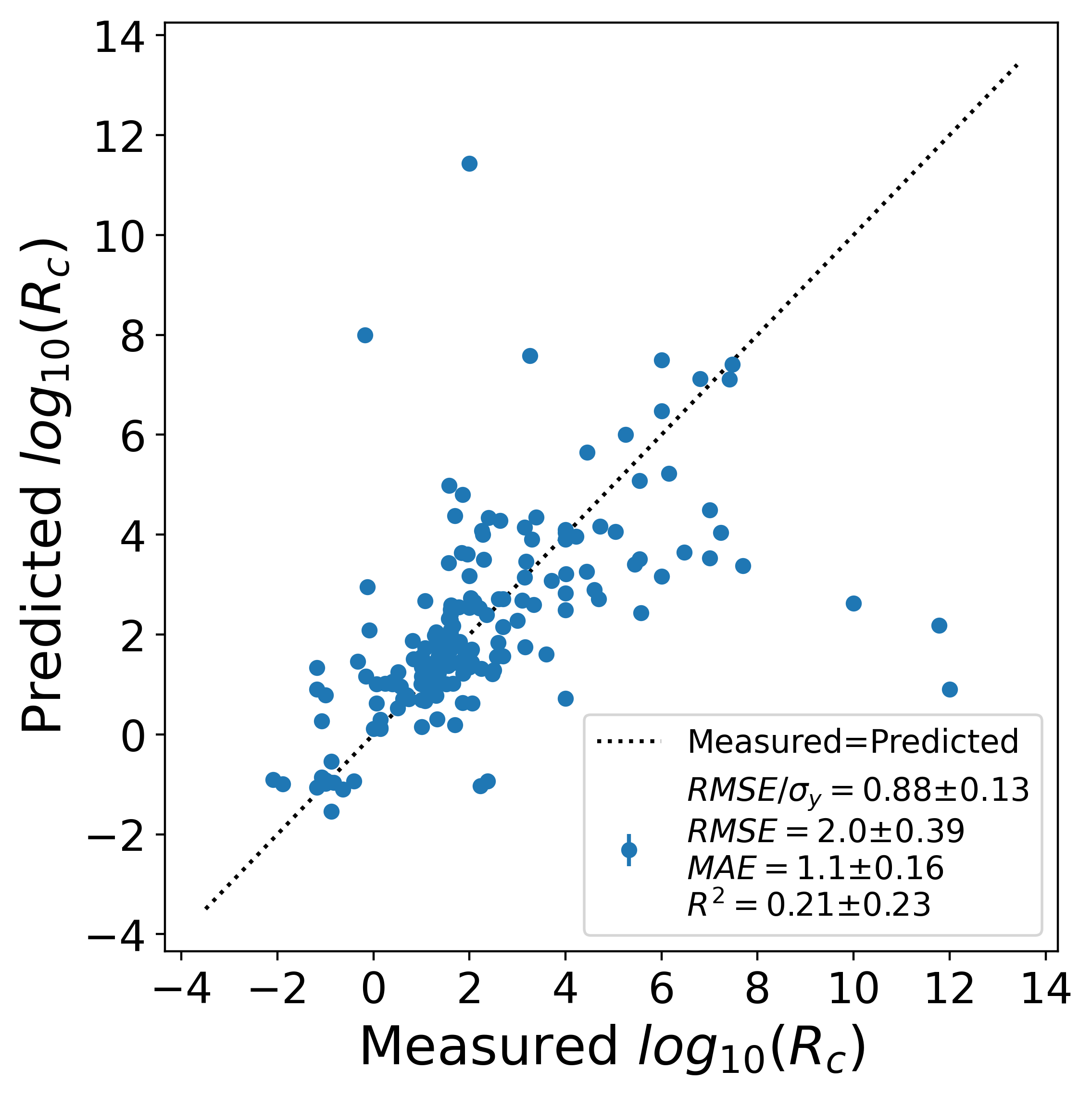}
	\caption{The parity plot for XGBoost models fit to $X_{mastml}$ for 177 materials. Note that test data were produced by 5-fold CV repeated 10 times.}
\end{figure}

\begin{figure}[H]
	\centering
	\includegraphics[width=0.65\textwidth]{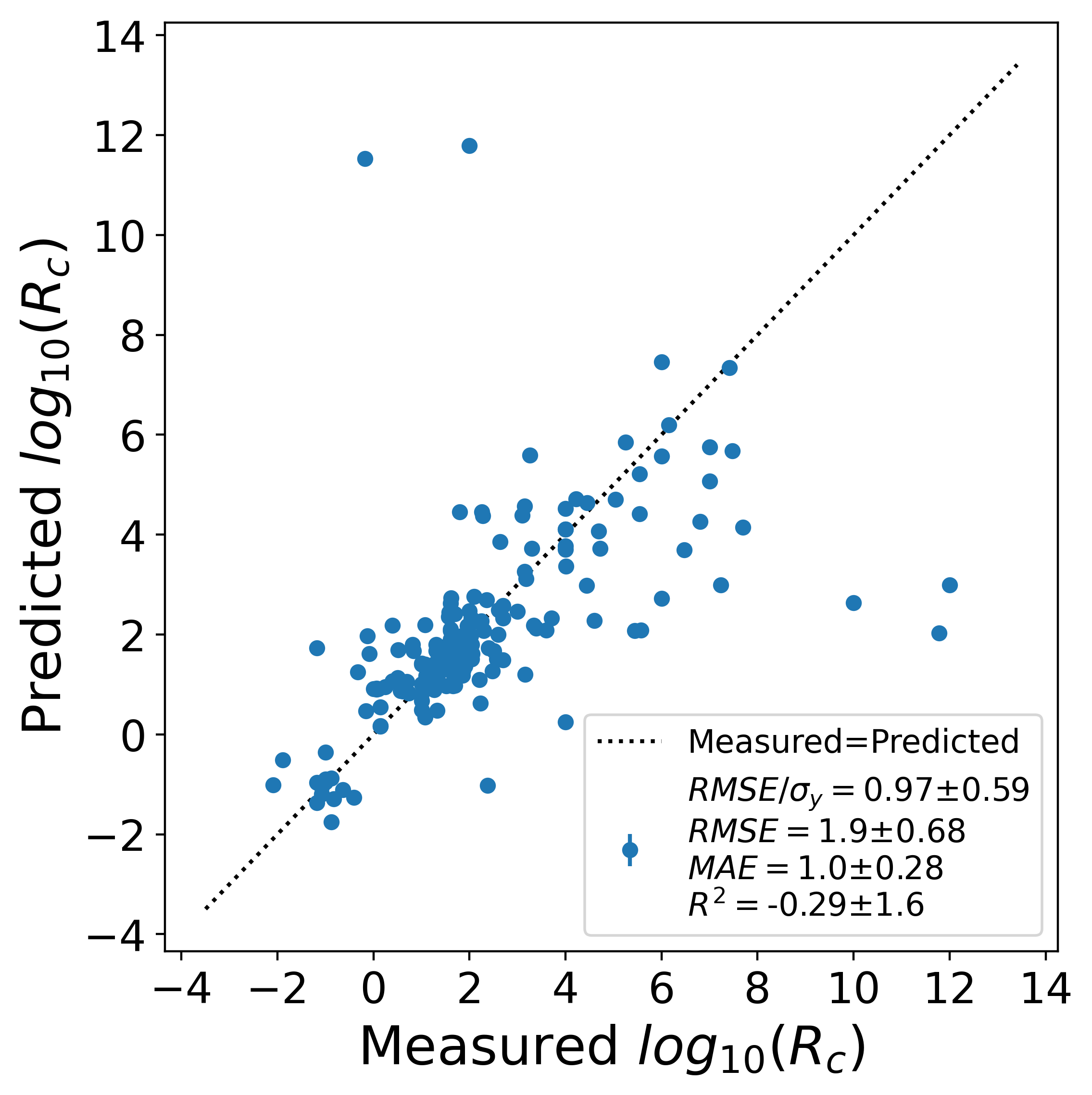}
	\caption{The parity plot for XGBoost models fit to $X_{mastml} \cup X_{long}$ for 177 materials. Note that test data were produced by 5-fold CV repeated 10 times.}
\end{figure}

\begin{figure}[H]
	\centering
	\includegraphics[width=0.65\textwidth]{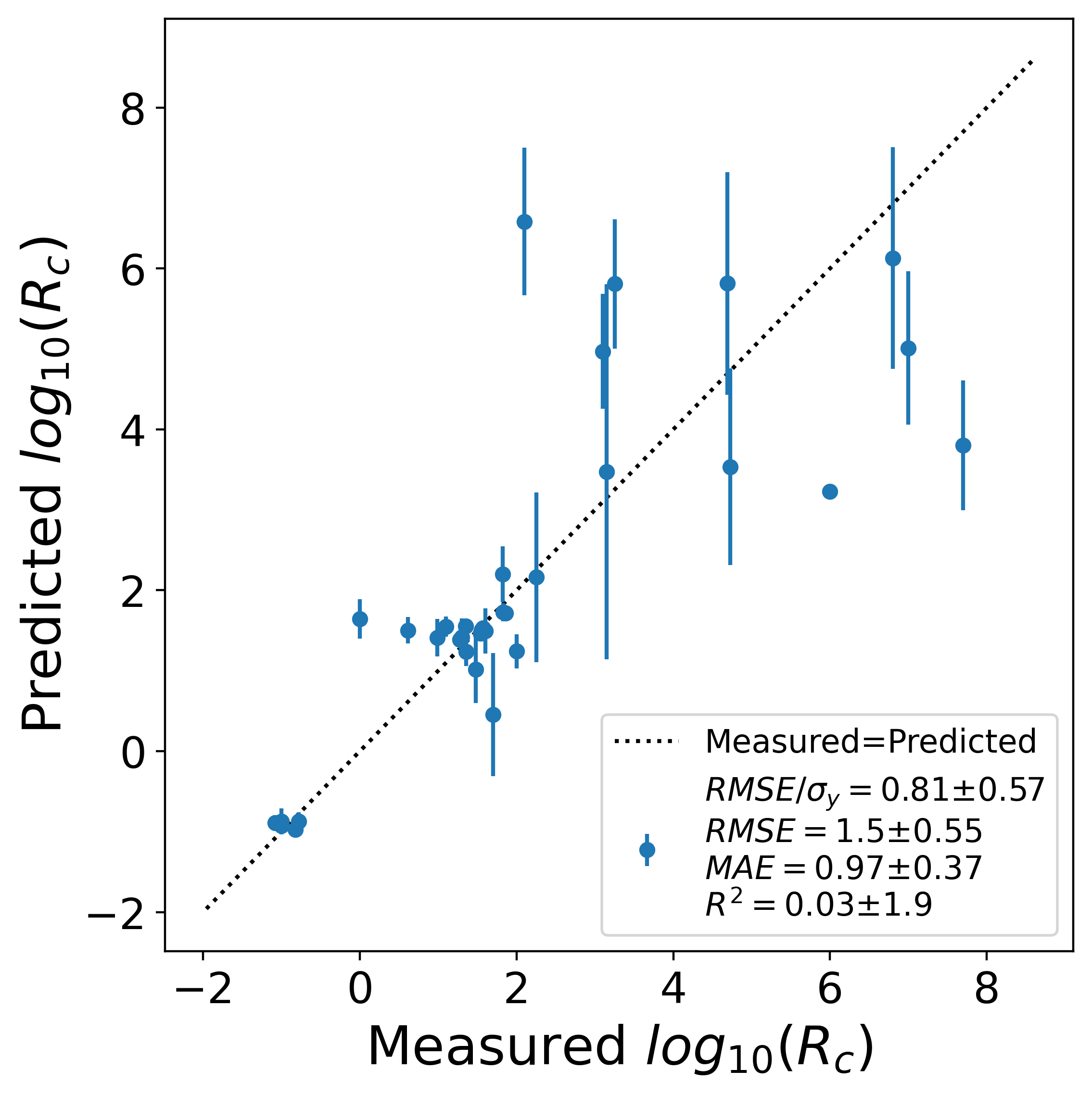}
	\caption{The parity plot for XGBoost models fit to $X_{long}$ for 34 (materials with a built MLP). Note that test data were produced by 5-fold CV repeated 10 times.}
\end{figure}

\begin{figure}[H]
	\centering
	\includegraphics[width=0.65\textwidth]{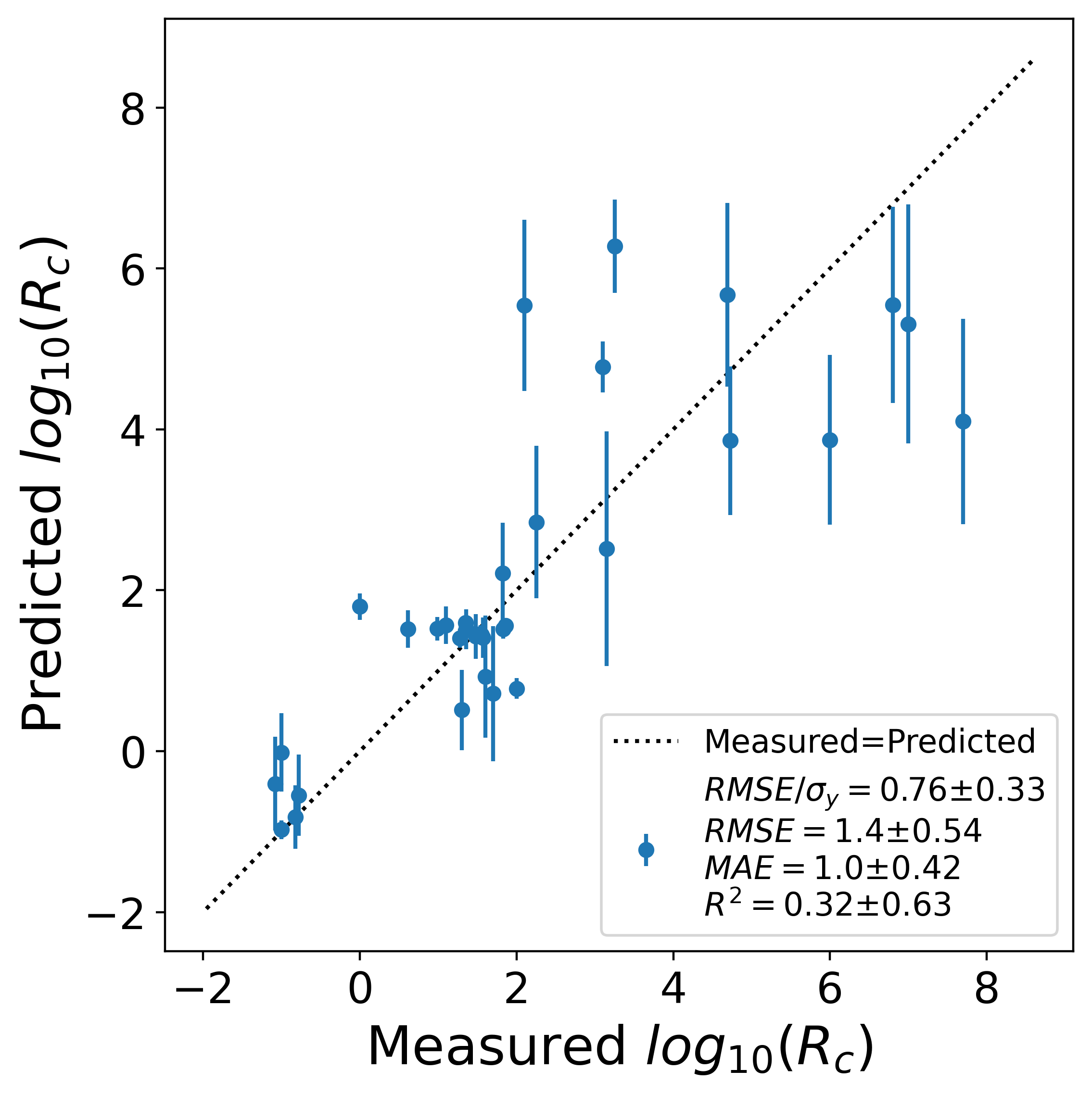}
	\caption{The parity plot for XGBoost models fit to $X_{tot}$ for 34 (materials with a built MLP). Note that test data were produced by 5-fold CV repeated 10 times.}
\end{figure}

\begin{figure}[H]
	\centering
	\includegraphics[width=0.65\textwidth]{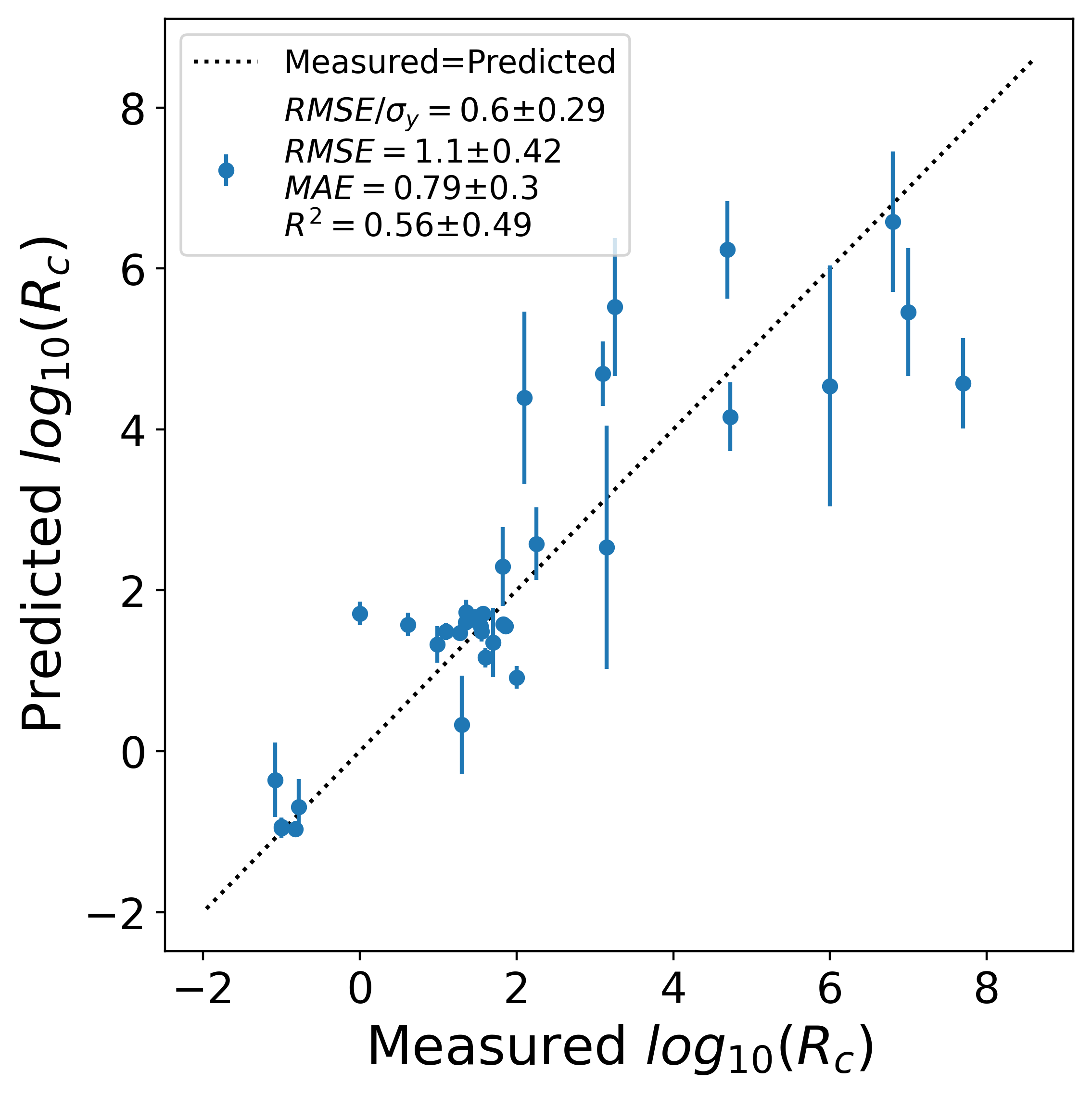}
	\caption{The parity plot for XGBoost models fit to $X_{best}$ for 34 (materials with a built MLP). Note that test data were produced by 5-fold CV repeated 10 times.}
\end{figure}

\subsection{Leave One Out CV}

\begin{figure}[H]
	\centering
	\includegraphics[width=0.65\textwidth]{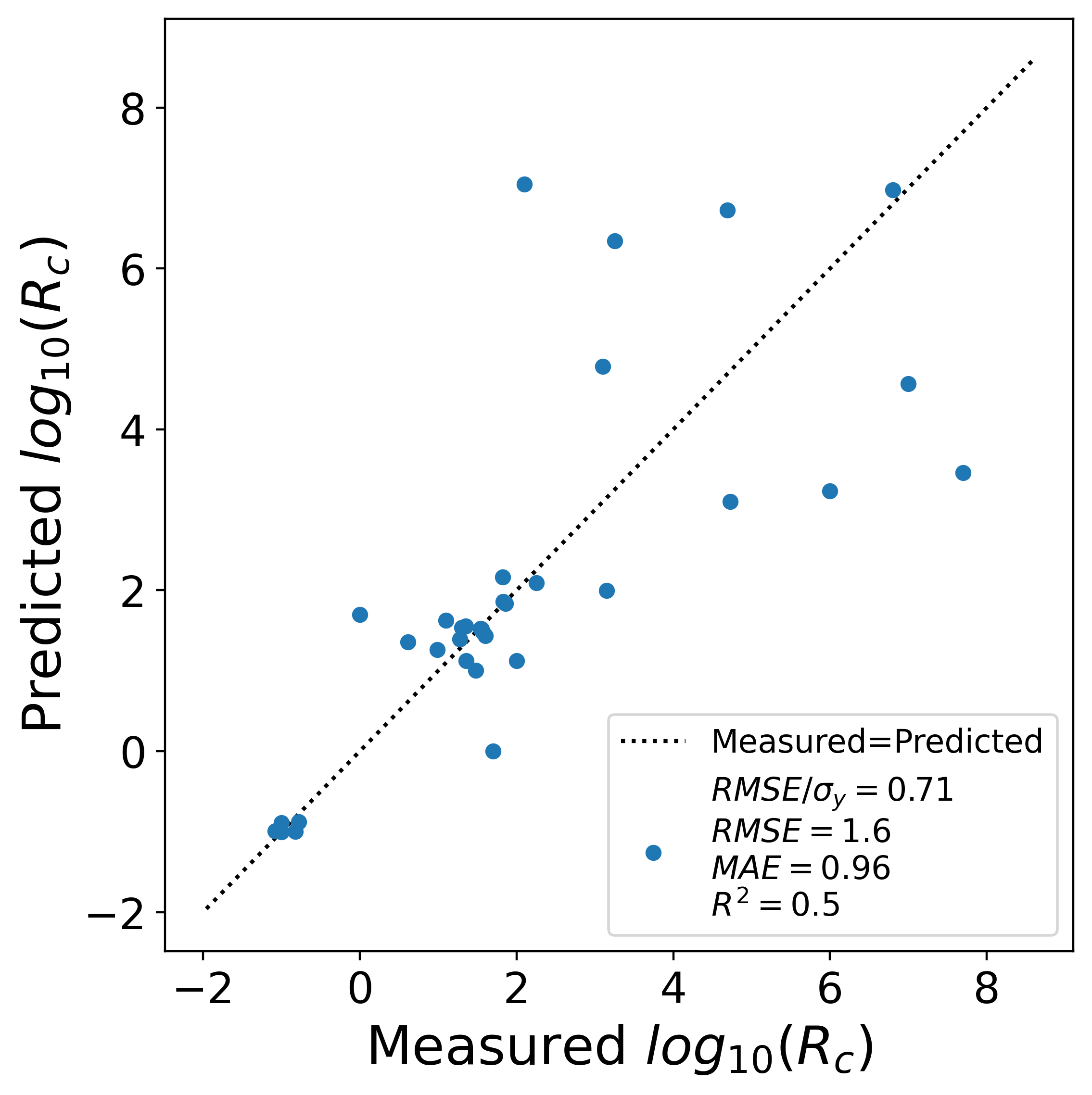}
	\caption{The parity plot for XGBoost models fit to $X_{long}$ for 34 (materials with a built MLP). Note that test data were produced by Leave One Out CV.}
\end{figure}

\begin{figure}[H]
	\centering
	\includegraphics[width=0.65\textwidth]{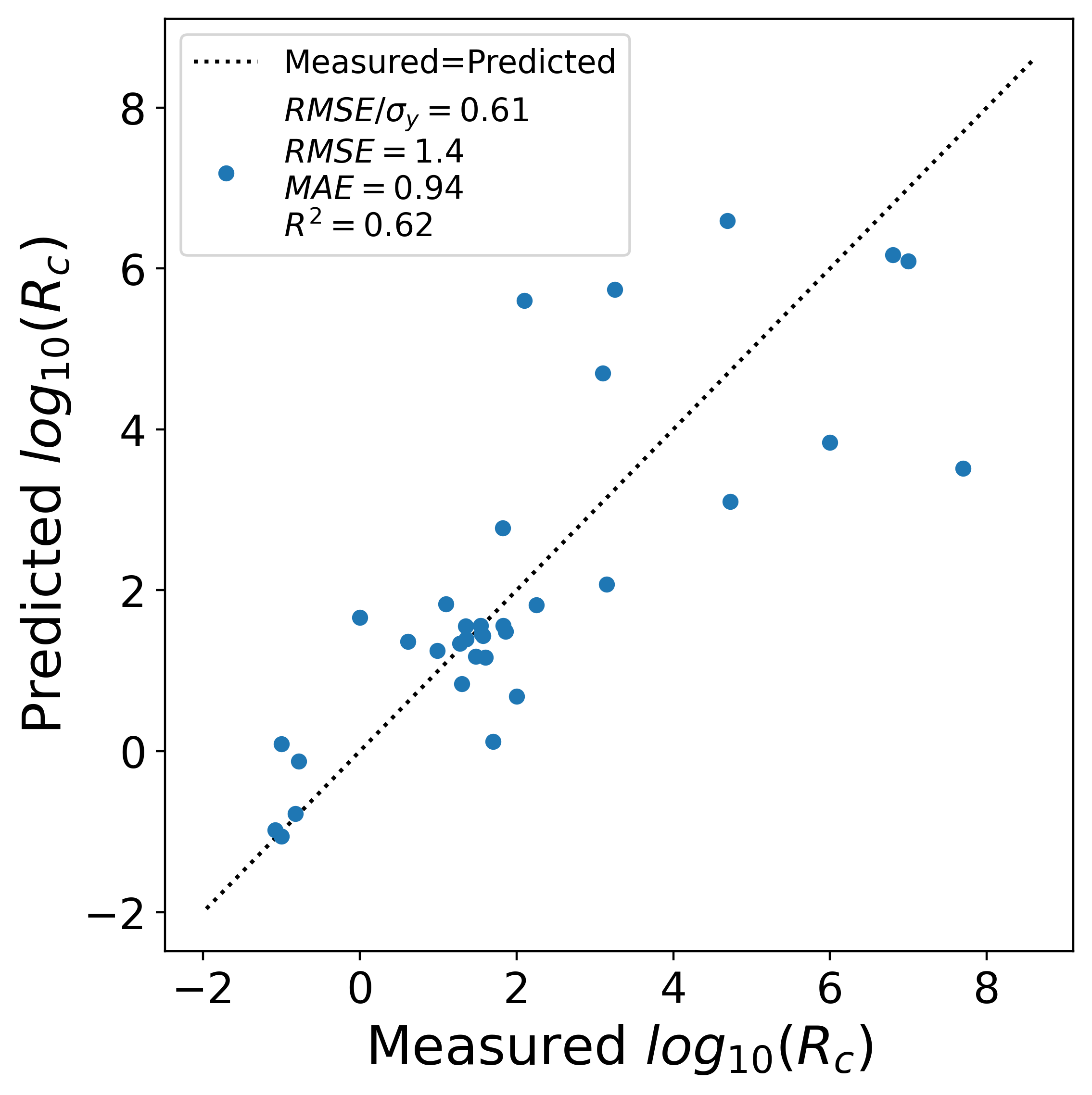}
	\caption{The parity plot for XGBoost models fit to $X_{tot}$ for 34 (materials with a built MLP). Note that test data were produced by Leave One Out CV.}
\end{figure}

\begin{figure}[H]
	\centering
	\includegraphics[width=0.65\textwidth]{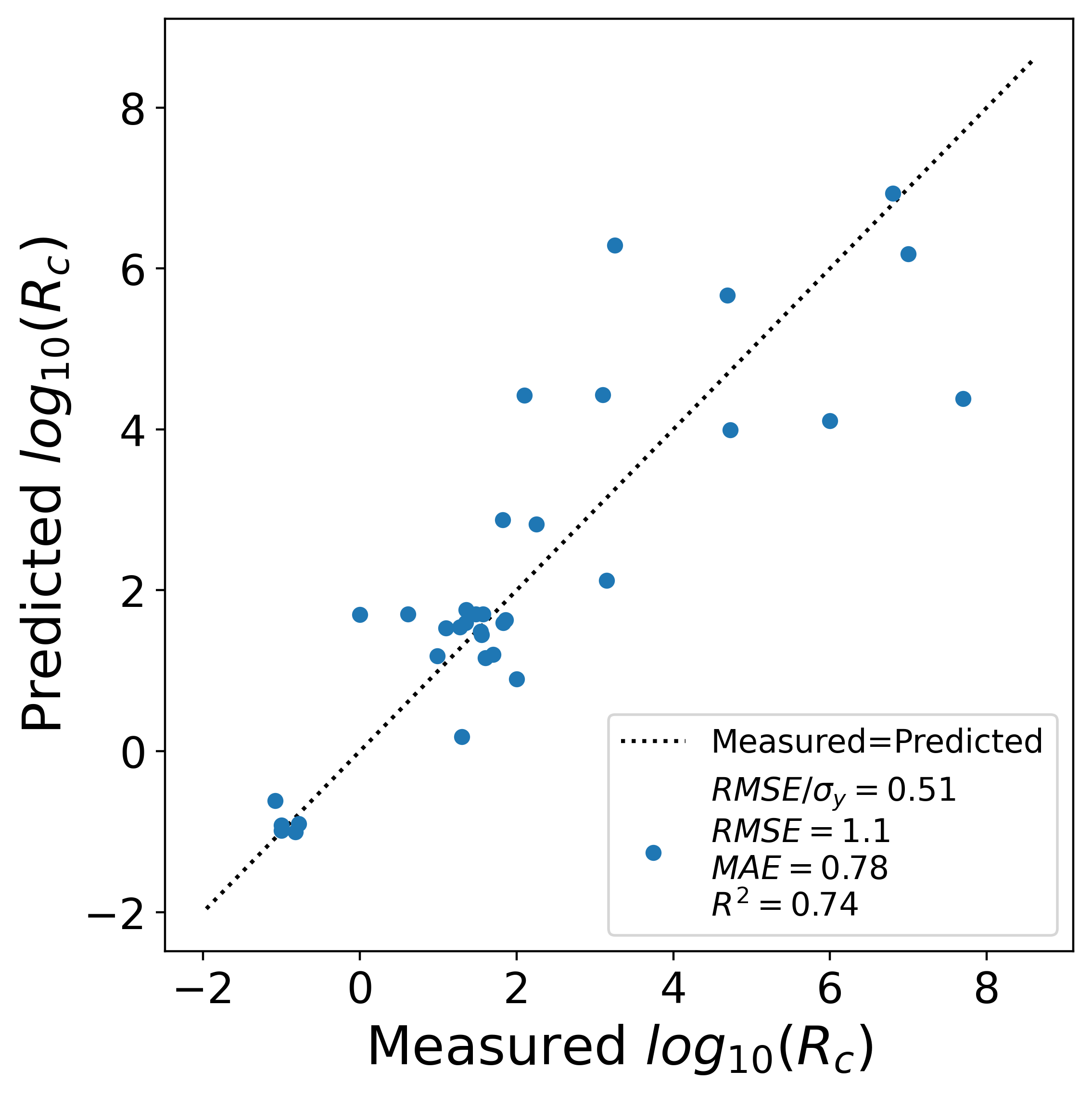}
	\caption{The parity plot for XGBoost models fit to $X_{best}$ for 34 (materials with a built MLP). Note that test data were produced by Leave One Out CV.}
\end{figure}

\end{document}